\documentclass{emulateapj}
\usepackage{subfigure}

\def\lsim{\mathrel{\rlap{\lower4pt\hbox{\hskip1pt$\sim$}}
    \raise1pt\hbox{$<$}}}                
\def\gsim{\mathrel{\rlap{\lower4pt\hbox{\hskip1pt$\sim$}}
    \raise1pt\hbox{$>$}}}                

\def\Kepler{\textit{Kepler}}

\shorttitle{Robo-AO Imaging of 715 Kepler Exoplanet Candidates}
\shortauthors{N.M. Law et al.}
\begin{document}
\title{Robotic Laser-Adaptive-Optics Imaging of 715 Kepler Exoplanet Candidates using Robo-AO}

\author{Nicholas M. Law\altaffilmark{1}, Tim Morton\altaffilmark{2}, Christoph Baranec\altaffilmark{3}, Reed Riddle\altaffilmark{2}, Ganesh Ravichandran\altaffilmark{4}, Carl Ziegler\altaffilmark{1}, John Asher Johnson\altaffilmark{5}, Shriharsh P. Tendulkar\altaffilmark{2}, Khanh Bui\altaffilmark{2}, Mahesh P. Burse\altaffilmark{6}, H. K. Das\altaffilmark{6}, Richard G. Dekany\altaffilmark{2}, Shrinivas Kulkarni\altaffilmark{2}, Sujit Punnadi\altaffilmark{2}, A. N. Ramaprakash\altaffilmark{6}}

\altaffiltext{1}{Department of Physics and Astronomy, University of North Carolina at Chapel Hill, Chapel Hill, NC 27599-3255, USA}
\altaffiltext{2}{Division of Physics, Mathematics, and Astronomy, California Institute of Technology, Pasadena, CA, 91125, USA}
\altaffiltext{3}{Institute for Astronomy, University of Hawai$\textquoteleft$i at M\={a}noa, Hilo, HI 96720-2700, USA}
\altaffiltext{4}{W. Tresper Clarke High School, East Meadow School District, 740 Edgewood Dr  Westbury, NY 11590, USA}
\altaffiltext{5}{Harvard-Smithsonian Center for Astrophysics, 60 Garden St., Cambridge, MA 02138}
\altaffiltext{6}{Inter-University Centre for Astronomy \& Astrophysics, Ganeshkhind, Pune, 411007, India}

\begin{abstract}
The Robo-AO \Kepler\, Planetary Candidate Survey is observing every \Kepler\, planet candidate host star with laser adaptive optics imaging to search for blended nearby stars, which may be physically associated companions and/or responsible for transit false positives. In this paper we present the results from the 2012 observing season, searching for stars close to 715 \Kepler\, planet candidate hosts. We find 53 companions, 43 of which are new discoveries. We detail the Robo-AO survey data reduction methods including a method of using the large ensemble of target observations as mutual point-spread-function references, along with a new automated companion-detection algorithm designed for large adaptive optics surveys. Our survey is sensitive to objects from $\approx$0\farcs15 to 2\farcs5 separation, with magnitude differences up to $\rm\Delta m\approx6$. We measure an overall nearby-star-probability for \Kepler\, planet candidates of 7.4\%$\pm$1.0\%, and calculate the effects of each detected nearby star on the \Kepler-measured planetary radius. We discuss several KOIs of particular interest, including KOI-191 and KOI-1151, which are both multi-planet systems with detected stellar companions whose unusual planetary system architecture might be best explained if they are ``coincident multiple'' systems, with several transiting planets shared between the two stars. Finally, we find 98\%-confidence evidence that short-period giant planets are 2-3$\times$ more likely than longer period planets to be found in wide stellar binaries.

\end{abstract}


\maketitle

\section{Introduction}

The \Kepler\ mission, which has searched approximately 190,000 stars for the tiny periodic dips in stellar brightness indicative of transiting planets, is unprecedented in both sensitivity and scale among transiting planet surveys \citep{Koch2010}.  Never before has a survey been able to detect such small planets---down to even the size of the Earth's moon \citep{barclay2013}---and never before has a survey delivered so many planet candidates, with over 3500 planet candidates (candidate \Kepler\ Objects of Interest; KOIs) found in a search of the first twelve quarters of \Kepler\ photometry (\citealt{Borucki2010, Borucki2011, Batalha2012, Tenenbaum2013}). 

All exoplanet transit surveys require follow-up observations of the detected candidates.  The purpose of this follow-up is twofold: first to confirm that the detected photometric dimmings are in fact truly transiting planets rather than astrophysical false positives; and second to characterize the host stellar system.  High-angular-resolution imaging is a crucial ingredient of the follow-up effort, as many astrophysical false positive scenarios involve nearby stellar systems whose light is blended with the target star (e.g. \citealt{Odonovan2006}).  Even if a transit candidate is a true planet, identifying whether it is in a binary stellar system has potentially important implications for determining the planet's detailed properties. For example, if there is considerable diluting flux from a companion star within the photometric aperture, even if the planet interpretation of the signal is secure, the planet will be larger than implied by the light curve alone under the assumption of a single host star (e.g. \citealt{Johnson2011}).  The presence or absence of third bodies in the systems can also have broader implications about the processes of planetary system formation and evolution; stellar binarity has been hypothesized to be important in shaping the architectures of planetary systems, both by regulating planet formation and by dynamically sculpting planets’ final orbits, such as forcing Kozai oscillations that cause planet migration (e.g. \citealt{Fabrycky2007, Katz2011, Naoz2012}) or tilting the circumstellar disk \citep{Batygin2012}.

The vast majority of the individual \Kepler\ candidates remain unconfirmed ($<$3\% currently confirmed according to the NASA Exoplanet Archive [NEA]). Current predictions based on models of the expected population of confusion sources suggest that at least 10-15\% of \Kepler's planetary candidates may be astrophysical false positives and that a large fraction of confirmed planets also have incorrectly determined planetary parameters because of confusing sources \citep{Morton2011, Fressin2013, Dressing2013, Santerne2013}. The possible false-positive scenario probabilities change with the brightness of the \Kepler\ target, the details of its \Kepler\, light curve, its spectral type, and the properties of the detected planetary system (e.g. \citealt{Morton2012}). The false positives thus limit our ability to interpret individual objects, to evaluate differences in planetary statistics between different stellar populations, and to generate fully robust statistical studies of the planetary population seen by \Kepler.

In order to fully validate the individual \Kepler\ planets and search for correlations between planetary systems and stellar multiplicity properties, we need to search for companions around every \Kepler\ Object of Interest. There have been several high-angular-resolution surveys of selected samples of KOIs to detect stellar companions and assess the false-positive probability \citep{Adams2012, LilloBox2012, Horch2012, Adams2013, Marcy2014}. However, many of these surveys are performed with adaptive optics systems, and the overheads typically associated with ground-based adaptive optics imaging have limited the number of targets which can be observed.

In this paper we present the first results from a laser-adaptive-optics survey that is taking short snapshot high-angular-resolution images of every \Kepler\, planet candidate. The survey uses Robo-AO, the first robotic laser adaptive optics system \citep{Baranec:12, Baranec:13}. We designed the automated system for relatively high time-efficiency, allowing the \Kepler\ target list to be completed in $\sim$36 hours of observing time. 

This paper presents the 2012-observing-season results of the ongoing Robo-AO KOI survey, covering 715 targets and finding 53 companions\footnote{for brevity we denote stars which we found within our detection radius of KOIs as ``companions'', in the sense that they are asterisms associated on the sky. In \S\ref{sec:discuss} we evaluate the probability that the detected objects are actually physically associated.}, 43 of them new discoveries.

The paper is organized as follows: in \S\ref{sec:obs} we describe the Robo-AO system and the KOI survey target selection and observations. \S\ref{sec:pipeline} describes the Robo-AO data reduction and companion-detection pipeline. In \S\ref{sec:discoveries} we describe the survey's results, including the discovered companions. We discuss the results in \S\ref{sec:discuss}, including detailing the effects of the survey's discoveries on the interpretation and veracity of the observed KOIs, and a brief discussion of the \Kepler\ planet candidates' overall binarity statistics. We conclude in \S\ref{sec:concs}.

\section{Survey targets \& observations}
\label{sec:obs}

\subsection{Target selection}

We selected targets from the \Kepler\ Objects of Interest (KOIs) catalog based on a Q1-Q6 \textit{Kepler} data search \citep{Batalha2012}. Our initial targets were selected randomly from the Q1-Q6 KOIs, requiring only that the targets are brighter than $\rm m_i=16.0$, a restriction which removed only 2\% of the KOIs. While it is our intent to observe every KOI with Robo-AO, this initial target selection provides a wide coverage of the range of KOI properties. Given Robo-AO's low time overheads, we took the time to re-observe KOIs which already had detected companions, to produce a complete and homogenous survey.

\begin{figure*}
    \subfigure{\resizebox{\columnwidth}{!}{{\includegraphics{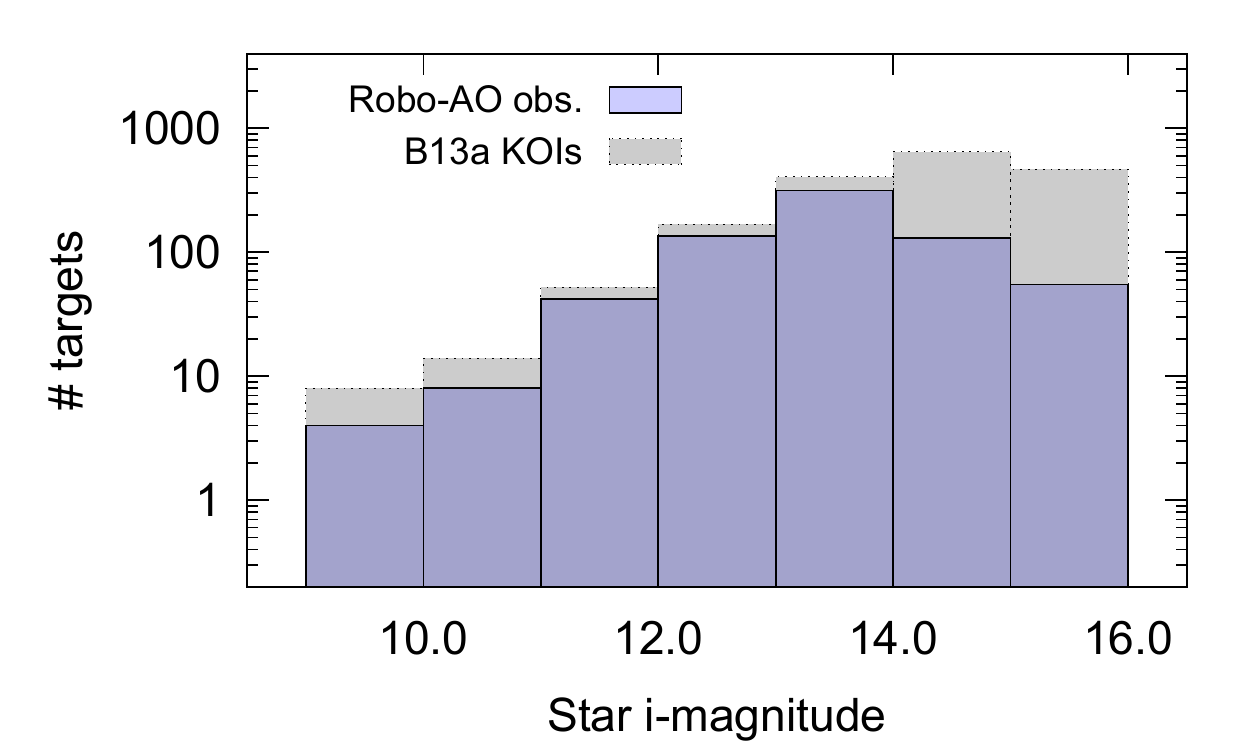}}}}\hspace{0.15in}
    \subfigure{\resizebox{\columnwidth}{!}{{\includegraphics{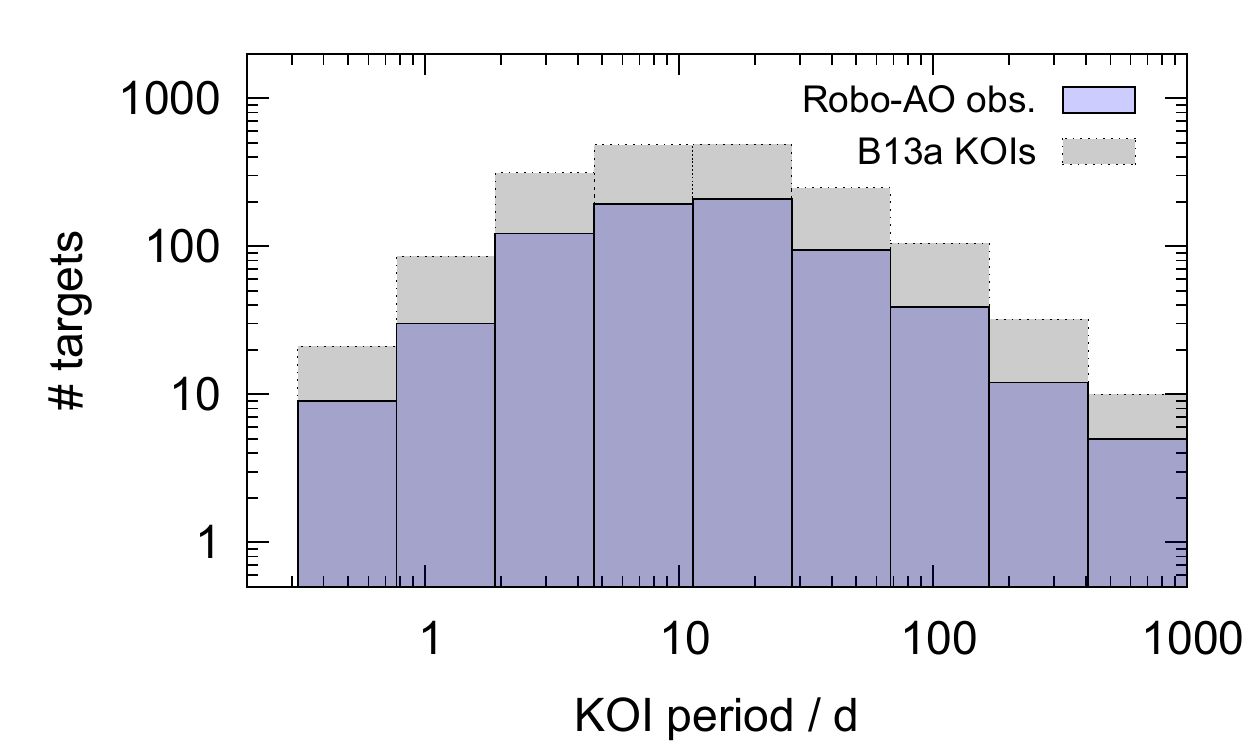}}}}\hspace{0.15in}
    \subfigure{\resizebox{\columnwidth}{!}{{\includegraphics{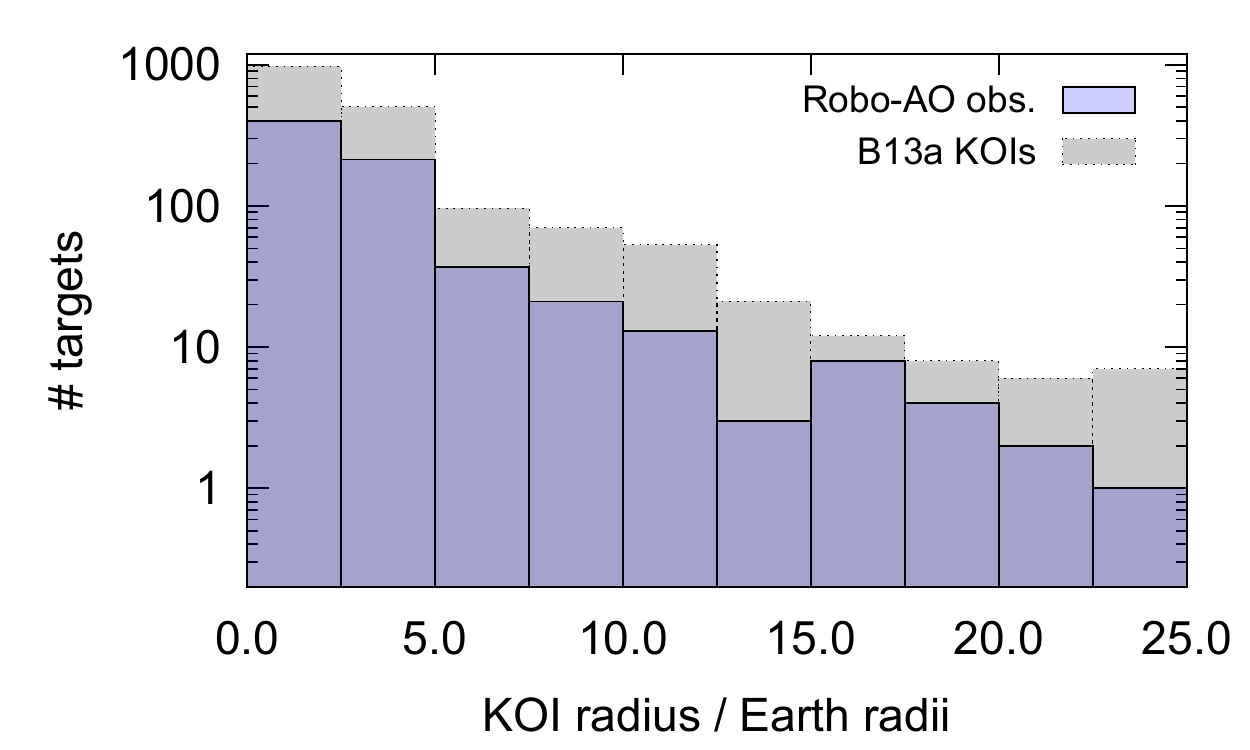}}}}\hspace{0.15in}
    \subfigure{\resizebox{\columnwidth}{!}{{\includegraphics{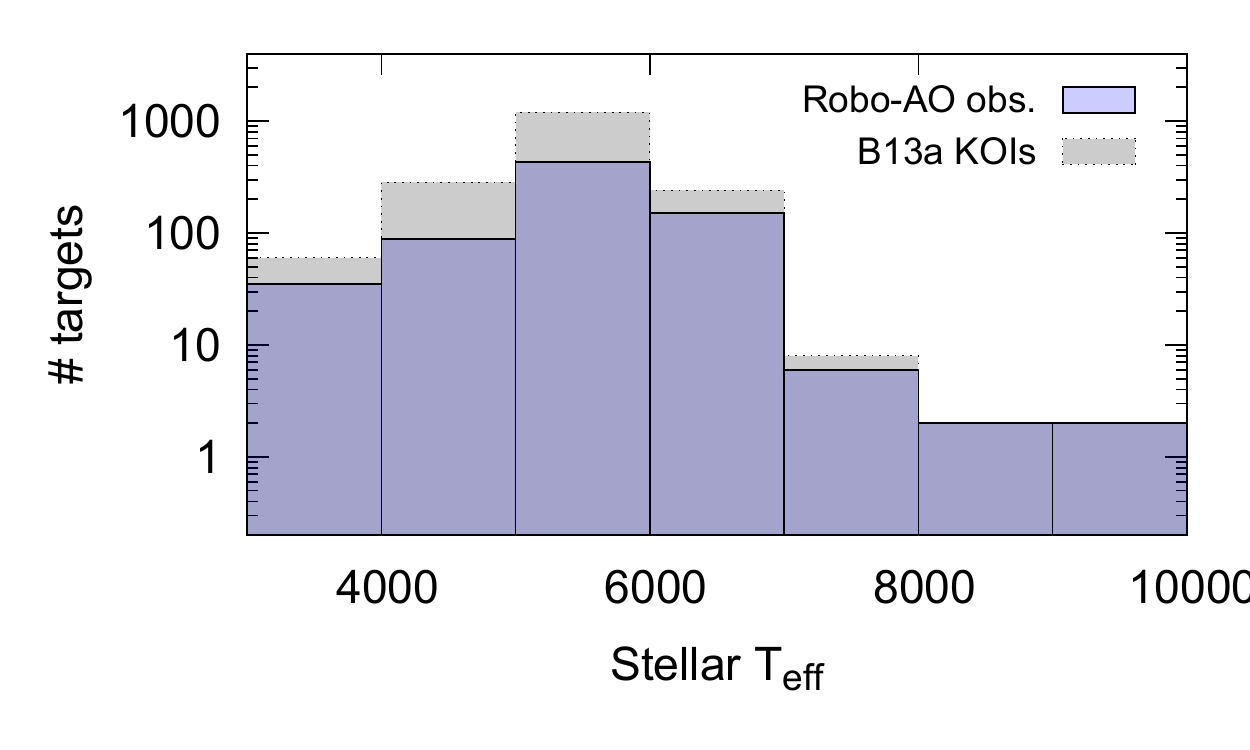}}}}\hspace{0.15in}

     \caption{The distribution of the Robo-AO sample compared to the B13a (\citealt{Batalha2012}) KOIs.}
   \label{fig:target_pops}
\end{figure*}

In Figure \ref{fig:target_pops} we compare the Robo-AO imaged KOIs to the distribution of all \citet{Batalha2012} KOIs in magnitude, planetary period, planetary radius and stellar temperature. The Robo-AO list closely follows the KOI list in the range of magnitude covered, with the exception of the three brightest stars (which have already been covered in detail by other non-laser adaptive optics systems), and a reduced coverage of the faintest KOIs, which Robo-AO requires excellent weather conditions to reach. Robo-AO's target distribution closely matches the full KOI list in planetary radius, planetary orbital period, and stellar temperature. 

\subsection{Observations}

We obtained high-angular-resolution images of the 715 \Kepler\ targeted planet candidate host stars in summer 2012. We performed all the observations in a queue-scheduled mode with the Robo-AO laser adaptive optics system \citep{Baranec:12, Baranec:13, Riddle2012} mounted on the robotic Palomar 60-inch telescope \citep{Cenko:06}. The survey and system specifications are summarized in table \ref{tab:survey_specs}.

\begin{figure}
  \centering
  \resizebox{1.0\columnwidth}{!}
   {
    \includegraphics[angle=-90]{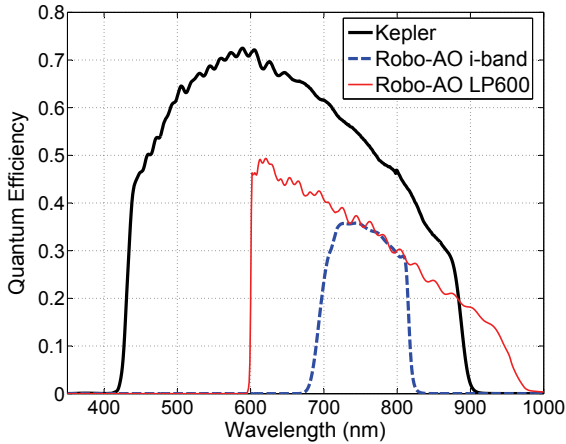}

   }
   \caption{The \Kepler\ and Robo-AO passbands. The Robo-AO curves are generated from measured reflection and transmission data from all optical components with the exception of the primary and secondary of the 60-inch telescope which are assumed to be ideal bare aluminium. The \Kepler\ curve is adapted from the \Kepler\ Instrument Handbook. }

   \label{fig:passbands}
\end{figure}

Robo-AO observed the targets between June 17 2012 and October 6 2012, on 23 separate nights (detailed in the table in the appendix). We chose a standardized 90-second exposure time to provide a snapshot image which would contain all sources likely to affect the \Kepler\ light curve, including close-in sources up to $\sim$5 magnitudes fainter than the \Kepler\ target.  For the observations described here we used either a Sloan i'-band filter \citep{York2000} or a long-pass filter cutting on at 600nm (LP600 hereafter). The latter filter roughly matches the \Kepler\ passband (Figure \ref{fig:passbands}) at the redder wavelengths while suppressing the blue wavelengths which have reduced adaptive optics performance (except in the very best seeing conditions). Compared to near-infrared adaptive optics observations, this filter more closely approximates direct measurement of the effects of unresolved companions on the \Kepler\ light curves.

\begin{deluxetable}{ll}
\tablecaption{\label{tab:survey_specs}The specifications of the Robo-AO KOI survey}
\tabletypesize{\footnotesize}

\startdata
\bf KOI survey specifications &\\
\hline
\noalign{\vskip 1mm}   
KOI targets observed        & 715 \\
Exposure time & 90 seconds \\
Observation wavelengths & 600-950nm \\
FWHM resolution       & 0\farcs12 -- 0\farcs15 \\
Field of view & 44\arcsec $\times$ 44\arcsec\\
Pixel scale & 43.1 mas / pix\\
Detector format & 1024$^2$ pixels\\
Detectable magnitude ratio & $\Delta m=5$ mag. at 0\farcs5 (typical)\\
Observation date range & June 17 2012 -- October 6 2012\\
Targets observed / hour & 20\\
\end{deluxetable}

Two dominant factors affect Robo-AO's imaging performance: the seeing and the brightness of the target. During the 23 nights of observing the median seeing was 1\farcs2, with minimum and maximum values of 0\farcs8 and 1\farcs9 respectively. We developed an automated routine to measure the actual imaging performance and to classify the targets into the imaging-performance classes given in the full observations list; this classification can be used with the contrast curve for each class to estimate the companion-detection performance for each target(\S\ref{sec:perf_metrics}).

\section{Data reduction}
\label{sec:pipeline}

To search the large dataset for companions we developed a fully-automated pipeline for data reduction, PSF subtraction, companion detection and companion measurements in Robo-AO data. The pipeline first takes the short-exposure data cubes recorded by the EMCCD camera and produces dark, flat-field and tip-tilt-corrected co-added output images (\S\ref{sec:pipe_imaging}). We then subtract a locally-optimized point-spread-function (PSF) estimate from the image of the \Kepler\ target in each field (\S\ref{sec:PSF_fit}), and either detect companions around the target stars or place limits on their existence (\S\ref{sec:comp_detect}). Finally, we measure the properties of the detected companions (\S\ref{sec:comp_char}).

\subsection{Imaging pipeline}
\label{sec:pipe_imaging}
The Robo-AO imaging pipeline \citep{Law2012, Terziev2013} is based on the Lucky Imaging reduction system described in \citet{Law2006, Law2009}. The recorded EMCCD-frames are dark-subtracted and flat-fielded, and are then corrected for image motion using a bright star in the field. For the KOI observations the relatively crowded fields often led to the automatic selection of a different guide star from the KOI. To avoid having to account for the effects of tip/tilt anisoplanatasism, we manually checked the location of the KOI in Digital Sky Survey images and selected the KOI itself as the guide star in each observation. To produce more consistent and predictable imaging performance for groups of similar KOIs, we used the KOI even if a brighter guide star was nearby and offered potentially increased performance.

\subsection{PSF subtraction using the large set of Robo-AO target observations}
\label{sec:PSF_fit}

The KOI target stars are all in similar parts of the sky, have similar brightness, and were observed at similar airmasses. Because it is unlikely that a companion would be found in the same position for two different targets, we can use each night's ensemble of (at least 20) KOI observations as PSF references without requiring separate observations.

We use a custom locally-optimized PSF subtraction routine based on the LOCI algorithm (Locally Optimized Combination of Images; \citep{Lafreniere2007}. For each KOI target we select 20 other KOI observations obtained in the same filter and closest to the target observation in time. We divide the region around the target star into sections based on polar coordinates: 5 upsampled pixels (110 mas) in radius and 45 degrees in angle. Similar sections are extracted from each PSF reference image. 

We then generate a locally-optimized estimate of the PSF in each section by generating linear combinations of the reference PSFs. In each section, an initial PSF is generated by averaging all the reference PSFs. We then use a downhill simplex algorithm to optimize the contribution from each PSF image, searching for the combination which provides the best fit to the target image.  This optimization is done on several sections simultaneously (in a region 3 sections in radius and 2 sections in angle) to minimize the probability of the algorithm artificially subtracting out real companions. After optimization in the large region, only the central section is output to the final PSF. This provides smooth transitions between adjacent PSF sections because they share many of the image pixels used for the optimization.

\begin{figure}
  \centering
  \resizebox{1.0\columnwidth}{!}
   {
    \includegraphics{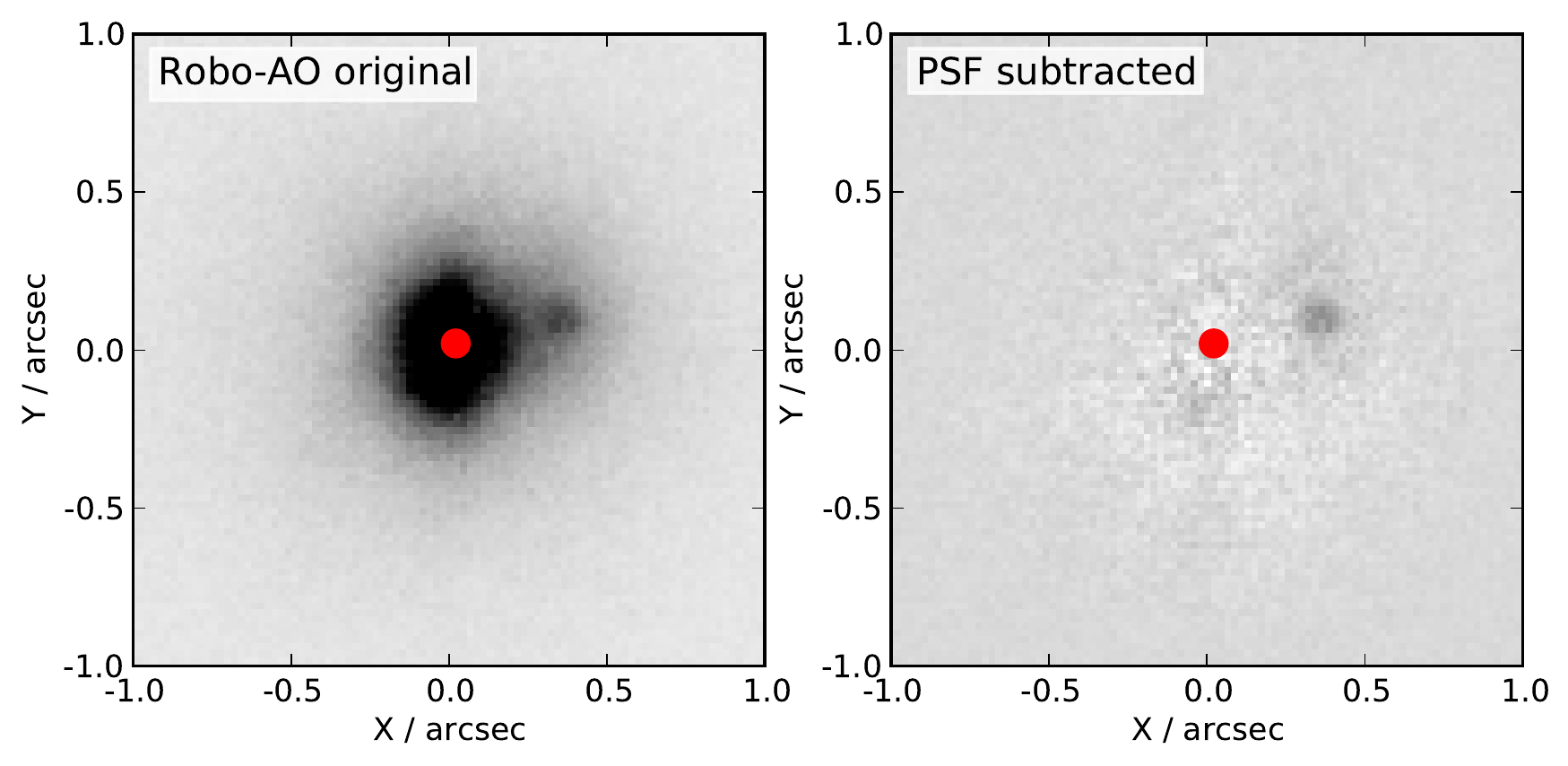}

   }
   \caption{A typical Robo-AO target before and after PSF subtraction using the locally-optimized ensemble of PSF references described in the text. The red circle shows the position of the primary star's PSF peak.}
   \label{fig:psf_sub}
\end{figure}

This procedure is iterated across all the sections of the image, producing a PSF which is an optimal local combination of the reference PSFs and which can then be subtracted from the target star's PSF. The PSF subtraction typically leaves residuals that are consistent with photon noise only (for these relatively short exposures). Figure \ref{fig:psf_sub} shows an example of the PSF subtraction performance.

\subsection{Automated Companion Detection}
\label{sec:comp_detect}
 We limited the detection radius of this initial search to a 2\farcs5 radius from the target KOIs, covering the range of separations between seeing-limited surveys and $\approx$0\farcs15 (subsequent papers will present an analysis of wider-radius companions in Robo-AO imaging). 
 
To more easily and robustly find companions in this large dataset, we developed a new automated companion detection algorithm for Robo-AO data. We first measure the local image noise as a function of distance from the target star, by covering the PSF-subtracted target image with 4-pixel-diameter apertures and measuring the RMS of the pixel values in each aperture, along with the average PSF-subtraction residual signal. We then fit a quadratic to interpolate the changes in noise and residual values as a function of radius from the target star position. For each pixel in the PSF-subtracted image we then use the noise and residual fits to estimate the significance of that pixel's signal level. This procedure generates a significance image where bright pixels in regions of high photon noise (i.e. in the core of the star) are down-weighted compared to those in lower-noise areas. 

The significance image yields the pixels which have some chance of denoting detections of stars, but does not take into account the shapes of the detections -- a single bright pixel surrounded by insignificant pixels is more likely to be due to a cosmic ray hit than a stellar companion, and a tens-of-pixels-wide blob is likely due to imperfect PSF subtraction. We quantify this by cross-correlating the significance image by a Gaussian corresponding to the diffraction limit of the Robo-AO observation. We then select the pixels which show the most significant detections ($>5\sigma$) as possible detections, and amalgamate groups of multiple significant pixels into single detections. 

After automated companion detection we also manually checked each image for companions, to check the performance of the automated system and to search for faint but real companions which could have been fit and removed by spurious speckles in the PSF references. The automated system picked up every manually-flagged companion, and had a 3.5$\%$ false-positive rate from all the images, mainly due imperfect PSF subtraction.

\subsection{Imaging Performance Metrics}
\label{sec:perf_metrics}
We evaluated the contrast-vs.-radius detection performance of the PSF-subtraction and automated companion detection code by performing Monte-Carlo companion-detection simulations. The time-consuming simulations could only be performed on a group of representative targets, and so we established a quantitative image quality metric that allows each of our observations to be tied into the contrast curves for a particular test target. We first parametrized the performance of each observation of our dataset by fitting a two-component model to the PSF based on two Moffat functions tuned to separately measure the widths of the core and halo of the PSF. We then picked 12 single-star observations to represent the variety of PSF parameter space in our dataset. For each test star, we added a simulated companion into the observation at a random separation, position angle and contrast, ran the PSF subtraction and automated companion detection routines, and measured the detection significance (if any) of the simulated companion. We repeated this for 1000 simulated companions\footnote{For each simulated companion PSF we removed the central spike introduced by shifted-and-added photon-noise-limited detectors by averaging with nearby pixels \citep{Law2006, Law2006bins, Law2009}; this conservative correction reduces our claimed detectable contrast by up to 25\%.}. We then binned the simulated detections as a function of separation from the target star, and in each radial bin fit a linear significance-vs.-contrast relation. We use the intersection of the fitted relation with a 5-$\sigma$ detection to provide the minimum-detectable contrast in each radial bin. 

We found that the PSF core size was an excellent predictor of contrast performance, while the halo size did not affect the contrast significantly. The halo is effectively removed by the PSF subtraction, and the contrast is thus chiefly limited by the companion SNR, which scales with the achieved PSF core size (rather than the image FWHM, which we found is a weak predictor of contrast performance in Robo-AO data). On this basis we use the PSF core size to assign targets to contrast-performance groups (low, medium and high). As the imaging performance degrades, we found that the relative contribution of the fitted core PSF decreases, while the core itself shrinks. The somewhat counter-intuitive size decrease is because poor imaging quality inevitably corresponds with poor SNR on the shift-and-add image alignment used by Robo-AO's EMCCD detector. This leads to the frame alignments locking onto photon noise spikes, and thus produces a single-pixel-sized spike in the images \citep{Law2006, Law2009}. We therefore assign images with a diffraction-limited-sizes core ($\sim$0\farcs15) to the high-performance groups; smaller cores, where the imaging performance is degraded, were assigned to the lower-performance groups.

\begin{figure}
  \centering
  \resizebox{1.0\columnwidth}{!}
   {
    \includegraphics{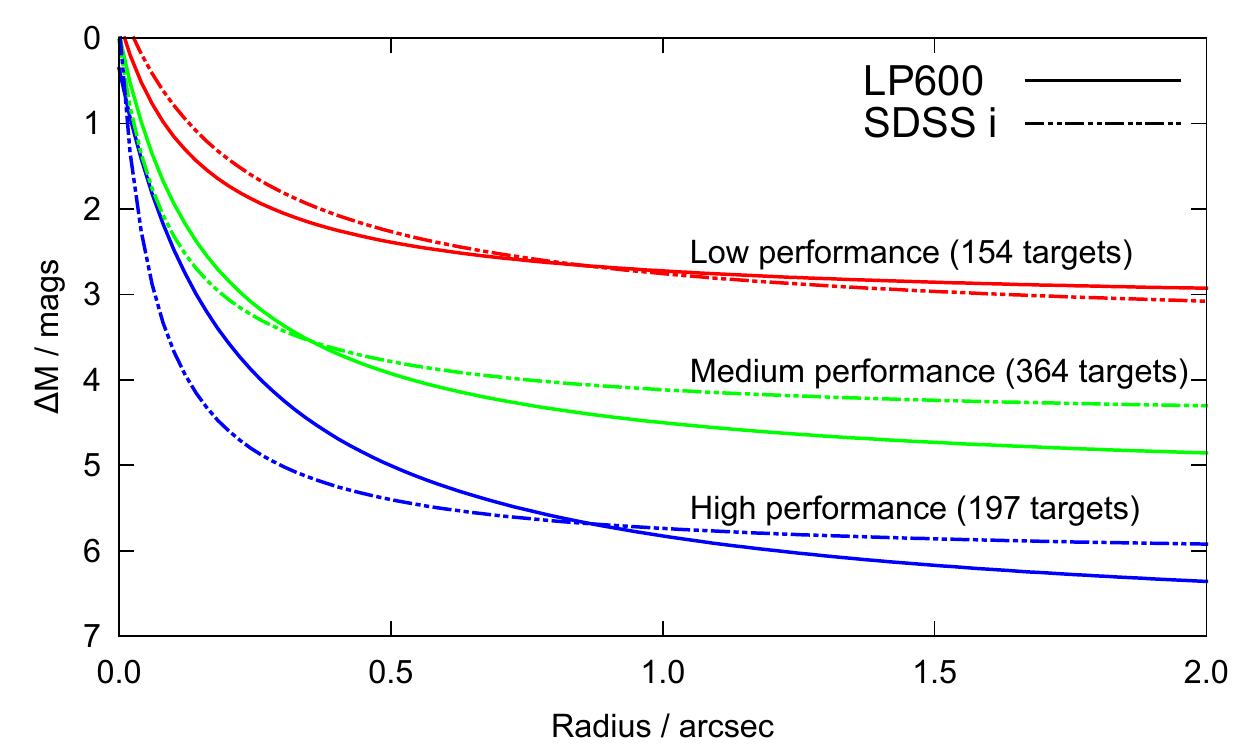}

   }
   \caption{Detectable magnitude ratios for three representative targets observed in the LP600 and SDSS i' filters (smoothed with fitting curves generated as described in \S\ref{sec:perf_metrics}).}

   \label{fig:cr_all}
\end{figure}

Figure \ref{fig:cr_all} shows the contrast curves resulting from this procedure, for clarity smoothed with fitting functions of form $a-b/(r-c)$ (where r is the radius from the target star and $a$,$b$,$c$ are fitting variables). The i-band observations obtain better contrast close-in than the LP600 filter, because of their improved Strehl ratios, while the broader LP600 filter allows somewhat improved contrast at wider radii under all but the poorest conditions.

\subsection{Companion characterization}
\label{sec:comp_char}
\subsubsection{Contrast ratios}
We determined the binaries' contrast ratio in two ways: for the widest separations we performed aperture photometry on the original images; for the closer systems we used the estimated PSF to remove the blended contributions of each of the stars before performing aperture photometry. In all cases the aperture sizes were optimized for the system separation and the available signal.

The locally-optimized PSF subtraction will attempt to remove flux associated with companions by using other PSFs with (non-astrophysical) excess brightness in those areas, because it is trying to achieve the best fit to the target images without discrimination between real companions and speckles. By selecting an optimization over a region containing many PSF core sizes, we reduce the algorithm's ability to subtract away companion light for detection purposes. However, the companion will still be artificially faint in PSF-subtracted images, leading to errors in flux ratio measurements. To avoid this we re-run the PSF fit excluding a 6-pixel-diameter region around any detected companion. The PSF-fit regions are large enough to provide a good estimate for the PSF underneath the companion, and the companion brightness is not artificially reduced by this procedure.

We calculated the contrast ratio uncertainty on the basis of the difference between the injected and measured contrasts of the fake companions injected during the contrast-curve calculations (\S\ref{sec:comp_detect}). We found that the detection significance of the companion was the best predictor of the contrast ratio accuracy, and so we use a fit to that relation to estimate the contrast ratio uncertainty for each companion. We note that the uncertainties (5-30\%) are much higher than would be naively expected from the SNR of the companion detection, as they include an estimate of the systematic errors resulting from the AO imaging, PSF-subtraction and contrast-measurement processes. 

\subsubsection{Separations and position angles}
To obtain the separation and position angle of the binaries we centroided the PSF-subtracted images of the companion and primary, as above. We converted the raw pixel positions to on-sky separations and position angles using a distortion solution produced from Robo-AO measurements of globular clusters observed during the same timeframe as the Robo-AO KOI survey\footnote{S. Hildebrandt, private communication}. 

 We calculated the uncertainties of the companion separation and position angles using estimated systematic errors in the position measurements due to blending between components, depending on the separation of the companion (typically 1-2 pixels uncertainty in the position of each star). We also included an estimate of the maximal changes in the Robo-AO orientation throughout the observation period ($\pm1.5^{\circ}$), as verified using the globular cluster measurements above.  Finally, we verified the measured positions and contrast ratios in direct measurement from non-PSF-subtracted images. 

\section{Discoveries}
\label{sec:discoveries}
\begin{figure*}
  \centering
  \resizebox{1.0\textwidth}{!}
   {
    \includegraphics{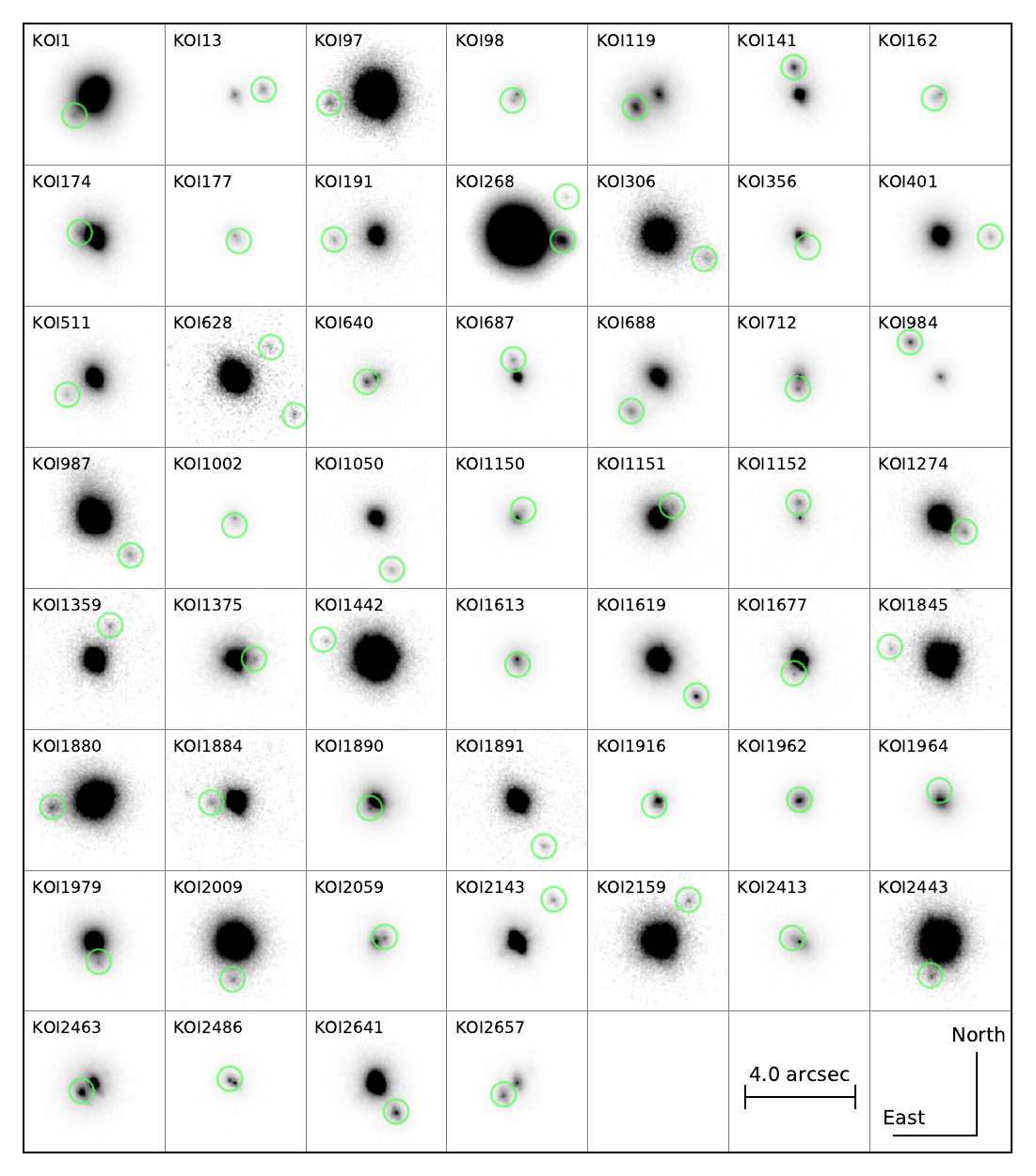}
   }
   \caption{The \Kepler\ planet candidates resolved into multiple stars by Robo-AO. The grayscale of each 4" cutout is selected to show the companion; the angular scale and orientation is identical for each cutout.}

   \label{fig:binaries_summary}
\end{figure*}

\begin{figure}
  \centering
  \resizebox{1.0\columnwidth}{!}
   {
    \includegraphics{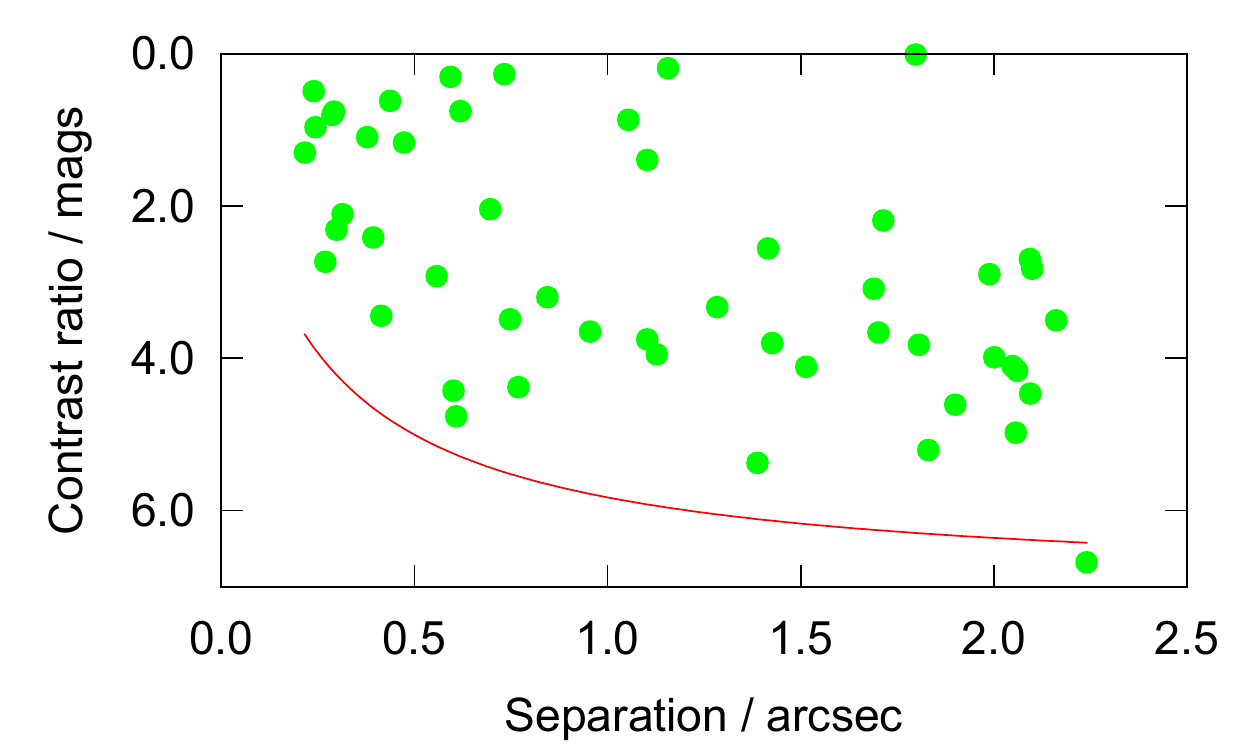}
   }
   \caption{The separations and magnitude differences of the detected companions compared to the survey's typical high-performance $5\sigma$ contrast curve (one very faint companion was detected around a bright KOI in exceptional conditions). The distribution of companion properties has no evidence for unaccounted incompleteness effects, although there is an excess of bright companions at close separations, suggesting that those companions are more likely to be physically associated.}

   \label{fig:koi_comps}
\end{figure}

We resolved 53 \Kepler\ planet candidate hosts into multiple stars; the discovery images are summarized in Figure \ref{fig:binaries_summary} and the separations and contrast ratios are shown in figure \ref{fig:koi_comps}. 
\S\ref{sec:discuss} addresses the probability of physical association for these objects. The measured companion properties for the targets with secure detections are detailed in Table \ref{tab:secure_detections}. Table  \ref{tab:likely_detections} describes 15  probable companions which fell just below our formal 5-$\sigma$ detection criteria. We consider these very likely to be real (indeed, three have been previously detected by other groups), but in the present data we cannot exclude the possibility that one or two of these detections are spurious speckles.

\begin{figure}
  \centering
  \resizebox{1.0\columnwidth}{!}
   {
    \includegraphics{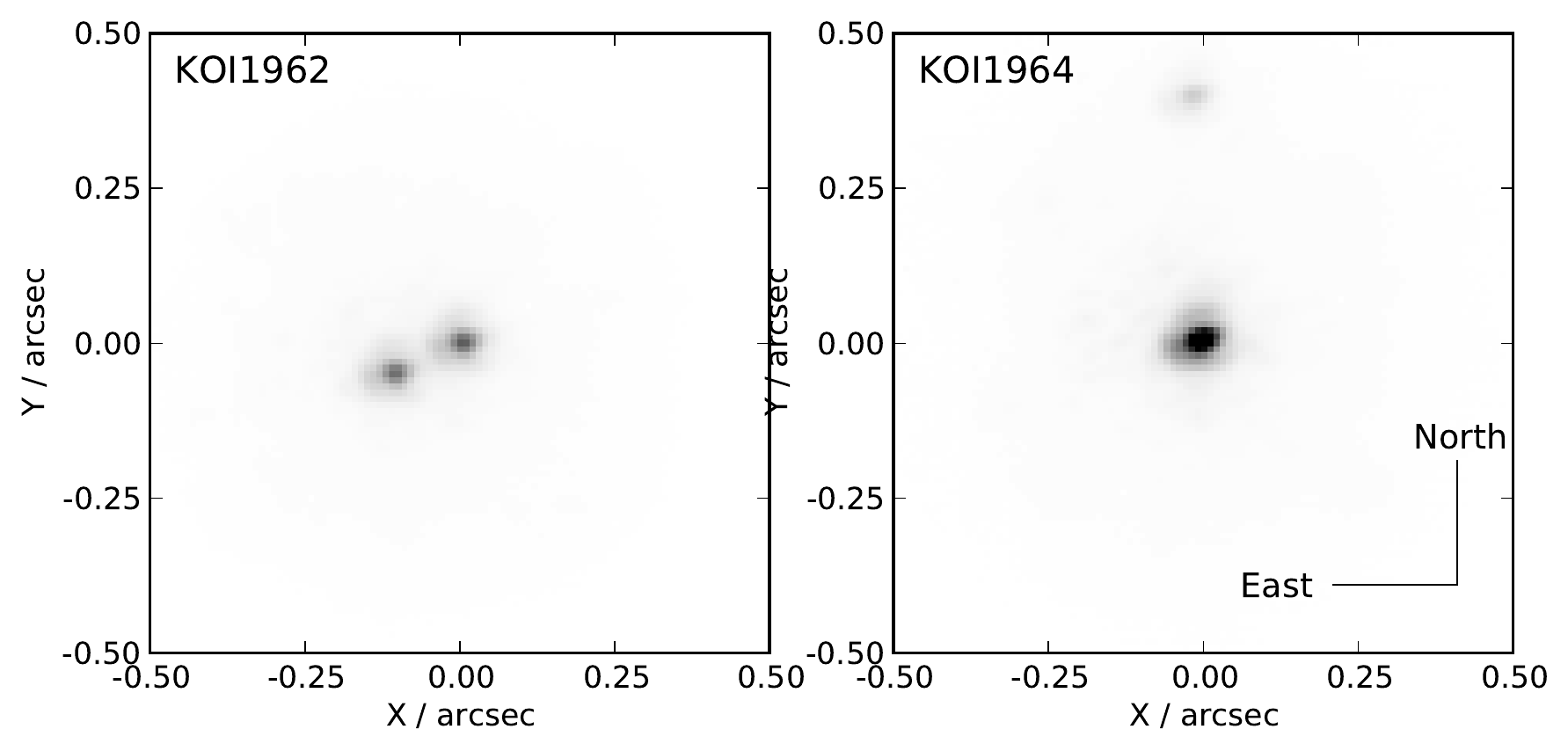}

   }
   \caption{Keck-AO NIRC2 J-band images confirming two Robo-AO companion detections.}
   \label{fig:keck_detections}
\end{figure}

Two of the targets showed potential companions that were not well-resolved by Robo-AO but were suggestive of interesting companions. KOI-1962 showed PSF-core-elongation indicative of a $<$0\farcs15-separation nearly-equal-magnitude binary. KOI-1964 has a probable faint companion at a separation of 0\farcs4; dynamic speckle noise reduces the detection significance to $\approx3\sigma$. We confirmed the Robo-AO detections with NIRC2-NGS \citep{Wizinowich2000} on Keck II on 23 July 2013 (Figure \ref{fig:keck_detections}).

\begin{deluxetable*}{lllllllllll}
\tablecaption{\label{tab:secure_detections}Secure detections of objects within 2\farcs5 of \Kepler\ planet candidates}
\tablehead{ \colhead{KOI} &\colhead{$m_i$} & \colhead{Obs.date} &  \colhead{Filter} & \colhead{Signf.} & \colhead{Separation} & \colhead{PA} & \colhead{Mag. diff.} & \colhead{Previous detection?}\\ &  & \colhead{mag} &   & \colhead{$\sigma$} & \colhead{arcsec} & \colhead{deg.} & \colhead{mag}  & }
\startdata
KOI-1 & 11.2 & 2012/07/16 & i & 13&1.13$\pm$0.06&135$\pm$2&3.95$\pm$0.33 & D09 \\
KOI-13 & 10.5 & 2012/10/06 & i & 950&1.16$\pm$0.06&279$\pm$2&0.19$\pm$0.06 & H11, A12 \\
KOI-98 & 12.0 & 2012/07/17 & i & 80&0.29$\pm$0.06&140$\pm$6&0.76$\pm$0.16 & B11, H11, A12, H12\\
KOI-119 & 12.5 & 2012/07/16 & i & 38&1.05$\pm$0.06&118$\pm$2&0.87$\pm$0.22 &  \\
KOI-141 & 13.4 & 2012/07/18 & i & 34&1.10$\pm$0.06&11$\pm$2&1.39$\pm$0.23 & A12\\
KOI-162 & 13.6 & 2012/07/18 & LP600 & 19&0.29$\pm$0.06&117$\pm$7&0.81$\pm$0.29 &  \\
KOI-174 & 13.4 & 2012/07/18 & LP600 & 7&0.60$\pm$0.06&77$\pm$3&4.43$\pm$0.44 & A13 \\
KOI-177 & 13.0 & 2012/07/18 & i & 12&0.24$\pm$0.06&215$\pm$8&0.97$\pm$0.35  &  \\
KOI-191 & 14.7 & 2012/09/01 & LP600 & 5&1.69$\pm$0.06&94$\pm$2&3.09$\pm$0.49 &  \\
KOI-268 &  & 2012/09/14 & LP600 & 23&1.81$\pm$0.06&265$\pm$2&3.82$\pm$0.27 & A12 \\
KOI-356 & 13.5 & 2012/07/28 & LP600 & 17&0.56$\pm$0.06&218$\pm$4&2.92$\pm$0.30 &  \\
KOI-401 & 13.7 & 2012/08/05 & LP600 & 19&1.99$\pm$0.06&268$\pm$2&2.90$\pm$0.29  & L12 \\
KOI-511 & 14.0 & 2012/09/01 & LP600 & 7&1.28$\pm$0.06&123$\pm$2&3.33$\pm$0.43 &  \\
KOI-640 & 13.1 & 2012/07/28 & i & 16&0.44$\pm$0.06&117$\pm$4&0.62$\pm$0.31  &  \\
KOI-687 & 13.6 & 2012/08/04 & i & 21&0.70$\pm$0.06&13$\pm$3&2.04$\pm$0.28 &  \\
KOI-688 & 13.8 & 2012/09/14 & LP600 & 19&1.71$\pm$0.06&141$\pm$2&2.19$\pm$0.29 &  \\
KOI-712 & 13.5 & 2012/08/05 & i & 21&0.47$\pm$0.06&173$\pm$4&1.17$\pm$0.28 &  \\
KOI-984 & 11.4 & 2012/08/03 & i & 120&1.80$\pm$0.06&42$\pm$2&0.01$\pm$0.14 &  \\
KOI-1002 & 13.4 & 2012/08/03 & i & 9&0.30$\pm$0.06&173$\pm$6&2.31$\pm$0.38 &  \\
KOI-1050 & 13.7 & 2012/08/03 & i & 8&2.09$\pm$0.06&197$\pm$2&2.70$\pm$0.40 &  \\
KOI-1150 & 13.1 & 2012/08/05 & i & 9&0.39$\pm$0.06&322$\pm$5&2.41$\pm$0.39 &  \\
KOI-1152 & 13.6 & 2012/09/14 & LP600 & 16&0.59$\pm$0.06&2$\pm$3&0.31$\pm$0.31 &  \\
KOI-1274 & 13.1 & 2012/08/06 & i & 7&1.10$\pm$0.06&241$\pm$2&3.75$\pm$0.44 &  \\
KOI-1613 &  & 2012/08/29 & i & 36&0.22$\pm$0.06&184$\pm$9&1.30$\pm$0.22 &  \\
KOI-1619 & 11.4 & 2012/08/29 & i & 60&2.10$\pm$0.06&226$\pm$2&2.82$\pm$0.18 &  \\
KOI-1677 & 14.1 & 2012/09/04 & LP600 & 7&0.61$\pm$0.06&159$\pm$3&4.76$\pm$0.44 &  \\
KOI-1880 & 13.8 & 2012/07/15 & LP600 & 6&1.70$\pm$0.06&100$\pm$2&3.66$\pm$0.45 &  \\
KOI-1890 & 11.6 & 2012/08/29 & i & 42&0.41$\pm$0.06&142$\pm$5&3.44$\pm$0.21 &  \\
KOI-1916 & 13.4 & 2012/09/13 & LP600 & 31&0.27$\pm$0.06&143$\pm$7&2.73$\pm$0.24 &  \\
KOI-1962 &       & 2012/08/30 & i     & \nodata & 0.12$\pm$0.03 & \nodata & 0.04 ($\rm K_s$)\\
KOI-1964 & 10.5 & 2012/08/30 & i     & \nodata & 0.39$\pm$0.03 & \nodata & 1.9 ($\rm K_s$)\\
KOI-1979 & 12.8 & 2012/08/30 & i & 9&0.84$\pm$0.06&192$\pm$3&3.20$\pm$0.39  &  \\
KOI-2059 & 12.6 & 2012/10/06 & LP600 & 120&0.38$\pm$0.06&291$\pm$5&1.10$\pm$0.14 &  \\
KOI-2143 & 13.9 & 2012/10/06 & LP600 & 19&2.16$\pm$0.06&317$\pm$2&3.50$\pm$0.29 &  \\
KOI-2463 & 12.6 & 2012/08/31 & i & 70&0.62$\pm$0.06&125$\pm$3&0.75$\pm$0.17 &  \\
KOI-2486 & 12.9 & 2012/08/31 & i & 18&0.24$\pm$0.06&63$\pm$8&0.49$\pm$0.30 &  \\
KOI-2641 & 13.6 & 2012/10/06 & LP600 & 36&1.42$\pm$0.06&214$\pm$2&2.56$\pm$0.22 &  \\
KOI-2657 & 12.7 & 2012/10/06 & LP600 & 62&0.73$\pm$0.06&131$\pm$3&0.27$\pm$0.18 &  \\

\enddata
\tablenotetext{}{References for previous detections are denoted with the following codes: \citealt{Adams2012} (A12); \citealt{Adams2013} (A13); \citealt{Buchhave2011} (B11); \citealt{Daemgen2009} (D09); \citealt{Horch2012} (H12); \citealt{Howell2011} (H11); \citealt{LilloBox2012} (L12).}
\end{deluxetable*}

\begin{deluxetable*}{lllllllllll}
\tablecaption{\label{tab:likely_detections}Likely detections of objects within 2\farcs5 of \Kepler\ planet candidates}
\tablehead{ \colhead{KOI} &\colhead{$m_i$} & \colhead{Obs.date} &  \colhead{Filter} & \colhead{Signf.} & \colhead{Separation} & \colhead{PA} & \colhead{Mag. diff.} & \colhead{Previous detection?}\\ & &\colhead{mag}  &   & \colhead{$\sigma$} & \colhead{arcsec} & \colhead{deg.} & \colhead{mag}& }
\startdata
KOI-97 & 12.7 & 2012/07/17 & i & 4.2&1.90$\pm$0.06&99$\pm$2&4.61$\pm$0.52 & A12 \\
KOI-306 & 12.4 & 2012/07/18 & i & 3.6&2.06$\pm$0.06&243$\pm$2&4.16$\pm$0.56 & A12 \\
KOI-628 & 13.7 & 2012/08/03 & i & 1.4&1.83$\pm$0.06&309$\pm$2&5.20$\pm$0.80 & L12 \\
KOI-987 & 12.3 & 2012/08/03 & i & 2.4&2.05$\pm$0.06&225$\pm$2&4.10$\pm$0.66 &  \\
KOI-1151 & 13.2 & 2012/08/05 & i & 3.2&0.75$\pm$0.06&309$\pm$3&3.49$\pm$0.58 &  \\
KOI-1359 & 15.0 & 2012/09/04 & LP600 & 3.4&1.43$\pm$0.06&333$\pm$2&3.80$\pm$0.57 &  \\
KOI-1375 & 13.5 & 2012/08/06 & i & 4.0&0.77$\pm$0.06&269$\pm$3&4.38$\pm$0.53 &  \\
KOI-1442 & 12.3 & 2012/08/06 & i & 3.3&2.24$\pm$0.06&70$\pm$2&6.68$\pm$0.57 &  \\
KOI-1845 & 14.1 & 2012/09/13 & LP600 & 2.9&2.06$\pm$0.06&77$\pm$2&4.97$\pm$0.60 &  \\
KOI-1884 & 15.2 & 2012/09/13 & LP600 & 2.5&0.95$\pm$0.06&96$\pm$2&3.65$\pm$0.64 &  \\
KOI-1891 & 15.0 & 2012/09/13 & LP600 & 3.0&2.09$\pm$0.06&210$\pm$2&4.46$\pm$0.60 &  \\
KOI-2009 & 13.6 & 2012/09/14 & LP600 & 4.9&1.51$\pm$0.06&176$\pm$2&4.11$\pm$0.49 &  \\
KOI-2159 & 13.3 & 2012/08/31 & i & 4.0&2.00$\pm$0.06&323$\pm$2&3.99$\pm$0.53 &  \\
KOI-2413 & 14.7 & 2012/09/14 & LP600 & 2.4&0.31$\pm$0.06&67$\pm$6&2.11$\pm$0.66 &  \\
KOI-2443 & 13.8 & 2012/10/06 & LP600 & 3.7&1.39$\pm$0.06&163$\pm$2&5.37$\pm$0.55 &  \\
\enddata
\tablenotetext{}{References for previous detections are denoted with the following codes: \citealt{Adams2012} (A12); \citealt{LilloBox2012} (L12).}
\end{deluxetable*}

\subsection{Comparison to other surveys}

\citet{LilloBox2012} (hereafter L12) observed 98 KOIs using a Lucky Imaging system. Seven of the targets for which they discovered companions within a 2\farcs5 radius are also in our survey. Both surveys detect KOI-401 at a separation of 2\farcs0 and at a contrast of 2.6 magnitudes (L12 i-band) or 2.9 magnitudes (Robo-AO LP600).  The companions to KOI-628 were visible in our survey but at contrasts that placed them in the ``likely detections'' group. L12 detected a companion to KOI-658 at 1\farcs9 radius and a contrast of 4.6 magnitudes in i-band. At that radius, for the performance achieved on KOI-658, the Robo-AO snapshot-survey limiting magnitude ratio is $\sim$4.0 magnitudes and so we do not re-detect that companion. For the same reason we also do not re-detect the companions to KOI-703 (6.4 magnitudes contrast), KOI-704 (5.0 magnitudes contrast) and KOI-721 (3.9 magnitudes contrast). The 0\farcs13-radius companion to KOI-1537 detected in \citet{Adams2013} is at too close a separation to be detectable in our survey. The L12 companion to KOI-1375 is visible in our dataset, but has a contrast ratio of 4.0 magnitudes, under our formal detection limit and well below the 2.75 magnitude i-band contrast measured by L12. The target is not strongly coloured according to L12 and it is not obvious why the companion is so much fainter in our survey. 

\section{Discussion}
\label{sec:discuss}

\subsection{Implications for \Kepler\, Planet Candidates}

The detection of a previously unknown star within the photometric aperture of a KOI host star will affect the derived radius of any planet candidate around that host star, because the \Kepler\, observed transit depth is shallower than the true depth due to dilution.  The degree of this effect depends upon the relative brightness of the target and secondary star, and which star is actually being transited.  In particular, if there is more than one star in the photometric aperture and the transiting object is around a star that contributes a fraction $F_i$ to the total light in the aperture, then
\begin{equation}
\label{eq:dilution}
\delta_{\rm true} = \delta_{\rm obs} \left(\frac{1}{F_i} \right),
\end{equation}
where $\delta_{\rm true}$ is the true intrinsic fractional transit depth and $\delta_{\rm obs}$ is the observed, diluted depth.  Since $\delta \propto (R_p/R_\star)^2$, the true planet radius in the case where the transit is around star $i$ is
\begin{equation}
\label{eq:rptrue}
R_{p,i} = R_{\star,i} \left(\frac{R_p}{R_\star}\right)_0 \sqrt{\frac{1}{F_i}},
\end{equation}
where $R_{\star,i}$ is the radius of star $i$, and the $0$ subscript represents the radius ratio implied by the diluted transit, or what would be inferred by ignoring the presence of any blending flux. 

Thus, for each planet candidate in KOI systems observed to have close stellar companions, the derived planet radius must be corrected---and there are two potential scenarios for each candidate: the eclipsed star is either star A (the brighter target star) or star B (the fainter companion).

In case A, the corrected planet radius is
\begin{equation}
\label{eq:rpA}
R_{p,A} = R_{p,0} \sqrt{\frac{1}{F_A}}, 
\end{equation}
and in case B,
\begin{equation}
\label{eq:rpB}
R_{p,B} = R_{p,0} \frac{R_B}{R_A} \sqrt{\frac{1}{F_B}}. 
\end{equation}
Case A is straightforward, with nothing needed except the observed contrast ratio (in order to calculate $F_A$).  It should be noted, however, that this assumes that the estimated host stellar radius $R_A$ is unchanged by the detection of the companion star.  As the radii for most \Kepler\ stars are inferred photometrically, this may not be strictly true, as light from the companion might cause the primary stellar type to be misidentified.  We do not attempt to quantify the extent of this effect in this paper. We do, however, note that it is likely to be negligible for larger contrast ratios where the colors of the blended system are dominated by light from the primary.  

Case B, in addition to needing $F_B$, needs also the ratio $R_B/R_A$.  If the observed companion is an unassociated background star, then the single-band Robo-AO observation does not constrain $R_B$.  However, under the assumption that the companion is physically bound, then we can estimate its size and spectral type, given assumed knowledge about the primary star A.   

In order to accomplish this, we use the Dartmouth stellar models \citep{Dotter2008} and the measured primary KOI star properties listed in the NASA Exoplanet Archive.  For the mass and age of the primary, we use the Dartmouth isochrones to find an absolute magnitude in the observed band (approximating the LP600 bandpass as \Kepler\ band), then we inspect the isochrone to find the mass of a star that is the appropriate amount fainter (according to the observed contrast ratio), and assign the stellar radius $R_B$ accordingly.

Table \ref{table:koidetails} summarizes how the planet radii change under both case $A$ and $B$ for each KOI in all the systems in which we detect companions.  We also list an additional case $B_{bg}$ for the situation in which the eclipsed star is not physically bound---since we do not have a constraint on $R_B$ in this situation, we simply list the planet radii for the case of $R_B = 1 R_\odot$, which allows for simple scaling.  

Interestingly, under case $B$ where the transit is assumed to be around a bound companion, in many cases the implied planet radius is not indicative of a false positive.  This is because in order to get a large radius correction there must be a large contrast ratio, which then (in the physically associated scenario) implies that the secondary is a small star, which shrinks the radius correction factor.  In fact, the only candidates which attain clearly non-planetary radii under case $B$ are those which already have radii comparable to or larger than Jupiter to begin with.  On the other hand, case $B_{bg}$ often suggests a non-planetary radius, as the stellar radius in this case is not bound to shrink as the contrast ratio grows.

We leave a quantitative analysis exploring the relative probability of scenario B being a physically bound or chance-aligned companion to future work.  However, we note qualitatively that relatively bright, small-separation companions are more likely to be physically associated, whereas more distant and higher contrast-ratio companions are more likely to be foreground/background objects.

\tabletypesize{\scriptsize}
\def\arraystretch{0.9}

\begin{deluxetable*}{llll|llllll}
    \tablecolumns{10}
    \tablecaption{Implications on derived radius of \Kepler\ planet candidates}
    \tablehead{ \colhead{KOI} &\colhead{$P$}\tablenotemark{a} & \colhead{$R_p$}\tablenotemark{a} &  \colhead{$R_\star$}\tablenotemark{a} &  \colhead{$\Delta m$} & \colhead{sep} & \colhead{$R_{\star,B}$}\tablenotemark{b} & \colhead{$R_{p,A}$}\tablenotemark{c} & \colhead{$R_{p,B}$} & \colhead{$R_{p,B_{bg}}$}\tablenotemark{d}\\
\colhead{--} & \colhead{d} &\colhead{$R_\oplus$} & \colhead{$R_\odot$} & \colhead{mag} & \colhead{\arcsec} & \colhead{$R_\odot$} & \colhead{$R_\oplus$} & \colhead{$R_\oplus$} & \colhead{$R_\oplus$}}
    \startdata
1.01 & 2.471 & 14.40 & 1.06 & 4.0 & 1.13 & 0.50 & 14.6 & 42.0 & 84.9\\
13.01 & 1.764 & 23.00 & 2.70 & 0.2 & 1.16 & 2.70 & 31.2 & 34.0 & 12.6\\
97.01 & 4.885 & 16.10 & 1.78 & 4.6 & 1.90 & 0.57 & 16.2 & 43.6 & 76.1\\
98.01 & 6.790 & 10.00 & 1.63 & 0.8 & 0.29 & 1.26 & 12.2 & 13.4 & 10.7\\
119.01 & 49.184 & 3.90 & 0.94 & 0.9 &  1.05 & 0.76 & 4.7 & 5.6 & 7.5\\
119.02 & 190.313 & 3.40 & --- & --- & --- & --- & 4.1 & 4.9 & 6.5\\
141.01 & 2.624 & 5.43 & 0.93 & 1.4 & 1.10 & 0.72 & 6.1 & 9.0 & 12.5\\
162.01 & 14.006 & 2.54 & 0.96 & 0.8 & 0.29 & 0.79 & 3.1 & 3.7 & 4.7\\
174.01 & 56.354 & 1.94 & 0.63 & 4.4 & 0.60 & 0.21 & 2.0 & 5.1 & 24.0\\
177.01 & 21.060 & 1.84 & 1.06 & 1.0 & 0.24 & 0.84 & 2.2 & 2.7 & 3.2\\
191.01 & 15.359 & 11.00 & 0.88 & 3.1 & 1.69 & 0.55 & 11.3 & 29.3 & 53.4\\
191.02 & 2.418 & 2.30 & --- & --- & --- & --- & 2.4 & 6.1 & 11.2\\
191.03 & 0.709 & 1.24 & --- & --- & --- & --- & 1.3 & 3.3 & 6.0\\
191.04 & 38.652 & 2.30 & --- & --- & --- & --- & 2.4 & 6.1 & 11.2\\
268.01 & 110.379 & 1.73 & 0.79 & 3.8 & 1.81 & 0.33 & 1.8 & 4.3 & 13.0\\
306.01 & 24.308 & 2.29 & 0.87 & 4.2 & 2.06 & 0.39 & 2.3 & 7.0 & 18.0\\
356.01 & 1.827 & 5.73 & 1.60 & 2.9 & 0.56 & 0.66 & 5.9 & 9.4 & 14.2\\
401.01 & 29.199 & 7.23 & 1.58 & 2.9 & 1.99 & 0.66 & 7.5 & 11.9 & 18.0\\
401.02 & 160.017 & 7.31 & --- & --- & --- & --- & 7.6 & 12.0 & 18.2\\
401.03 & 55.328 & 2.66 & --- & --- & --- & --- & 2.8 & 4.4 & 6.6\\
511.01 & 8.006 & 2.80 & 1.08 & 3.3 & 1.28 & 0.61 & 2.9 & 7.5 & 12.3\\
511.02 & 4.264 & 1.58 & --- & --- & --- & --- & 1.6 & 4.2 & 6.9\\
628.01 & 14.486 & 3.10 & 1.29 & 5.2 & 1.83 & 0.38 & 3.1 & 10.1 & 26.5\\
640.01 & 30.996 & 2.44 & 0.89 & 0.6 & 0.44 & 0.80 & 3.1 & 3.6 & 4.5\\
687.01 & 4.178 & 1.46 & 0.93 & 2.0 & 0.70 & 0.64 & 1.6 & 2.8 & 4.3\\
688.01 & 3.276 & 2.28 & 1.35 & 2.2 & 1.71 & 0.77 & 2.4 & 3.8 & 4.9\\
712.01 & 2.178 & 1.08 & 0.84 & 1.2 & 0.47 & 0.76 & 1.3 & 1.9 & 2.6\\
984.01 & 4.287 & 3.19 & 0.92 & 0.0 & 1.80 & 0.92 & 4.5 & 4.5 & 4.9\\
987.01 & 3.179 & 1.28 & 0.92 & 4.1 & 2.05 & 0.42 & 1.3 & 3.9 & 9.3\\
1002.01 & 3.482 & 1.36 & 1.01 & 2.3 & 0.30 & 0.66 & 1.4 & 2.7 & 4.1\\
1050.01 & 1.269 & 1.40 & 0.76 & 2.7 & 2.09 & 0.46 & 1.5 & 3.1 & 6.6\\
1050.02 & 2.853 & 1.40 & --- & --- & --- & --- & 1.5 & 3.1 & 6.6\\
1150.01 & 0.677 & 1.10 & 1.09 & 2.4 & 0.39 & 0.66 & 1.2 & 2.1 & 3.2\\
1151.01 & 10.435 & 1.46 & 0.97 & 3.5 & 0.75 & 0.50 & 1.5 & 3.8 & 7.7\\
1151.02 & 7.411 & 1.15 & --- & --- & --- & --- & 1.2 & 3.0 & 6.0\\
1151.03 & 5.249 & 0.70 & --- & --- & --- & --- & 0.7 & 1.8 & 3.7\\
1151.04 & 17.453 & 0.87 & --- & --- & --- & --- & 0.9 & 2.3 & 4.6\\
1151.05 & 21.720 & 0.97 & --- & --- & --- & --- & 1.0 & 2.5 & 5.1\\
1152.01 & 4.722 & 19.56 & 0.65 & 0.3 & 0.59 & 0.54 & 25.9 & 24.9 & 45.7\\
1274.01 & 362.000 & 4.73 & 0.79 & 3.8 & 1.10 & 0.37 & 4.8 & 12.6 & 34.3\\
1359.01 & 37.101 & 3.50 & 0.92 & 3.8 & 1.43 & 0.58 & 3.6 & 13.0 & 22.2\\
1359.02 & 104.820 & 7.30 & --- & --- & --- & --- & 7.4 & 27.1 & 46.3\\
1375.01 & 321.214 & 6.78 & 1.17 & 4.4 & 0.77 & 0.50 & 6.8 & 22.0 & 44.0\\
1442.01 & 0.669 & 1.23 & 1.00 & 6.7 & 2.24 & 0.20 & 1.2 & 5.2 & 26.8\\
1613.01 & 15.866 & 1.07 & 1.04 & 1.3 & 0.22 & 0.78 & 1.2 & 1.7 & 2.1\\
1613.02 & 94.091 & 1.08 & --- & --- & --- & --- & 1.2 & 1.7 & 2.2\\
1619.01 & 20.666 & 0.80 & 0.62 & 2.8 & 2.10 & 0.33 & 0.8 & 1.6 & 4.9\\
1677.01 & 52.070 & 2.18 & 0.85 & 4.8 & 0.61 & 0.43 & 2.2 & 9.8 & 23.1\\
1677.02 & 8.512 & 0.81 & --- & --- & --- & --- & 0.8 & 3.7 & 8.6\\
1845.01 & 1.970 & 1.50 & 0.70 & 5.0 & 2.06 & 0.19 & 1.5 & 4.0 & 21.2\\
1845.02 & 5.058 & 21.00 & --- & --- & --- & --- & 21.1 & 56.2 & 297.4\\
1880.01 & 1.151 & 1.49 & 0.52 & 3.7 & 1.70 & 0.18 & 1.5 & 2.8 & 15.6\\
1884.01 & 23.120 & 5.00 & 0.92 & 3.6 & 0.95 & 0.55 & 5.1 & 16.3 & 29.7\\
1884.02 & 4.775 & 2.63 & --- & --- & --- & --- & 2.7 & 8.6 & 15.6\\
1890.01 & 4.336 & 1.50 & 1.32 & 3.4 & 0.41 & 0.62 & 1.5 & 3.5 & 5.7\\
1891.01 & 15.955 & 1.85 & 0.69 & 4.5 & 2.09 & 0.33 & 1.9 & 6.9 & 21.1\\
1891.02 & 8.260 & 1.26 & --- & --- & --- & --- & 1.3 & 4.7 & 14.4\\
1916.01 & 20.679 & 2.16 & 0.96 & 2.7 & 0.27 & 0.67 & 2.2 & 5.5 & 8.2\\
1916.02 & 9.600 & 1.89 & --- & --- & --- & --- & 2.0 & 4.8 & 7.2\\
1916.03 & 2.025 & 0.92 & --- & --- & --- & --- & 1.0 & 2.4 & 3.5\\
1979.01 & 2.714 & 1.13 & 0.94 & 3.2 & 0.84 & 0.52 & 1.2 & 2.8 & 5.4\\
2009.01 & 86.749 & 2.20 & 0.97 & 4.1 & 1.51 & 0.48 & 2.2 & 7.3 & 15.2\\
2059.01 & 6.147 & 0.83 & 0.67 & 1.1 & 0.38 & 0.60 & 1.0 & 1.4 & 2.4\\
2059.02 & 2.186 & 0.60 & --- & --- & --- & --- & 0.7 & 1.0 & 1.7\\
2143.01 & 4.790 & 1.14 & 0.81 & 3.5 & 2.16 & 0.54 & 1.2 & 3.9 & 7.2\\
2159.01 & 7.597 & 1.07 & 0.88 & 4.0 & 2.00 & 0.48 & 1.1 & 3.8 & 7.7\\
2159.02 & 2.393 & 0.99 & --- & --- & --- & --- & 1.0 & 3.5 & 7.2\\
2413.01 & 12.905 & 1.32 & 0.65 & 2.1 & 0.31 & 0.46 & 1.4 & 2.7 & 5.8\\
2413.02 & 31.200 & 1.26 & --- & --- & --- & --- & 1.3 & 2.6 & 5.5\\
2443.01 & 6.792 & 1.20 & 1.09 & 5.4 & 1.39 & 0.41 & 1.2 & 5.3 & 13.1\\
2443.02 & 11.837 & 1.02 & --- & --- & --- & --- & 1.0 & 4.5 & 11.1\\
2463.01 & 7.467 & 1.02 & 0.97 & 0.8 & 0.62 & 0.94 & 1.2 & 1.7 & 1.8\\
2486.01 & 4.268 & 2.71 & 1.17 & 0.5 & 0.24 & 1.08 & 3.5 & 4.0 & 3.7\\
2641.01 & 3.556 & 1.20 & 1.10 & 2.6 & 1.42 & 0.66 & 1.3 & 2.5 & 3.7\\
2657.01 & 5.224 & 0.60 & 0.80 & 0.3 & 0.73 & 0.89 & 0.8 & 1.0 & 1.1\\
\enddata
\tablenotetext{a}{Values taken from the NASA Exoplanet Archive.}
\tablenotetext{b}{Estimated radius of the stellar companion in the scenario where it is physically bound to the target star.  Estimate made according to the absolute magnitude difference in the Kepler band, according the Dartmouth stellar models \citep{Dotter2008}.}
\tablenotetext{c}{Eclipsing object radius in the scenario where the companion star is the eclipsed object and is physically bound to the target star, assuming the stellar radius of star $B$ as estimated in this table.}
\tablenotetext{d}{Eclipsing object radius in the scenario where the companion star is the eclipsed object and is a chance-aligned background star with radius 1 $R_\odot$.  We note that a background or foreground object is perhaps unlikely to be Solar-type, but this quantification allows for simple scaling of the implied eclipsing object radius.}
\label{table:koidetails}
\end{deluxetable*}

\subsection{Particularly interesting systems}

There are several KOIs with detected companions which we note as being of particular interest, some of which might represent rare false-positive scenarios. Future work will quantitatively assess the true nature of these particular KOIs (e.g. the probability that any given KOI is a false positive).

\subsubsection{KOI-191: A probable ``coincident multiple''}
\label{sec:koi191}
KOI-191 was identified by \citet{Batalha2012} to have four planet candidates, with periods of approximately 0.7, 2.4, 15.4, and 38.7 days.  The 15.4d candidate has an estimated radius of 11 $R_\oplus$, whereas all the rest are smaller than 1.5 $R_\oplus$.   This system is notable because in the entire current cumulative KOI catalog, there are only four multi-candidate systems that have a planet candidate (either ``CANDIDATE'' or ``NOT DISPOSITIONED'' in the NEA) with $10 R_\oplus < R < 20 R_\oplus$ and $P < 20$d.  Two of these four (KOI-199 and KOI-3627) are marked as 2-planet systems but the second candidate in each is identified as a FP in the Q1-Q12 activity table, making them effectively single-candidate systems.  The host star of KOI-338 has $R_\star = 19.2M_{\odot}$, and its two candidates have radii of 17 and 37 $R_\oplus$, making that system most likely a stellar multiple system.  This leaves KOI-191 as the only multiple-candidate \Kepler\ system including a Jupiter-like candidate with $P < 20$d.  By contrast, there are 62 single candidates that match these same radius and period cuts (64 including KOI-199 and KOI-3627).

Based on the apparent rarity of planetary systems with this architecture and the fact that we detect a stellar companion to the KOI-191 host star, we conclude that this is a likely ``coincident multiple'' system, with KOI-191.01 around one of the stars, and the other three around the other.  There are three possibilities: 1) Since the companion star (1\farcs69 separation) is 3.1 mag fainter, if it is the host of KOI-191.01, then it is most likely a stellar eclipsing binary; 2) if the primary star hosts .01, then the secondary likely hosts the three-candidate system, in which case .02-.04 are more likely all super-Earth/Neptune-sized; 3) it may be the case that all four planets are indeed around the same star, which would make KOI-191 a planetary system of unusual architecture, inviting further study.

\subsubsection{KOI-268: Habitable Zone Candidate?}

KOI-268 hosts a planet candidate in a 110-d orbit. The candidate has a radius of 1.7 $R_\oplus$ and an equilibrium temperature of 295 K, according to the NEA.  However, Robo-AO detects a stellar companion 3.8 mag fainter at a separation of 1\farcs81.  We also note the presence of a possible fainter companion at a 2.45'' separation, a position angle of 306$^{\circ}$ and a contrast ratio of $\approx$5.5 magnitudes. The equilibrium temperature calculation of the candidate is based on the estimated effective temperature of the host star and the planet is therefore unlikely to be in the habitable zone if it is around one of the companions.

\subsubsection{KOI-628: possible triple-system}
KOI-628 has a previously-detected faint companion at a separation of 1\farcs83 \citep{Barrado2013,LilloBox2012}. We also re-detect a further possible companion just beyond our detection-target radius, at 2.55” separation.

\subsubsection{KOI-1151: Another possible coincident multiple}
KOI-1151, discovered by this survey to have a companion with $\Delta i \approx 3.5$ at a separation of 0\farcs75, is another system with unusual architecture that might be best explained if the candidates were shared between the two stars.  This system has 5 detected planet candidates, with periods of 5.25, 7.41, 10.44, 17.45, and 21.72 days.\footnote{The NEA cumulative KOI table gives KOI-1151.01 a 5.22-d period rather than 10.44d, which would be clearly unphysical in the presence of another candidate with a 5.25-d period; the Q1-Q12 table corrects the period of 1151.01 to 10.44.} What makes this system appear unusual is the presence of the 7.41d candidate in between the 5.25d and 10.44d candidates, which have nearly exact 2:1 commensurability.  Of the 22 multi-KOI systems that have a pair of planets within 2\% of exact 2:1 commensurability, only KOI-1151 and KOI-2038 have another candidate between the pair (the inner two planets in this system have been confirmed via transit timing variations by \citealt{Ming2013}).  Migration can tend to deposit planets in or near resonant configurations, but it appears to be unusual for a planet to be stuck between two other planets that are near a strong resonance---perhaps this is an indication that the KOI-1151 system is not a single planetary system at all, but rather two separate systems. Another plausible configuration is that KOIs 1151.02 (the interloper at 7.41d) and 1151.05 (the 21.72d candidate) are separated from the other three as those two are near 3:1 commensurability.

\subsubsection{KOI-1442: Largest contrast-ratio companion}
We detect a likely companion to KOI-1442 (\Kepler\ magnitude of 12.52) at a separation of 2\farcs24 and a contrast ratio of $\sim$6.7 magnitudes. Because of the relatively large separation and large contrast ratio, this detection is more likely to be a background object rather than a physically bound companion.  KOI-1442.01 is a planet candidate with a period of 0.67d and a radius of 1.2 $R_\oplus$; however, if the fainter companion star is the source of the transit, the radius of the eclipsing object would be significantly larger---$\sim$20$\times$ larger if the companion has the same radius as KOI-1442. Especially since there are hints that very short-period systems may be more likely to be blended binaries \citep{Colon2012}, there might be concern that this candidate is a background eclipsing binary false positive.  However, against this hypothesis stands the centroid offset analysis of \citet{Bryson2013} as presented on the NEA, which suggests that the source of the transit could be at most maybe 0\farcs5 away from the target position.  Therefore, while this system is notable due to the faintness of its detected companion, the companion is unlikely to be the source of a false positive due to its large separation.

\subsubsection{KOI-1845: One likely false positive in a two-candidate system}
KOI-1845 hosts two planetary candidates: .01 is a 1.5 $R_\oplus$ candidate in a 1.97-d orbit, and .02 is a 21 $R_\oplus$ candidate in a 5.06-d orbit.  Without any AO observations this system would be suspicious because close-in giant planets are very unlikely to have other planets nearby (see \S\ref{sec:koi191}); in addition, candidate .02 has a very large \Kepler\,-estimated radius and appears to have a significantly V-shaped transit.  In this survey we detect a companion 5.0 mag fainter at a separation of 2\farcs06, and suggest that the most likely explanation for KOI-1845.02 is that this companion is a background eclipsing binary.

\subsubsection{Systems with secure small planets}
There are five systems that host planet candidates with $R_p < 2 R_\oplus$ in which we have detected stellar companions but whose interpretation as small planets ($<$$2 R_\oplus$) is nonetheless secure, as long as the companions are physically bound.  This happens when the candidates are small and the companion is of comparable brightness such that the potential effect of dilution is minimized, even if the eclipse is around the fainter star.  The specifics of these systems can be seen in Table \ref{table:koidetails} but we call attention to them here:  KOI-1613, KOI-1619, KOI-2059, KOI-2463, and KOI-2657.

\subsection{Stellar Multiplicity and \Kepler\, Planet Candidates}
\label{sec:bin_rates}
Our detection of 53 planetary candidates with nearby stars, from 715 targets, implies an overall nearby-star probability of 7.4\%$\pm$1.0\%, within the detectable separation range of our survey (0\farcs15 to 2.5\farcs, $\Delta m \lsim 6$). 

In this section we go on to search for broad-scale correlations between stellar multiplicity and planetary candidate properties. The companions we detect may not be physically bound, nor are we sensitive to binaries in all possible orbital locations around these KOIs.  This multiplicity rate, therefore, should not be expected give a full description of the physical stellar multiplicity of \Kepler\, planet candidates; however, we can use the current survey results to compare the multiplicity rates of different populations of planet candidates. Future papers from the ongoing Robo-AO survey will investigate the multiplicity properties of \Kepler\, candidates in more detail, including quantifying the effects of association probability and incompleteness.  

The above nearby-star probability calculation and the following sections use the binomial distribution to calculate the uncertainty ranges in the multiplicity fractions (e.g. \citealt{Burgasser2003}) and Fisher exact tests (e.g. \citealt{Feigelson2012}) to evaluate the significance of differences in multiplicity between different populations.

\subsubsection{Stellar multiplicity rates vs. host-star temperature}
\begin{figure}
  \centering
  \resizebox{1.0\columnwidth}{!}
   {
     \includegraphics{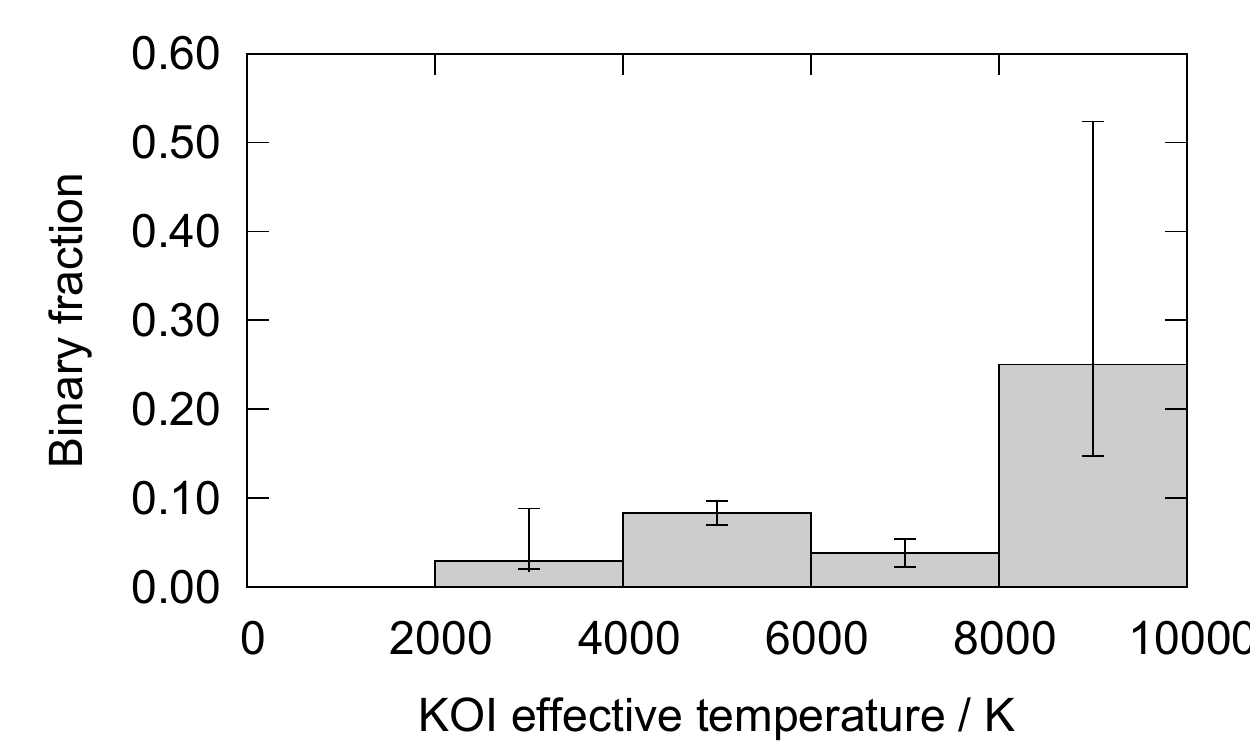}
   }
   \caption{The fraction of KOIs with detected nearby stars as a function of stellar effective temperature.}

   \label{fig:steff}
\end{figure}

Figure \ref{fig:steff} shows the fraction of multiple stellar systems around \Kepler\,-detected planetary systems as a function of stellar temperature from the \Kepler\ Input Catalog \citep{Brown2011}. The hottest stars appear to have an increased stellar multiplicity fraction, but there is a 16\% probability this is due to chance. We thus do not detect any significant change in the stellar multiplicity fraction with KOI temperature, although the initial survey presented here does not yet cover the entire \Kepler\, sample of non-solar-type stars.

\begin{figure}
  \centering
  \resizebox{1.0\columnwidth}{!}
   {
    \includegraphics{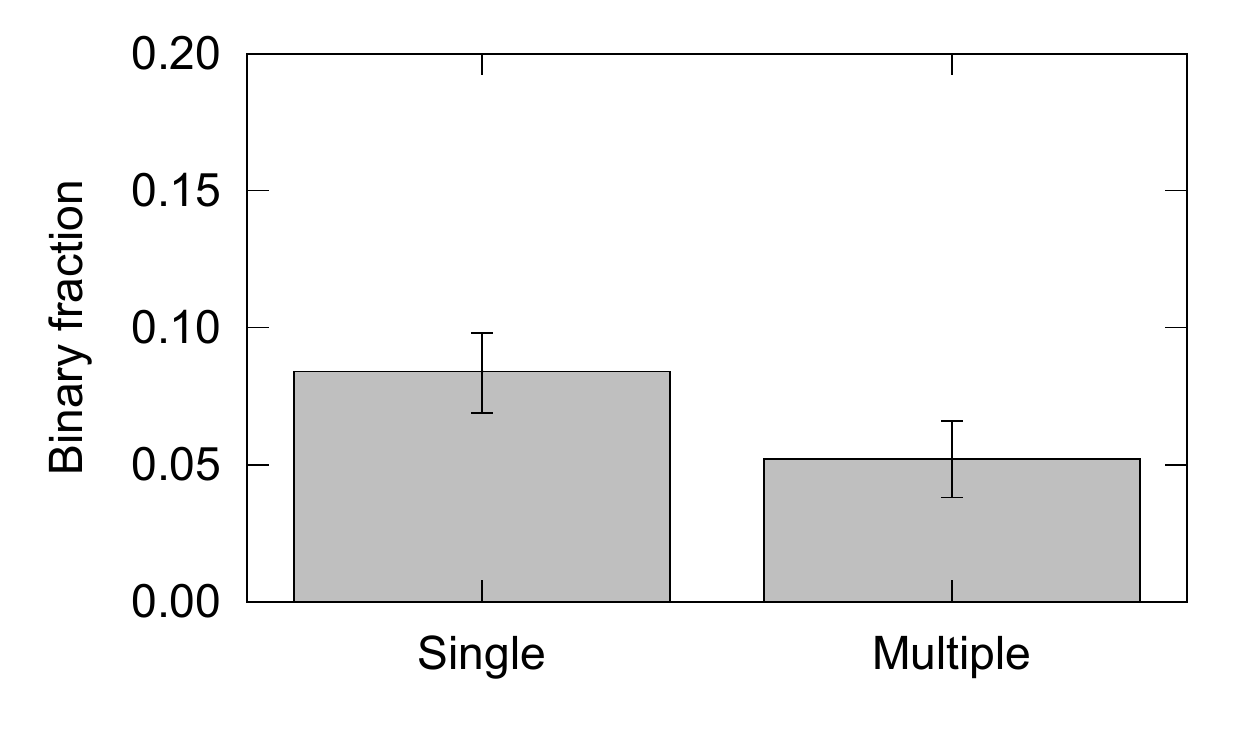}
   }
   \caption{The binarity fractions of KOIs hosting single and multiple detected planetary systems.}

   \label{fig:multiple_planets}
\end{figure}

\subsubsection{Stellar multiplicity and multiple-planet systems}
It is expected that multiple-planet systems detected by \Kepler\ are less likely to be false positives than single-planet systems because there are far fewer false-positive scenarios which can lead to multiple-period false-positives. In Figure \ref{fig:multiple_planets} we show the stellar multiplicity rates for single and multiple planet detections. There is a difference in stellar multiplicity between the single and multiple planet detections, but a Fisher exact test shows a 13\% probability of this being a chance difference due to small-number statistics. At least in the current dataset we cannot distinguish stellar multiplicity between single and multiple planet systems.

\subsubsection{Stellar multiplicity and close-in planets}
Stellar binarity has been hypothesized to be important in shaping the architectures of planetary systems, both by regulating planet formation and by dynamically sculpting planets’ final orbits, such as forcing Kozai oscillations that cause planet migration \citep{Fabrycky2007, Katz2011, Naoz2012} or by tilting the circumstellar disk \citep{Batygin2012}. If planetary migration is induced by a third body, one would expect to find a correlation between the presence of a detected third body and the presence of short-period planets.

\begin{figure}
  \resizebox{1.0\columnwidth}{!}
   {
    \includegraphics{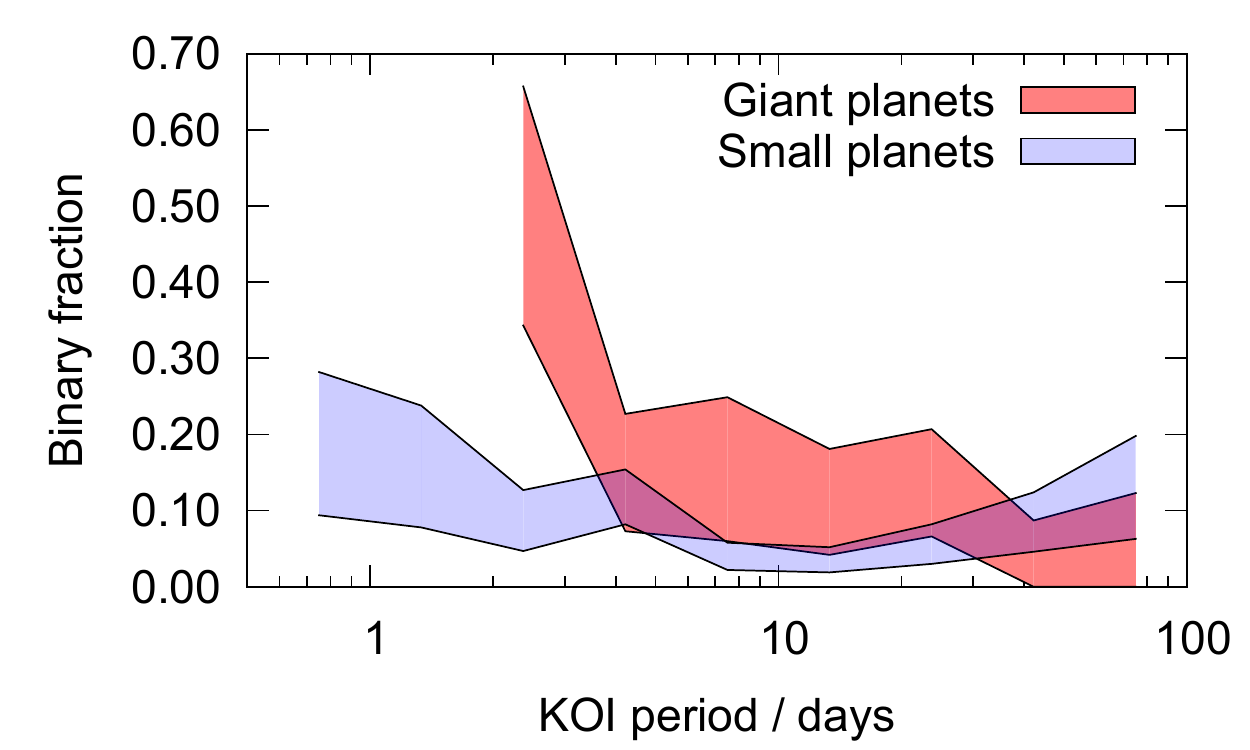}
   }
   \caption{1$\sigma$ uncertainty regions for binarity fraction as a function of KOI period for two different planetary populations (we split ``small'' from ``giant'' at Neptune's radius (3.9 $R_{\earth}$), but the exact value of the split does not significantly affect the uncertainty region shape). The gas giants cut off for shorter periods because of insufficient targets for acceptable statistics}.

   \label{fig:period_hist_uncert_regions}
\end{figure}

\begin{figure}
  \centering
  \resizebox{1.0\columnwidth}{!}
   {
     \includegraphics{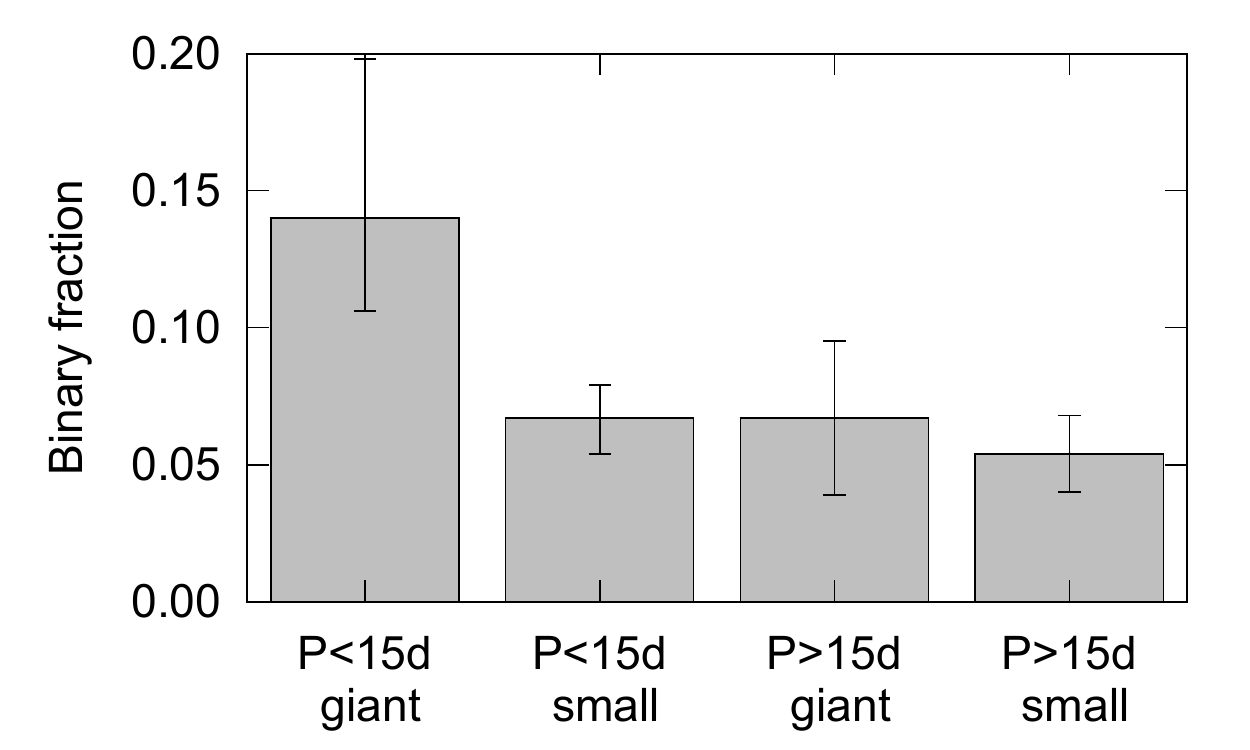}

   }
   \caption{Fraction of KOIs with nearby stars for four different planetary populations. Giant here is shorthand for a radius equal to or larger than that of Neptune. We assign KOIs to these populations if any planet in the system meets the requirements; a small number of multiple-planet systems are therefore assigned to multiple populations.}

   \label{fig:planet_pops_barchart}
\end{figure}

Figure \ref{fig:period_hist_uncert_regions} shows the fraction of \Kepler\ planet candidates with nearby stars as a function of the period of the closest-in planet, grouping the planets into two different size ranges. From these raw binarity fractions, where we have not accounted for the probability of physical association, it appears that while small planets do not show a significant change in third-body probability with the orbital period of the \Kepler\ candidate, giant planets show a significant increase at periods less than $\sim$15 days. Binning all our targets into only four population groups allows us to search for smaller changes in the binarity statistics (Figure \ref{fig:planet_pops_barchart}). We arbitrarily split ``small'' planets from ``giant'' planets at Neptune's radius (3.9 $R_{\earth}$), but the exact value of the split does not significantly affect the results; only two of the detected systems have planetary radii within 20\% of the cutoff value. We see that small planets at short periods share the same binarity fraction as all sizes of planets with $>$15d periods (within statistical errors). However, the short period giant planets again show a significantly increased binarity fraction. A Fisher exact test rejects the hypothesis that the two planetary populations have the same binarity fraction, at the 95\% level.

We can attempt to remove the background asterisms by selecting on the basis of magnitude ratio, as faint background stars are more likely to be chance alignments than roughly-equal-brightness companions. Our survey displayed an excess of close-separation bright companions: there are 13 companions with $\rm\Delta m < 2$ with separations $<$1.5\arcsec, and only one at larger radii (Figure \ref{fig:koi_comps}), while the numbers of fainter companions do not show such a bias. We suggest that this excess reveals a bright-companion population which is more likely to be physically associated than an average companion in the survey. 

\begin{figure}
  \centering
  \resizebox{1.0\columnwidth}{!}
   {
     \includegraphics{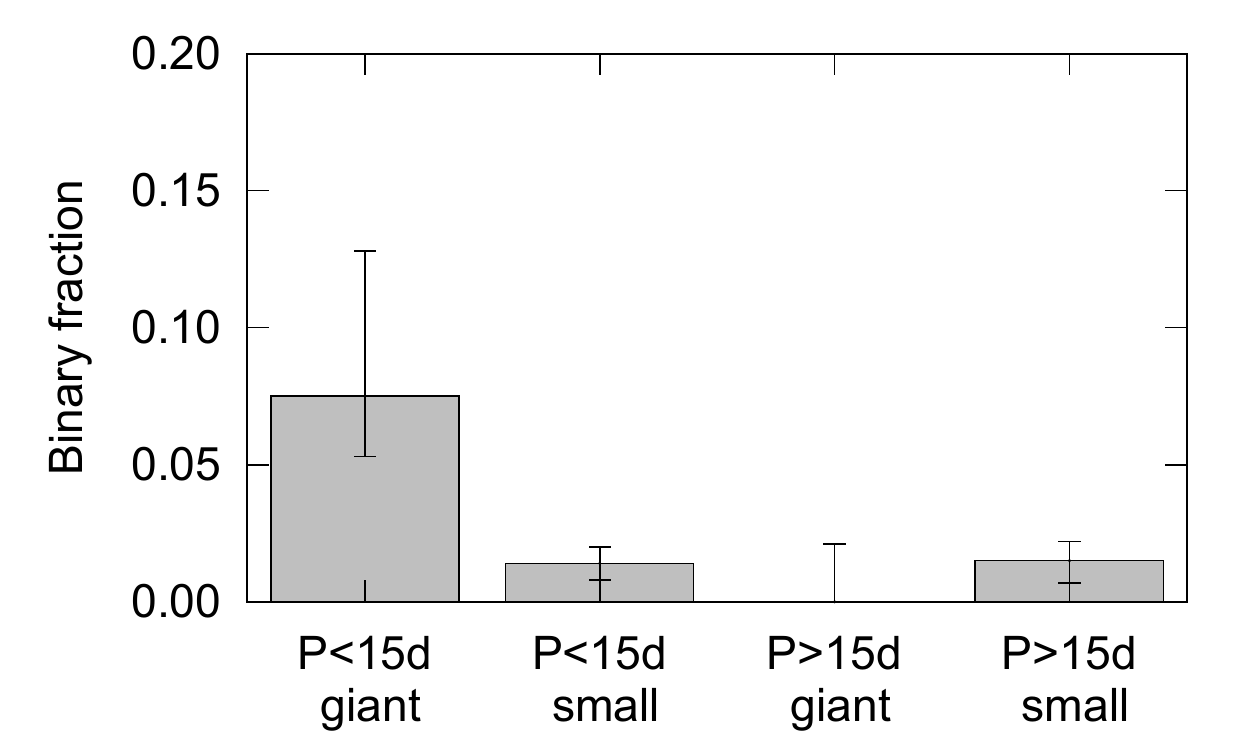}

   }
   \caption{Fraction of KOIs with nearby stars for four different planetary populations -- as figure \ref{fig:planet_pops_barchart} with only companions with $\rm\Delta m<2$ and separations $<$1\farcs5, removing faint nearby stars which are less likely to be physically associated (we did not detect any bright companions around the 84 longer-period giant planet KOIs in our survey, so we only show an upper limit). There is a 98\%-confidence detection of a difference in stellar multiplicity rates for close-in giant planets compared to further-out giants.}

   \label{fig:planet_pops_barchart_contrastsel}
\end{figure}

Selecting the companions with $\rm\Delta m<2$ and separation $<$1\farcs5 leads an increased difference in stellar multiplicity between the planetary populations (Figure \ref{fig:planet_pops_barchart_contrastsel}), increasing the significance to 98\%. This approach does not fully account for the probability of each companion being physically associated, and so its results should be interpreted with caution. For example, close-in companions are less likely to be rejected by the \Kepler\, centroid-based false-positive tests, but it is not obvious why this rejection would be different for planetary systems with short-period ($<$15d) and longer-period KOIs (with a median period of 54d for the KOIs we surveyed). In fact, the shorter-period systems have more eclipse events in the \Kepler\, dataset and it should therefore be easier to detect a small centroid shift from close-in companions. 

On the basis of our current analysis, we suggest that the difference of multiplicity rates between the planetary populations may be tentative evidence for third bodies in stellar systems producing an excess of close-in giant planets. We expect the full Robo-AO surveys to be able to evaluate this possibility at more than the 3$\sigma$ confidence level.

\section{Conclusions}
\label{sec:concs}

We observed 715 \Kepler\ planetary system candidates with the Robo-AO robotic laser adaptive optics system.  Our detection of 53 planetary candidates with nearby stars from 715 targets implies an overall nearby-star probability of 7.4\%$\pm$1.0\% at separations between 0\farcs1 and 2\farcs5 and $\Delta m \lsim 6$. We have detailed the effects of the detected nearby stars on the interpretation of the \Kepler\, planetary candidates, including the detection of probable ''co-incident'' multiples (KOI-191 and KOI-1151), multiple-planet systems likely containing false positives (KOI-1845), and the confirmation of five KOIs as roughly Earth-radius planets in multiple stellar systems (KOI-1613, KOI-1619, KOI-2059, KOI-2463, and KOI 2657). We have also found tentative, 98\%-confidence, evidence for stellar third bodies leading to a 2-3$\times$ increased rate of close-in giant planets.

We expect the ongoing Robo-AO surveys to complete observations of every \Kepler\ planet candidate by the end of 2014. The increased survey numbers will allow us to search for stellar multiplicity correlations only in multiple-detected-planet systems, which are expected to have a much lower false-positive probability, and thus will improve our ability to disentangle false-positives from astrophysical effects. The number of multiple systems in our current sample is not large enough to verify our tentative conclusions on the effects of stellar multiplicity on short-period giant planets (in particular, we have only covered one multiple-planet system with a short-period giant planet), but we plan to investigate these possibilities in future data releases. We are also continuing observations of our detected companions to search for common-proper-motion pairs. The completed Robo-AO survey will also allow us to confirm many more \Kepler\ planet candidates and likely find more exotic planetary systems.

\acknowledgments
\section*{Acknowledgements}
We thank the anonymous referee for careful analysis and useful comments on the manuscript. The Robo-AO system is supported by collaborating partner institutions, the California Institute of Technology and the Inter-University Centre for Astronomy and Astrophysics, and by the National Science Foundation under Grant Nos. AST-0906060 and AST-0960343, by the Mount Cuba Astronomical Foundation, by a gift from Samuel Oschin. We are grateful to the Palomar Observatory staff for their ongoing support of Robo-AO on the 60-inch telescope, particularly S. Kunsman, M. Doyle, J. Henning, R. Walters, G. Van Idsinga, B. Baker, K. Dunscombe and D. Roderick. We recognize and acknowledge the very significant cultural role and reverence that the summit of Mauna Kea has always had within the indigenous Hawaiian community. We are most fortunate to have the opportunity to conduct observations from this mountain.  C.B and J.A.J. acknowledge support from the Alfred P. Sloan Foundation. J.A.J acknowledges support from the David and Lucile Packard Foundation.

{\it Facilities:} \facility{PO:1.5m (Robo-AO)}, \facility{Keck:II (NIRC2-NGS)}
\bibliographystyle{apj}
\bibliography{roboao_kois}

\begin{thebibliography}{45}
\expandafter\ifx\csname natexlab\endcsname\relax\def\natexlab#1{#1}\fi

\bibitem[{{Adams} {et~al.}(2012){Adams}, {Ciardi}, {Dupree}, {Gautier},
  {Kulesa}, \& {McCarthy}}]{Adams2012}
{Adams}, E.~R., {Ciardi}, D.~R., {Dupree}, A.~K., {Gautier}, III, T.~N.,
  {Kulesa}, C., \& {McCarthy}, D. 2012, \aj, 144, 42

\bibitem[{{Adams} {et~al.}(2013){Adams}, {Dupree}, {Kulesa}, \&
  {McCarthy}}]{Adams2013}
{Adams}, E.~R., {Dupree}, A.~K., {Kulesa}, C., \& {McCarthy}, D. 2013, \aj,
  146, 9

\bibitem[{Baranec {et~al.}(2013)Baranec, Riddle, Law, Ramaprakash, Tendulkar,
  Bui, Burse, Chordia, Das, Davis, Dekany, Kasliwal, Kulkarni, Morton, Ofek, \&
  Punnadi}]{Baranec:13}
Baranec, C., Riddle, R., Law, N.~M., Ramaprakash, A., Tendulkar, S.~P., Bui,
  K., Burse, M.~P., Chordia, P., Das, H.~K., Davis, J.~T., Dekany, R.~G.,
  Kasliwal, M.~M., Kulkarni, S.~R., Morton, T.~D., Ofek, E.~O., \& Punnadi, S.
  2013, Journal of Visualized Experiments, 72, e50021

\bibitem[{Baranec {et~al.}(2012)Baranec, Riddle, Ramaprakash, Law, Tendulkar,
  Kulkarni, Dekany, Bui, Davis, Burse, Das, Hildebrandt, Punnadi, \&
  Smith}]{Baranec:12}
Baranec, C., Riddle, R., Ramaprakash, A.~N., Law, N., Tendulkar, S., Kulkarni,
  S., Dekany, R., Bui, K., Davis, J., Burse, M., Das, H., Hildebrandt, S.,
  Punnadi, S., \& Smith, R. 2012, Proc. SPIE 8447, Adaptive Optics Systems III,
  8447, 844704

\bibitem[{{Barclay} {et~al.}(2013){Barclay}, {Rowe}, {Lissauer}, {Huber},
  {Fressin}, {Howell}, {Bryson}, {Chaplin}, {D{\'e}sert}, {Lopez}, {Marcy},
  {Mullally}, {Ragozzine}, {Torres}, {Adams}, {Agol}, {Barrado}, {Basu},
  {Bedding}, {Buchhave}, {Charbonneau}, {Christiansen},
  {Christensen-Dalsgaard}, {Ciardi}, {Cochran}, {Dupree}, {Elsworth},
  {Everett}, {Fischer}, {Ford}, {Fortney}, {Geary}, {Haas}, {Handberg},
  {Hekker}, {Henze}, {Horch}, {Howard}, {Hunter}, {Isaacson}, {Jenkins},
  {Karoff}, {Kawaler}, {Kjeldsen}, {Klaus}, {Latham}, {Li}, {Lillo-Box},
  {Lund}, {Lundkvist}, {Metcalfe}, {Miglio}, {Morris}, {Quintana}, {Stello},
  {Smith}, {Still}, \& {Thompson}}]{barclay2013}
{Barclay}, T., {Rowe}, J.~F., {Lissauer}, J.~J., {Huber}, D., {Fressin}, F.,
  {Howell}, S.~B., {Bryson}, S.~T., {Chaplin}, W.~J., {D{\'e}sert}, J.-M.,
  {Lopez}, E.~D., {Marcy}, G.~W., {Mullally}, F., {Ragozzine}, D., {Torres},
  G., {Adams}, E.~R., {Agol}, E., {Barrado}, D., {Basu}, S., {Bedding}, T.~R.,
  {Buchhave}, L.~A., {Charbonneau}, D., {Christiansen}, J.~L.,
  {Christensen-Dalsgaard}, J., {Ciardi}, D., {Cochran}, W.~D., {Dupree}, A.~K.,
  {Elsworth}, Y., {Everett}, M., {Fischer}, D.~A., {Ford}, E.~B., {Fortney},
  J.~J., {Geary}, J.~C., {Haas}, M.~R., {Handberg}, R., {Hekker}, S., {Henze},
  C.~E., {Horch}, E., {Howard}, A.~W., {Hunter}, R.~C., {Isaacson}, H.,
  {Jenkins}, J.~M., {Karoff}, C., {Kawaler}, S.~D., {Kjeldsen}, H., {Klaus},
  T.~C., {Latham}, D.~W., {Li}, J., {Lillo-Box}, J., {Lund}, M.~N.,
  {Lundkvist}, M., {Metcalfe}, T.~S., {Miglio}, A., {Morris}, R.~L.,
  {Quintana}, E.~V., {Stello}, D., {Smith}, J.~C., {Still}, M., \& {Thompson},
  S.~E. 2013, \nat, 494, 452

\bibitem[{{Barrado} {et~al.}(2013){Barrado}, {Lillo-Box}, {Bouy}, {Aceituno},
  \& {S{\'a}nchez}}]{Barrado2013}
{Barrado}, D., {Lillo-Box}, J., {Bouy}, H., {Aceituno}, J., \& {S{\'a}nchez},
  S. 2013, in European Physical Journal Web of Conferences, Vol.~47, European
  Physical Journal Web of Conferences, 5008

\bibitem[{{Batalha} {et~al.}(2013){Batalha}, {Rowe}, {Bryson}, {Barclay},
  {Burke}, {Caldwell}, {Christiansen}, {Mullally}, {Thompson}, {Brown},
  {Dupree}, {Fabrycky}, {Ford}, {Fortney}, {Gilliland}, {Isaacson}, {Latham},
  {Marcy}, {Quinn}, {Ragozzine}, {Shporer}, {Borucki}, {Ciardi}, {Gautier},
  {Haas}, {Jenkins}, {Koch}, {Lissauer}, {Rapin}, {Basri}, {Boss}, {Buchhave},
  {Carter}, {Charbonneau}, {Christensen-Dalsgaard}, {Clarke}, {Cochran},
  {Demory}, {Desert}, {Devore}, {Doyle}, {Esquerdo}, {Everett}, {Fressin},
  {Geary}, {Girouard}, {Gould}, {Hall}, {Holman}, {Howard}, {Howell},
  {Ibrahim}, {Kinemuchi}, {Kjeldsen}, {Klaus}, {Li}, {Lucas}, {Meibom},
  {Morris}, {Pr{\v s}a}, {Quintana}, {Sanderfer}, {Sasselov}, {Seader},
  {Smith}, {Steffen}, {Still}, {Stumpe}, {Tarter}, {Tenenbaum}, {Torres},
  {Twicken}, {Uddin}, {Van Cleve}, {Walkowicz}, \& {Welsh}}]{Batalha2012}
{Batalha}, N.~M., {Rowe}, J.~F., {Bryson}, S.~T., {Barclay}, T., {Burke},
  C.~J., {Caldwell}, D.~A., {Christiansen}, J.~L., {Mullally}, F., {Thompson},
  S.~E., {Brown}, T.~M., {Dupree}, A.~K., {Fabrycky}, D.~C., {Ford}, E.~B.,
  {Fortney}, J.~J., {Gilliland}, R.~L., {Isaacson}, H., {Latham}, D.~W.,
  {Marcy}, G.~W., {Quinn}, S.~N., {Ragozzine}, D., {Shporer}, A., {Borucki},
  W.~J., {Ciardi}, D.~R., {Gautier}, III, T.~N., {Haas}, M.~R., {Jenkins},
  J.~M., {Koch}, D.~G., {Lissauer}, J.~J., {Rapin}, W., {Basri}, G.~S., {Boss},
  A.~P., {Buchhave}, L.~A., {Carter}, J.~A., {Charbonneau}, D.,
  {Christensen-Dalsgaard}, J., {Clarke}, B.~D., {Cochran}, W.~D., {Demory},
  B.-O., {Desert}, J.-M., {Devore}, E., {Doyle}, L.~R., {Esquerdo}, G.~A.,
  {Everett}, M., {Fressin}, F., {Geary}, J.~C., {Girouard}, F.~R., {Gould}, A.,
  {Hall}, J.~R., {Holman}, M.~J., {Howard}, A.~W., {Howell}, S.~B., {Ibrahim},
  K.~A., {Kinemuchi}, K., {Kjeldsen}, H., {Klaus}, T.~C., {Li}, J., {Lucas},
  P.~W., {Meibom}, S., {Morris}, R.~L., {Pr{\v s}a}, A., {Quintana}, E.,
  {Sanderfer}, D.~T., {Sasselov}, D., {Seader}, S.~E., {Smith}, J.~C.,
  {Steffen}, J.~H., {Still}, M., {Stumpe}, M.~C., {Tarter}, J.~C., {Tenenbaum},
  P., {Torres}, G., {Twicken}, J.~D., {Uddin}, K., {Van Cleve}, J.,
  {Walkowicz}, L., \& {Welsh}, W.~F. 2013, \apjs, 204, 24

\bibitem[{{Batygin}(2012)}]{Batygin2012}
{Batygin}, K. 2012, \nat, 491, 418

\bibitem[{{Borucki} {et~al.}(2010){Borucki}, {Koch}, {Basri}, {Batalha},
  {Brown}, {Caldwell}, {Caldwell}, {Christensen-Dalsgaard}, {Cochran},
  {DeVore}, {Dunham}, {Dupree}, {Gautier}, {Geary}, {Gilliland}, {Gould},
  {Howell}, {Jenkins}, {Kondo}, {Latham}, {Marcy}, {Meibom}, {Kjeldsen},
  {Lissauer}, {Monet}, {Morrison}, {Sasselov}, {Tarter}, {Boss}, {Brownlee},
  {Owen}, {Buzasi}, {Charbonneau}, {Doyle}, {Fortney}, {Ford}, {Holman},
  {Seager}, {Steffen}, {Welsh}, {Rowe}, {Anderson}, {Buchhave}, {Ciardi},
  {Walkowicz}, {Sherry}, {Horch}, {Isaacson}, {Everett}, {Fischer}, {Torres},
  {Johnson}, {Endl}, {MacQueen}, {Bryson}, {Dotson}, {Haas}, {Kolodziejczak},
  {Van Cleve}, {Chandrasekaran}, {Twicken}, {Quintana}, {Clarke}, {Allen},
  {Li}, {Wu}, {Tenenbaum}, {Verner}, {Bruhweiler}, {Barnes}, \&
  {Prsa}}]{Borucki2010}
{Borucki}, W.~J., {Koch}, D., {Basri}, G., {Batalha}, N., {Brown}, T.,
  {Caldwell}, D., {Caldwell}, J., {Christensen-Dalsgaard}, J., {Cochran},
  W.~D., {DeVore}, E., {Dunham}, E.~W., {Dupree}, A.~K., {Gautier}, T.~N.,
  {Geary}, J.~C., {Gilliland}, R., {Gould}, A., {Howell}, S.~B., {Jenkins},
  J.~M., {Kondo}, Y., {Latham}, D.~W., {Marcy}, G.~W., {Meibom}, S.,
  {Kjeldsen}, H., {Lissauer}, J.~J., {Monet}, D.~G., {Morrison}, D.,
  {Sasselov}, D., {Tarter}, J., {Boss}, A., {Brownlee}, D., {Owen}, T.,
  {Buzasi}, D., {Charbonneau}, D., {Doyle}, L., {Fortney}, J., {Ford}, E.~B.,
  {Holman}, M.~J., {Seager}, S., {Steffen}, J.~H., {Welsh}, W.~F., {Rowe}, J.,
  {Anderson}, H., {Buchhave}, L., {Ciardi}, D., {Walkowicz}, L., {Sherry}, W.,
  {Horch}, E., {Isaacson}, H., {Everett}, M.~E., {Fischer}, D., {Torres}, G.,
  {Johnson}, J.~A., {Endl}, M., {MacQueen}, P., {Bryson}, S.~T., {Dotson}, J.,
  {Haas}, M., {Kolodziejczak}, J., {Van Cleve}, J., {Chandrasekaran}, H.,
  {Twicken}, J.~D., {Quintana}, E.~V., {Clarke}, B.~D., {Allen}, C., {Li}, J.,
  {Wu}, H., {Tenenbaum}, P., {Verner}, E., {Bruhweiler}, F., {Barnes}, J., \&
  {Prsa}, A. 2010, Science, 327, 977

\bibitem[{{Borucki} {et~al.}(2011){Borucki}, {Koch}, {Basri}, {Batalha},
  {Brown}, {Bryson}, {Caldwell}, {Christensen-Dalsgaard}, {Cochran}, {DeVore},
  {Dunham}, {Gautier}, {Geary}, {Gilliland}, {Gould}, {Howell}, {Jenkins},
  {Latham}, {Lissauer}, {Marcy}, {Rowe}, {Sasselov}, {Boss}, {Charbonneau},
  {Ciardi}, {Doyle}, {Dupree}, {Ford}, {Fortney}, {Holman}, {Seager},
  {Steffen}, {Tarter}, {Welsh}, {Allen}, {Buchhave}, {Christiansen}, {Clarke},
  {Das}, {D{\'e}sert}, {Endl}, {Fabrycky}, {Fressin}, {Haas}, {Horch},
  {Howard}, {Isaacson}, {Kjeldsen}, {Kolodziejczak}, {Kulesa}, {Li}, {Lucas},
  {Machalek}, {McCarthy}, {MacQueen}, {Meibom}, {Miquel}, {Prsa}, {Quinn},
  {Quintana}, {Ragozzine}, {Sherry}, {Shporer}, {Tenenbaum}, {Torres},
  {Twicken}, {Van Cleve}, {Walkowicz}, {Witteborn}, \& {Still}}]{Borucki2011}
{Borucki}, W.~J., {Koch}, D.~G., {Basri}, G., {Batalha}, N., {Brown}, T.~M.,
  {Bryson}, S.~T., {Caldwell}, D., {Christensen-Dalsgaard}, J., {Cochran},
  W.~D., {DeVore}, E., {Dunham}, E.~W., {Gautier}, III, T.~N., {Geary}, J.~C.,
  {Gilliland}, R., {Gould}, A., {Howell}, S.~B., {Jenkins}, J.~M., {Latham},
  D.~W., {Lissauer}, J.~J., {Marcy}, G.~W., {Rowe}, J., {Sasselov}, D., {Boss},
  A., {Charbonneau}, D., {Ciardi}, D., {Doyle}, L., {Dupree}, A.~K., {Ford},
  E.~B., {Fortney}, J., {Holman}, M.~J., {Seager}, S., {Steffen}, J.~H.,
  {Tarter}, J., {Welsh}, W.~F., {Allen}, C., {Buchhave}, L.~A., {Christiansen},
  J.~L., {Clarke}, B.~D., {Das}, S., {D{\'e}sert}, J.-M., {Endl}, M.,
  {Fabrycky}, D., {Fressin}, F., {Haas}, M., {Horch}, E., {Howard}, A.,
  {Isaacson}, H., {Kjeldsen}, H., {Kolodziejczak}, J., {Kulesa}, C., {Li}, J.,
  {Lucas}, P.~W., {Machalek}, P., {McCarthy}, D., {MacQueen}, P., {Meibom}, S.,
  {Miquel}, T., {Prsa}, A., {Quinn}, S.~N., {Quintana}, E.~V., {Ragozzine}, D.,
  {Sherry}, W., {Shporer}, A., {Tenenbaum}, P., {Torres}, G., {Twicken}, J.~D.,
  {Van Cleve}, J., {Walkowicz}, L., {Witteborn}, F.~C., \& {Still}, M. 2011,
  \apj, 736, 19

\bibitem[{{Brown} {et~al.}(2011){Brown}, {Latham}, {Everett}, \&
  {Esquerdo}}]{Brown2011}
{Brown}, T.~M., {Latham}, D.~W., {Everett}, M.~E., \& {Esquerdo}, G.~A. 2011,
  \aj, 142, 112

\bibitem[{{Bryson} {et~al.}(2013){Bryson}, {Jenkins}, {Gilliland}, {Twicken},
  {Clarke}, {Rowe}, {Caldwell}, {Batalha}, {Mullally}, {Haas}, \&
  {Tenenbaum}}]{Bryson2013}
{Bryson}, S.~T., {Jenkins}, J.~M., {Gilliland}, R.~L., {Twicken}, J.~D.,
  {Clarke}, B., {Rowe}, J., {Caldwell}, D., {Batalha}, N., {Mullally}, F.,
  {Haas}, M.~R., \& {Tenenbaum}, P. 2013, \pasp, 125, 889

\bibitem[{{Buchhave} {et~al.}(2011){Buchhave}, {Latham}, {Carter},
  {D{\'e}sert}, {Torres}, {Adams}, {Bryson}, {Charbonneau}, {Ciardi}, {Kulesa},
  {Dupree}, {Fischer}, {Fressin}, {Gautier}, {Gilliland}, {Howell}, {Isaacson},
  {Jenkins}, {Marcy}, {McCarthy}, {Rowe}, {Batalha}, {Borucki}, {Brown},
  {Caldwell}, {Christiansen}, {Cochran}, {Deming}, {Dunham}, {Everett}, {Ford},
  {Fortney}, {Geary}, {Girouard}, {Haas}, {Holman}, {Horch}, {Klaus},
  {Knutson}, {Koch}, {Kolodziejczak}, {Lissauer}, {Machalek}, {Mullally},
  {Still}, {Quinn}, {Seager}, {Thompson}, \& {Van Cleve}}]{Buchhave2011}
{Buchhave}, L.~A., {Latham}, D.~W., {Carter}, J.~A., {D{\'e}sert}, J.-M.,
  {Torres}, G., {Adams}, E.~R., {Bryson}, S.~T., {Charbonneau}, D.~B.,
  {Ciardi}, D.~R., {Kulesa}, C., {Dupree}, A.~K., {Fischer}, D.~A., {Fressin},
  F., {Gautier}, III, T.~N., {Gilliland}, R.~L., {Howell}, S.~B., {Isaacson},
  H., {Jenkins}, J.~M., {Marcy}, G.~W., {McCarthy}, D.~W., {Rowe}, J.~F.,
  {Batalha}, N.~M., {Borucki}, W.~J., {Brown}, T.~M., {Caldwell}, D.~A.,
  {Christiansen}, J.~L., {Cochran}, W.~D., {Deming}, D., {Dunham}, E.~W.,
  {Everett}, M., {Ford}, E.~B., {Fortney}, J.~J., {Geary}, J.~C., {Girouard},
  F.~R., {Haas}, M.~R., {Holman}, M.~J., {Horch}, E., {Klaus}, T.~C.,
  {Knutson}, H.~A., {Koch}, D.~G., {Kolodziejczak}, J., {Lissauer}, J.~J.,
  {Machalek}, P., {Mullally}, F., {Still}, M.~D., {Quinn}, S.~N., {Seager}, S.,
  {Thompson}, S.~E., \& {Van Cleve}, J. 2011, \apjs, 197, 3

\bibitem[{{Burgasser} {et~al.}(2003){Burgasser}, {Kirkpatrick}, {Reid},
  {Brown}, {Miskey}, \& {Gizis}}]{Burgasser2003}
{Burgasser}, A.~J., {Kirkpatrick}, J.~D., {Reid}, I.~N., {Brown}, M.~E.,
  {Miskey}, C.~L., \& {Gizis}, J.~E. 2003, \apj, 586, 512

\bibitem[{Cenko {et~al.}(2006)Cenko, Fox, Moon, Harrison, Kulkarni, Henning,
  Guzman, Bonati, Smith, Thicksten, Doyle, Petrie, Gal-Yam, Soderberg,
  Anagnostou, \& Laity}]{Cenko:06}
Cenko, S.~B., Fox, D.~B., Moon, D.-S., Harrison, F.~A., Kulkarni, S.~R.,
  Henning, J.~R., Guzman, C.~D., Bonati, M., Smith, R.~M., Thicksten, R.~P.,
  Doyle, M.~W., Petrie, H.~L., Gal-Yam, A., Soderberg, A.~M., Anagnostou,
  N.~L., \& Laity, A.~C. 2006, The Publications of the Astronomical Society of
  the Pacific, 118, 1396

\bibitem[{{Col{\'o}n} {et~al.}(2012){Col{\'o}n}, {Ford}, \&
  {Morehead}}]{Colon2012}
{Col{\'o}n}, K.~D., {Ford}, E.~B., \& {Morehead}, R.~C. 2012, \mnras, 426, 342

\bibitem[{{Daemgen} {et~al.}(2009){Daemgen}, {Hormuth}, {Brandner}, {Bergfors},
  {Janson}, {Hippler}, \& {Henning}}]{Daemgen2009}
{Daemgen}, S., {Hormuth}, F., {Brandner}, W., {Bergfors}, C., {Janson}, M.,
  {Hippler}, S., \& {Henning}, T. 2009, \aap, 498, 567

\bibitem[{{Dotter} {et~al.}(2008){Dotter}, {Chaboyer}, {Jevremovi{\'c}},
  {Kostov}, {Baron}, \& {Ferguson}}]{Dotter2008}
{Dotter}, A., {Chaboyer}, B., {Jevremovi{\'c}}, D., {Kostov}, V., {Baron}, E.,
  \& {Ferguson}, J.~W. 2008, \apjs, 178, 89

\bibitem[{{Dressing} \& {Charbonneau}(2013)}]{Dressing2013}
{Dressing}, C.~D. \& {Charbonneau}, D. 2013, \apj, 767, 95

\bibitem[{{Fabrycky} \& {Tremaine}(2007)}]{Fabrycky2007}
{Fabrycky}, D. \& {Tremaine}, S. 2007, \apj, 669, 1298

\bibitem[{{Feigelson} \& {Jogesh Babu}(2012)}]{Feigelson2012}
{Feigelson}, E.~D. \& {Jogesh Babu}, G. 2012, {Modern Statistical Methods for
  Astronomy}

\bibitem[{{Fressin} {et~al.}(2013){Fressin}, {Torres}, {Charbonneau}, {Bryson},
  {Christiansen}, {Dressing}, {Jenkins}, {Walkowicz}, \&
  {Batalha}}]{Fressin2013}
{Fressin}, F., {Torres}, G., {Charbonneau}, D., {Bryson}, S.~T.,
  {Christiansen}, J., {Dressing}, C.~D., {Jenkins}, J.~M., {Walkowicz}, L.~M.,
  \& {Batalha}, N.~M. 2013, \apj, 766, 81

\bibitem[{{Horch} {et~al.}(2012){Horch}, {Howell}, {Everett}, \&
  {Ciardi}}]{Horch2012}
{Horch}, E.~P., {Howell}, S.~B., {Everett}, M.~E., \& {Ciardi}, D.~R. 2012,
  \aj, 144, 165

\bibitem[{{Howell} {et~al.}(2011){Howell}, {Everett}, {Sherry}, {Horch}, \&
  {Ciardi}}]{Howell2011}
{Howell}, S.~B., {Everett}, M.~E., {Sherry}, W., {Horch}, E., \& {Ciardi},
  D.~R. 2011, \aj, 142, 19

\bibitem[{{Johnson} {et~al.}(2011){Johnson}, {Apps}, {Gazak}, {Crepp},
  {Crossfield}, {Howard}, {Marcy}, {Morton}, {Chubak}, \&
  {Isaacson}}]{Johnson2011}
{Johnson}, J.~A., {Apps}, K., {Gazak}, J.~Z., {Crepp}, J.~R., {Crossfield},
  I.~J., {Howard}, A.~W., {Marcy}, G.~W., {Morton}, T.~D., {Chubak}, C., \&
  {Isaacson}, H. 2011, \apj, 730, 79

\bibitem[{{Katz} {et~al.}(2011){Katz}, {Dong}, \& {Malhotra}}]{Katz2011}
{Katz}, B., {Dong}, S., \& {Malhotra}, R. 2011, Physical Review Letters, 107,
  181101

\bibitem[{{Koch} {et~al.}(2010){Koch}, {Borucki}, {Basri}, {Batalha}, {Brown},
  {Caldwell}, {Christensen-Dalsgaard}, {Cochran}, {DeVore}, {Dunham},
  {Gautier}, {Geary}, {Gilliland}, {Gould}, {Jenkins}, {Kondo}, {Latham},
  {Lissauer}, {Marcy}, {Monet}, {Sasselov}, {Boss}, {Brownlee}, {Caldwell},
  {Dupree}, {Howell}, {Kjeldsen}, {Meibom}, {Morrison}, {Owen}, {Reitsema},
  {Tarter}, {Bryson}, {Dotson}, {Gazis}, {Haas}, {Kolodziejczak}, {Rowe}, {Van
  Cleve}, {Allen}, {Chandrasekaran}, {Clarke}, {Li}, {Quintana}, {Tenenbaum},
  {Twicken}, \& {Wu}}]{Koch2010}
{Koch}, D.~G., {Borucki}, W.~J., {Basri}, G., {Batalha}, N.~M., {Brown}, T.~M.,
  {Caldwell}, D., {Christensen-Dalsgaard}, J., {Cochran}, W.~D., {DeVore}, E.,
  {Dunham}, E.~W., {Gautier}, III, T.~N., {Geary}, J.~C., {Gilliland}, R.~L.,
  {Gould}, A., {Jenkins}, J., {Kondo}, Y., {Latham}, D.~W., {Lissauer}, J.~J.,
  {Marcy}, G., {Monet}, D., {Sasselov}, D., {Boss}, A., {Brownlee}, D.,
  {Caldwell}, J., {Dupree}, A.~K., {Howell}, S.~B., {Kjeldsen}, H., {Meibom},
  S., {Morrison}, D., {Owen}, T., {Reitsema}, H., {Tarter}, J., {Bryson},
  S.~T., {Dotson}, J.~L., {Gazis}, P., {Haas}, M.~R., {Kolodziejczak}, J.,
  {Rowe}, J.~F., {Van Cleve}, J.~E., {Allen}, C., {Chandrasekaran}, H.,
  {Clarke}, B.~D., {Li}, J., {Quintana}, E.~V., {Tenenbaum}, P., {Twicken},
  J.~D., \& {Wu}, H. 2010, \apjl, 713, L79

\bibitem[{{Lafreni{\`e}re} {et~al.}(2007){Lafreni{\`e}re}, {Marois}, {Doyon},
  {Nadeau}, \& {Artigau}}]{Lafreniere2007}
{Lafreni{\`e}re}, D., {Marois}, C., {Doyon}, R., {Nadeau}, D., \& {Artigau},
  {\'E}. 2007, \apj, 660, 770

\bibitem[{{Law} {et~al.}(2006{\natexlab{a}}){Law}, {Hodgkin}, \&
  {Mackay}}]{Law2006}
{Law}, N.~M., {Hodgkin}, S.~T., \& {Mackay}, C.~D. 2006{\natexlab{a}}, \mnras,
  368, 1917

\bibitem[{{Law} {et~al.}(2006{\natexlab{b}}){Law}, {Hodgkin}, \&
  {Mackay}}]{Law2006bins}
---. 2006{\natexlab{b}}, \mnras, 368, 1917

\bibitem[{{Law} {et~al.}(2012){Law}, {Kraus}, {Street}, {Fulton},
  {Hillenbrand}, {Shporer}, {Lister}, {Baranec}, {Bloom}, {Bui}, {Burse},
  {Cenko}, {Das}, {Davis}, {Dekany}, {Filippenko}, {Kasliwal}, {Kulkarni},
  {Nugent}, {Ofek}, {Poznanski}, {Quimby}, {Ramaprakash}, {Riddle},
  {Silverman}, {Sivanandam}, \& {Tendulkar}}]{Law2012}
{Law}, N.~M., {Kraus}, A.~L., {Street}, R., {Fulton}, B.~J., {Hillenbrand},
  L.~A., {Shporer}, A., {Lister}, T., {Baranec}, C., {Bloom}, J.~S., {Bui}, K.,
  {Burse}, M.~P., {Cenko}, S.~B., {Das}, H.~K., {Davis}, J.~T.~C., {Dekany},
  R.~G., {Filippenko}, A.~V., {Kasliwal}, M.~M., {Kulkarni}, S.~R., {Nugent},
  P., {Ofek}, E.~O., {Poznanski}, D., {Quimby}, R.~M., {Ramaprakash}, A.~N.,
  {Riddle}, R., {Silverman}, J.~M., {Sivanandam}, S., \& {Tendulkar}, S.~P.
  2012, \apj, 757, 133

\bibitem[{{Law} {et~al.}(2009){Law}, {Mackay}, {Dekany}, {Ireland}, {Lloyd},
  {Moore}, {Robertson}, {Tuthill}, \& {Woodruff}}]{Law2009}
{Law}, N.~M., {Mackay}, C.~D., {Dekany}, R.~G., {Ireland}, M., {Lloyd}, J.~P.,
  {Moore}, A.~M., {Robertson}, J.~G., {Tuthill}, P., \& {Woodruff}, H.~C. 2009,
  \apj, 692, 924

\bibitem[{{Lillo-Box} {et~al.}(2012){Lillo-Box}, {Barrado}, \&
  {Bouy}}]{LilloBox2012}
{Lillo-Box}, J., {Barrado}, D., \& {Bouy}, H. 2012, \aap, 546, A10

\bibitem[{{Marcy} {et~al.}(2014){Marcy}, {Isaacson}, {Howard}, {Rowe},
  {Jenkins}, {Bryson}, {Latham}, {Howell}, {Gautier}, {Batalha}, {Rogers},
  {Ciardi}, {Fischer}, {Gilliland}, {Kjeldsen}, {Christensen-Dalsgaard},
  {Huber}, {Chaplin}, {Basu}, {Buchhave}, {Quinn}, {Borucki}, {Koch}, {Hunter},
  {Caldwell}, {Van Cleve}, {Kolbl}, {Weiss}, {Petigura}, {Seager}, {Morton},
  {Johnson}, {Ballard}, {Burke}, {Cochran}, {Endl}, {MacQueen}, {Everett},
  {Lissauer}, {Ford}, {Torres}, {Fressin}, {Brown}, {Steffen}, {Charbonneau},
  {Basri}, {Sasselov}, {Winn}, {Sanchis-Ojeda}, {Christiansen}, {Adams},
  {Henze}, {Dupree}, {Fabrycky}, {Fortney}, {Tarter}, {Holman}, {Tenenbaum},
  {Shporer}, {Lucas}, {Welsh}, {Orosz}, {Bedding}, {Campante}, {Davies},
  {Elsworth}, {Handberg}, {Hekker}, {Karoff}, {Kawaler}, {Lund}, {Lundkvist},
  {Metcalfe}, {Miglio}, {Silva Aguirre}, {Stello}, {White}, {Boss}, {Devore},
  {Gould}, {Prsa}, {Agol}, {Barclay}, {Coughlin}, {Brugamyer}, {Mullally},
  {Quintana}, {Still}, {Thompson}, {Morrison}, {Twicken}, {D{\'e}sert},
  {Carter}, {Crepp}, {H{\'e}brard}, {Santerne}, {Moutou}, {Sobeck}, {Hudgins},
  {Haas}, {Robertson}, {Lillo-Box}, \& {Barrado}}]{Marcy2014}
{Marcy}, G.~W., {Isaacson}, H., {Howard}, A.~W., {Rowe}, J.~F., {Jenkins},
  J.~M., {Bryson}, S.~T., {Latham}, D.~W., {Howell}, S.~B., {Gautier}, III,
  T.~N., {Batalha}, N.~M., {Rogers}, L., {Ciardi}, D., {Fischer}, D.~A.,
  {Gilliland}, R.~L., {Kjeldsen}, H., {Christensen-Dalsgaard}, J., {Huber}, D.,
  {Chaplin}, W.~J., {Basu}, S., {Buchhave}, L.~A., {Quinn}, S.~N., {Borucki},
  W.~J., {Koch}, D.~G., {Hunter}, R., {Caldwell}, D.~A., {Van Cleve}, J.,
  {Kolbl}, R., {Weiss}, L.~M., {Petigura}, E., {Seager}, S., {Morton}, T.,
  {Johnson}, J.~A., {Ballard}, S., {Burke}, C., {Cochran}, W.~D., {Endl}, M.,
  {MacQueen}, P., {Everett}, M.~E., {Lissauer}, J.~J., {Ford}, E.~B., {Torres},
  G., {Fressin}, F., {Brown}, T.~M., {Steffen}, J.~H., {Charbonneau}, D.,
  {Basri}, G.~S., {Sasselov}, D.~D., {Winn}, J., {Sanchis-Ojeda}, R.,
  {Christiansen}, J., {Adams}, E., {Henze}, C., {Dupree}, A., {Fabrycky},
  D.~C., {Fortney}, J.~J., {Tarter}, J., {Holman}, M.~J., {Tenenbaum}, P.,
  {Shporer}, A., {Lucas}, P.~W., {Welsh}, W.~F., {Orosz}, J.~A., {Bedding},
  T.~R., {Campante}, T.~L., {Davies}, G.~R., {Elsworth}, Y., {Handberg}, R.,
  {Hekker}, S., {Karoff}, C., {Kawaler}, S.~D., {Lund}, M.~N., {Lundkvist}, M.,
  {Metcalfe}, T.~S., {Miglio}, A., {Silva Aguirre}, V., {Stello}, D., {White},
  T.~R., {Boss}, A., {Devore}, E., {Gould}, A., {Prsa}, A., {Agol}, E.,
  {Barclay}, T., {Coughlin}, J., {Brugamyer}, E., {Mullally}, F., {Quintana},
  E.~V., {Still}, M., {Thompson}, S.~E., {Morrison}, D., {Twicken}, J.~D.,
  {D{\'e}sert}, J.-M., {Carter}, J., {Crepp}, J.~R., {H{\'e}brard}, G.,
  {Santerne}, A., {Moutou}, C., {Sobeck}, C., {Hudgins}, D., {Haas}, M.~R.,
  {Robertson}, P., {Lillo-Box}, J., \& {Barrado}, D. 2014, \apjs, 210, 20

\bibitem[{{Ming} {et~al.}(2013){Ming}, {Hui-Gen}, {Hui}, {Jia-Yi}, \&
  {Ji-Lin}}]{Ming2013}
{Ming}, Y., {Hui-Gen}, L., {Hui}, Z., {Jia-Yi}, Y., \& {Ji-Lin}, Z. 2013, \apj,
  778, 110

\bibitem[{{Morton}(2012)}]{Morton2012}
{Morton}, T.~D. 2012, \apj, 761, 6

\bibitem[{{Morton} \& {Johnson}(2011)}]{Morton2011}
{Morton}, T.~D. \& {Johnson}, J.~A. 2011, \apj, 738, 170

\bibitem[{{Naoz} {et~al.}(2012){Naoz}, {Farr}, \& {Rasio}}]{Naoz2012}
{Naoz}, S., {Farr}, W.~M., \& {Rasio}, F.~A. 2012, \apjl, 754, L36

\bibitem[{{O'Donovan} {et~al.}(2006){O'Donovan}, {Charbonneau}, {Torres},
  {Mandushev}, {Dunham}, {Latham}, {Alonso}, {Brown}, {Esquerdo}, {Everett}, \&
  {Creevey}}]{Odonovan2006}
{O'Donovan}, F.~T., {Charbonneau}, D., {Torres}, G., {Mandushev}, G., {Dunham},
  E.~W., {Latham}, D.~W., {Alonso}, R., {Brown}, T.~M., {Esquerdo}, G.~A.,
  {Everett}, M.~E., \& {Creevey}, O.~L. 2006, \apj, 644, 1237

\bibitem[{{Riddle} {et~al.}(2012){Riddle}, {Burse}, {Law}, {Tendulkar},
  {Baranec}, {Rudy}, {Sitt}, {Arya}, {Papadopoulos}, {Ramaprakash}, \&
  {Dekany}}]{Riddle2012}
{Riddle}, R.~L., {Burse}, M.~P., {Law}, N.~M., {Tendulkar}, S.~P., {Baranec},
  C., {Rudy}, A.~R., {Sitt}, M., {Arya}, A., {Papadopoulos}, A., {Ramaprakash},
  A.~N., \& {Dekany}, R.~G. 2012, in Society of Photo-Optical Instrumentation
  Engineers (SPIE) Conference Series, Vol. 8447, Society of Photo-Optical
  Instrumentation Engineers (SPIE) Conference Series

\bibitem[{{Santerne} {et~al.}(2013){Santerne}, {Fressin}, {D{\'{\i}}az},
  {Figueira}, {Almenara}, \& {Santos}}]{Santerne2013}
{Santerne}, A., {Fressin}, F., {D{\'{\i}}az}, R.~F., {Figueira}, P.,
  {Almenara}, J.-M., \& {Santos}, N.~C. 2013, \aap, 557, A139

\bibitem[{{Tenenbaum} {et~al.}(2013){Tenenbaum}, {Jenkins}, {Seader}, {Burke},
  {Christiansen}, {Rowe}, {Caldwell}, {Clarke}, {Li}, {Quintana}, {Smith},
  {Thompson}, {Twicken}, {Borucki}, {Batalha}, {Cote}, {Haas}, {Hunter},
  {Sanderfer}, {Girouard}, {Hall}, {Ibrahim}, {Klaus}, {McCauliff}, {Middour},
  {Sabale}, {Uddin}, {Wohler}, {Barclay}, \& {Still}}]{Tenenbaum2013}
{Tenenbaum}, P., {Jenkins}, J.~M., {Seader}, S., {Burke}, C.~J.,
  {Christiansen}, J.~L., {Rowe}, J.~F., {Caldwell}, D.~A., {Clarke}, B.~D.,
  {Li}, J., {Quintana}, E.~V., {Smith}, J.~C., {Thompson}, S.~E., {Twicken},
  J.~D., {Borucki}, W.~J., {Batalha}, N.~M., {Cote}, M.~T., {Haas}, M.~R.,
  {Hunter}, R.~C., {Sanderfer}, D.~T., {Girouard}, F.~R., {Hall}, J.~R.,
  {Ibrahim}, K., {Klaus}, T.~C., {McCauliff}, S.~D., {Middour}, C.~K.,
  {Sabale}, A., {Uddin}, A.~K., {Wohler}, B., {Barclay}, T., \& {Still}, M.
  2013, \apjs, 206, 5

\bibitem[{{Terziev} {et~al.}(2013){Terziev}, {Law}, {Arcavi}, {Baranec},
  {Bloom}, {Bui}, {Burse}, {Chorida}, {Das}, {Dekany}, {Kraus}, {Kulkarni},
  {Nugent}, {Ofek}, {Punnadi}, {Ramaprakash}, {Riddle}, {Sullivan}, \&
  {Tendulkar}}]{Terziev2013}
{Terziev}, E., {Law}, N.~M., {Arcavi}, I., {Baranec}, C., {Bloom}, J.~S.,
  {Bui}, K., {Burse}, M.~P., {Chorida}, P., {Das}, H.~K., {Dekany}, R.~G.,
  {Kraus}, A.~L., {Kulkarni}, S.~R., {Nugent}, P., {Ofek}, E.~O., {Punnadi},
  S., {Ramaprakash}, A.~N., {Riddle}, R., {Sullivan}, M., \& {Tendulkar}, S.~P.
  2013, \apjs, 206, 18

\bibitem[{{Wizinowich} {et~al.}(2000){Wizinowich}, {Acton}, {Shelton},
  {Stomski}, {Gathright}, {Ho}, {Lupton}, {Tsubota}, {Lai}, {Max}, {Brase},
  {An}, {Avicola}, {Olivier}, {Gavel}, {Macintosh}, {Ghez}, \&
  {Larkin}}]{Wizinowich2000}
{Wizinowich}, P., {Acton}, D.~S., {Shelton}, C., {Stomski}, P., {Gathright},
  J., {Ho}, K., {Lupton}, W., {Tsubota}, K., {Lai}, O., {Max}, C., {Brase}, J.,
  {An}, J., {Avicola}, K., {Olivier}, S., {Gavel}, D., {Macintosh}, B., {Ghez},
  A., \& {Larkin}, J. 2000, \pasp, 112, 315

\bibitem[{{York} {et~al.}(2000){York}, {Adelman}, {Anderson}, {Anderson},
  {Annis}, {Bahcall}, {Bakken}, {Barkhouser}, {Bastian}, {Berman}, {Boroski},
  {Bracker}, {Briegel}, {Briggs}, {Brinkmann}, {Brunner}, {Burles}, {Carey},
  {Carr}, {Castander}, {Chen}, {Colestock}, {Connolly}, {Crocker}, {Csabai},
  {Czarapata}, {Davis}, {Doi}, {Dombeck}, {Eisenstein}, {Ellman}, {Elms},
  {Evans}, {Fan}, {Federwitz}, {Fiscelli}, {Friedman}, {Frieman}, {Fukugita},
  {Gillespie}, {Gunn}, {Gurbani}, {de Haas}, {Haldeman}, {Harris}, {Hayes},
  {Heckman}, {Hennessy}, {Hindsley}, {Holm}, {Holmgren}, {Huang}, {Hull},
  {Husby}, {Ichikawa}, {Ichikawa}, {Ivezi{\'c}}, {Kent}, {Kim}, {Kinney},
  {Klaene}, {Kleinman}, {Kleinman}, {Knapp}, {Korienek}, {Kron}, {Kunszt},
  {Lamb}, {Lee}, {Leger}, {Limmongkol}, {Lindenmeyer}, {Long}, {Loomis},
  {Loveday}, {Lucinio}, {Lupton}, {MacKinnon}, {Mannery}, {Mantsch}, {Margon},
  {McGehee}, {McKay}, {Meiksin}, {Merelli}, {Monet}, {Munn}, {Narayanan},
  {Nash}, {Neilsen}, {Neswold}, {Newberg}, {Nichol}, {Nicinski}, {Nonino},
  {Okada}, {Okamura}, {Ostriker}, {Owen}, {Pauls}, {Peoples}, {Peterson},
  {Petravick}, {Pier}, {Pope}, {Pordes}, {Prosapio}, {Rechenmacher}, {Quinn},
  {Richards}, {Richmond}, {Rivetta}, {Rockosi}, {Ruthmansdorfer}, {Sandford},
  {Schlegel}, {Schneider}, {Sekiguchi}, {Sergey}, {Shimasaku}, {Siegmund},
  {Smee}, {Smith}, {Snedden}, {Stone}, {Stoughton}, {Strauss}, {Stubbs},
  {SubbaRao}, {Szalay}, {Szapudi}, {Szokoly}, {Thakar}, {Tremonti}, {Tucker},
  {Uomoto}, {Vanden Berk}, {Vogeley}, {Waddell}, {Wang}, {Watanabe},
  {Weinberg}, {Yanny}, {Yasuda}, \& {SDSS Collaboration}}]{York2000}
{York}, D.~G., {Adelman}, J., {Anderson}, Jr., J.~E., {Anderson}, S.~F.,
  {Annis}, J., {Bahcall}, N.~A., {Bakken}, J.~A., {Barkhouser}, R., {Bastian},
  S., {Berman}, E., {Boroski}, W.~N., {Bracker}, S., {Briegel}, C., {Briggs},
  J.~W., {Brinkmann}, J., {Brunner}, R., {Burles}, S., {Carey}, L., {Carr},
  M.~A., {Castander}, F.~J., {Chen}, B., {Colestock}, P.~L., {Connolly}, A.~J.,
  {Crocker}, J.~H., {Csabai}, I., {Czarapata}, P.~C., {Davis}, J.~E., {Doi},
  M., {Dombeck}, T., {Eisenstein}, D., {Ellman}, N., {Elms}, B.~R., {Evans},
  M.~L., {Fan}, X., {Federwitz}, G.~R., {Fiscelli}, L., {Friedman}, S.,
  {Frieman}, J.~A., {Fukugita}, M., {Gillespie}, B., {Gunn}, J.~E., {Gurbani},
  V.~K., {de Haas}, E., {Haldeman}, M., {Harris}, F.~H., {Hayes}, J.,
  {Heckman}, T.~M., {Hennessy}, G.~S., {Hindsley}, R.~B., {Holm}, S.,
  {Holmgren}, D.~J., {Huang}, C.-h., {Hull}, C., {Husby}, D., {Ichikawa},
  S.-I., {Ichikawa}, T., {Ivezi{\'c}}, {\v Z}., {Kent}, S., {Kim}, R.~S.~J.,
  {Kinney}, E., {Klaene}, M., {Kleinman}, A.~N., {Kleinman}, S., {Knapp},
  G.~R., {Korienek}, J., {Kron}, R.~G., {Kunszt}, P.~Z., {Lamb}, D.~Q., {Lee},
  B., {Leger}, R.~F., {Limmongkol}, S., {Lindenmeyer}, C., {Long}, D.~C.,
  {Loomis}, C., {Loveday}, J., {Lucinio}, R., {Lupton}, R.~H., {MacKinnon}, B.,
  {Mannery}, E.~J., {Mantsch}, P.~M., {Margon}, B., {McGehee}, P., {McKay},
  T.~A., {Meiksin}, A., {Merelli}, A., {Monet}, D.~G., {Munn}, J.~A.,
  {Narayanan}, V.~K., {Nash}, T., {Neilsen}, E., {Neswold}, R., {Newberg},
  H.~J., {Nichol}, R.~C., {Nicinski}, T., {Nonino}, M., {Okada}, N., {Okamura},
  S., {Ostriker}, J.~P., {Owen}, R., {Pauls}, A.~G., {Peoples}, J., {Peterson},
  R.~L., {Petravick}, D., {Pier}, J.~R., {Pope}, A., {Pordes}, R., {Prosapio},
  A., {Rechenmacher}, R., {Quinn}, T.~R., {Richards}, G.~T., {Richmond}, M.~W.,
  {Rivetta}, C.~H., {Rockosi}, C.~M., {Ruthmansdorfer}, K., {Sandford}, D.,
  {Schlegel}, D.~J., {Schneider}, D.~P., {Sekiguchi}, M., {Sergey}, G.,
  {Shimasaku}, K., {Siegmund}, W.~A., {Smee}, S., {Smith}, J.~A., {Snedden},
  S., {Stone}, R., {Stoughton}, C., {Strauss}, M.~A., {Stubbs}, C., {SubbaRao},
  M., {Szalay}, A.~S., {Szapudi}, I., {Szokoly}, G.~P., {Thakar}, A.~R.,
  {Tremonti}, C., {Tucker}, D.~L., {Uomoto}, A., {Vanden Berk}, D., {Vogeley},
  M.~S., {Waddell}, P., {Wang}, S.-i., {Watanabe}, M., {Weinberg}, D.~H.,
  {Yanny}, B., {Yasuda}, N., \& {SDSS Collaboration}. 2000, \aj, 120, 1579

\end{thebibliography}

\appendix
\label{sec:obs_list}

\begin{longtable}{lllllllllll}
\tablecaption{Full Robo-AO observation list}
\tablehead{
\bf KOI & \bf $\rm\bf m_i$ / mags & \bf Obs. date & \bf Filter & \bf Obs. qual. & \bf Companion? \\
}
\startdata
K00001.01 & 11.168 & 2012/07/16 & i & high & yes\\
K00002.01 &  & 2012/07/16 & i & high & \\
K00003.01 &  & 2012/07/16 & i & high & \\
K00005.01 & 11.485 & 2012/07/16 & i & high & \\
K00007.01 & 12.038 & 2012/07/16 & i & high & \\
K00010.01 & 13.424 & 2012/07/16 & i & medium & \\
K00012.01 & 11.245 & 2012/07/17 & i & high & \\
K00013.01 & 10.548 & 2012/10/06 & i & high & yes\\
K00017.01 & 13.094 & 2012/07/16 & i & medium & \\
K00018.01 & 13.148 & 2012/07/16 & i & medium & \\
K00022.01 & 13.265 & 2012/07/16 & i & medium & \\
K00041.01 & 11.03 & 2012/07/16 & i & high & \\
K00044.01 & 13.268 & 2012/07/16 & i & low & \\
K00046.01 & 13.497 & 2012/07/16 & i & medium & \\
K00049.01 & 13.508 & 2012/07/16 & i & low & \\
K00063.01 & 11.379 & 2012/07/16 & i & high & \\
K00064.01 & 12.866 & 2012/07/16 & i & medium & \\
K00069.01 & 9.739 & 2012/07/16 & i & high & \\
K00070.01 & 12.284 & 2012/07/16 & i & medium & \\
K00075.01 & 10.617 & 2012/07/16 & i & high & \\
K00082.01 & 11.15 & 2012/07/16 & i & high & \\
K00084.01 & 11.694 & 2012/07/16 & i & high & \\
K00085.01 & 10.882 & 2012/07/16 & i & high & \\
K00087.01 & 11.478 & 2012/07/16 & i & high & \\
K00089.01 & 11.649 & 2012/07/16 & i & medium & \\
K00092.01 & 11.506 & 2012/07/16 & i & high & \\
K00094.01 & 12.057 & 2012/07/16 & i & medium & \\
K00097.01 & 12.724 & 2012/07/17 & i & medium & yes\\
K00098.01 & 12.024 & 2012/07/17 & i & high & yes\\
K00099.01 & 12.68 & 2012/07/16 & i & medium & \\
K00100.01 & 12.466 & 2012/07/16 & i & medium & \\
K00102.01 & 12.384 & 2012/07/16 & i & medium & \\
K00103.01 & 12.399 & 2012/07/16 & i & medium & \\
K00105.01 & 12.649 & 2012/07/16 & i & medium & \\
K00107.01 & 12.53 & 2012/07/16 & i & medium & \\
K00108.01 & 12.132 & 2012/07/16 & i & high & \\
K00110.01 & 12.545 & 2012/07/16 & i & medium & \\
K00111.01 & 12.442 & 2012/07/16 & i & medium & \\
K00112.01 & 12.602 & 2012/07/18 & i & medium & \\
K00113.01 & 12.163 & 2012/07/17 & i & high & \\
K00115.01 & 12.654 & 2012/07/16 & i & medium & \\
K00117.01 & 12.309 & 2012/07/16 & i & medium & \\
K00118.01 & 12.195 & 2012/07/17 & i & medium & \\
K00119.01 & 12.452 & 2012/07/16 & i & low & yes\\
K00122.01 & 12.161 & 2012/07/16 & i & medium & \\
K00124.01 & 12.784 & 2012/07/16 & i & low & \\
K00128.01 & 13.54 & 2012/07/16 & i & low & \\
K00131.01 & 13.64 & 2012/07/16 & i & low & \\
K00137.01 & 13.287 & 2012/07/17 & i & low & \\
K00139.01 & 13.327 & 2012/07/17 & i & low & \\
K00141.01 & 13.441 & 2012/07/18 & i & medium & yes\\
K00142.01 & 12.895 & 2012/07/17 & i & low & \\
K00144.01 & 13.329 & 2012/07/17 & i & low & \\
K00148.01 & 12.761 & 2012/07/17 & i & low & \\
K00149.01 & 13.167 & 2012/07/17 & i & low & \\
K00152.01 & 13.761 & 2012/07/17 & i & low & \\
K00153.01 & 13.097 & 2012/07/17 & LP600 & medium & \\
K00156.01 & 13.334 & 2012/09/01 & LP600 & high & \\
K00157.01 & 13.508 & 2012/09/01 & LP600 & high & \\
K00159.01 & 13.243 & 2012/07/18 & LP600 & high & \\
K00161.01 & 12.99 & 2012/07/18 & LP600 & high & \\
K00162.01 & 13.626 & 2012/07/18 & LP600 & medium & yes\\
K00165.01 & 13.665 & 2012/07/17 & LP600 & medium & \\
K00166.01 & 13.315 & 2012/07/17 & LP600 & medium & \\
K00167.01 & 13.15 & 2012/07/17 & LP600 & high & \\
K00168.01 & 13.244 & 2012/07/17 & LP600 & high & \\
K00171.01 & 13.575 & 2012/07/17 & LP600 & high & \\
K00172.01 & 13.559 & 2012/07/18 & LP600 & high & \\
K00173.01 & 13.659 & 2012/07/18 & LP600 & high & \\
K00174.01 & 13.449 & 2012/07/18 & LP600 & high & yes\\
K00176.01 & 13.307 & 2012/07/18 & LP600 & high & \\
K00177.01 & 12.979 & 2012/07/18 & i & medium & yes\\
K00179.01 & 13.765 & 2012/07/18 & LP600 & high & \\
K00180.01 & 12.813 & 2012/07/18 & i & medium & \\
K00191.01 & 14.747 & 2012/09/01 & LP600 & low & yes\\
K00197.01 & 13.706 & 2012/07/18 & i & medium & \\
K00201.01 & 13.785 & 2012/07/18 & LP600 & high & \\
K00203.01 & 13.928 & 2012/07/18 & LP600 & high & \\
K00209.01 & 14.131 & 2012/09/01 & LP600 & medium & \\
K00211.01 & 14.82 & 2012/09/14 & LP600 & low & \\
K00214.01 & 14.003 & 2012/07/18 & LP600 & high & \\
K00216.01 & 14.4 & 2012/09/01 & LP600 & low & \\
K00219.01 & 13.925 & 2012/07/18 & LP600 & high & \\
K00220.01 & 14.011 & 2012/09/01 & LP600 & medium & \\
K00222.01 & 14.315 & 2012/09/01 & LP600 & low & \\
K00223.01 & 14.447 & 2012/09/01 & LP600 & medium & \\
K00232.01 & 14.067 & 2012/07/18 & LP600 & medium & \\
K00237.01 & 13.964 & 2012/07/18 & LP600 & medium & \\
K00238.01 & 13.891 & 2012/09/01 & LP600 & medium & \\
K00241.01 & 13.881 & 2012/07/18 & LP600 & medium & \\
K00244.01 &  & 2012/07/18 & i & high & \\
K00246.01 & 9.82 & 2012/07/18 & i & high & \\
K00247.01 & 13.585 & 2012/08/03 & LP600 & high & \\
K00248.01 & 14.68 & 2012/08/03 & LP600 & medium & \\
K00250.01 & 14.887 & 2012/08/03 & LP600 & low & \\
K00253.01 & 14.667 & 2012/08/03 & LP600 & medium & \\
K00254.01 & 15.364 & 2012/08/03 & LP600 & low & \\
K00256.01 & 14.636 & 2012/07/15 & LP600 & medium & \\
K00260.01 &  & 2012/07/18 & i & high & \\
K00261.01 & 10.109 & 2012/07/18 & i & high & \\
K00263.01 & 10.647 & 2012/07/18 & i & high & \\
K00268.01 &  & 2012/09/14 & LP600 & high & yes\\
K00269.01 & 10.823 & 2012/07/18 & i & high & \\
K00270.01 &  & 2012/07/17 & i & high & \\
K00273.01 & 11.262 & 2012/07/18 & i & high & \\
K00275.01 &  & 2012/07/18 & i & high & \\
K00276.01 & 11.711 & 2012/07/18 & i & high & \\
K00277.01 &  & 2012/07/18 & i & high & \\
K00279.01 & 11.563 & 2012/07/18 & i & medium & \\
K00281.01 & 11.77 & 2012/07/18 & i & high & \\
K00282.01 &  & 2012/07/18 & i & high & \\
K00283.01 & 11.334 & 2012/07/18 & i & high & \\
K00288.01 &  & 2012/07/18 & i & high & \\
K00291.01 & 12.642 & 2012/07/18 & i & high & \\
K00294.01 & 12.511 & 2012/07/18 & i & high & \\
K00296.01 & 12.77 & 2012/08/02 & i & medium & \\
K00297.01 & 12.042 & 2012/08/02 & i & high & \\
K00299.01 & 12.675 & 2012/07/18 & i & medium & \\
K00301.01 & 12.586 & 2012/07/18 & i & high & \\
K00302.01 & 11.969 & 2012/07/18 & i & high & \\
K00303.01 & 11.994 & 2012/07/18 & i & high & \\
K00305.01 & 12.606 & 2012/07/18 & i & medium & \\
K00306.01 & 12.363 & 2012/07/18 & i & low & yes\\
K00307.01 & 12.65 & 2012/08/02 & i & medium & \\
K00308.01 & 12.205 & 2012/07/18 & i & medium & \\
K00312.01 &  & 2012/07/28 & i & high & \\
K00313.01 & 12.736 & 2012/08/02 & i & high & \\
K00314.01 & 12.457 & 2012/08/03 & LP600 & high & \\
K00315.01 & 12.63 & 2012/07/28 & i & medium & \\
K00316.01 & 12.494 & 2012/08/02 & i & high & \\
K00317.01 & 12.751 & 2012/07/28 & i & medium & \\
K00319.01 &  & 2012/08/02 & i & high & \\
K00321.01 & 12.312 & 2012/08/02 & i & high & \\
K00323.01 & 12.24 & 2012/08/02 & i & high & \\
K00327.01 & 12.858 & 2012/08/02 & i & medium & \\
K00330.01 & 13.73 & 2012/07/28 & LP600 & high & \\
K00331.01 & 13.277 & 2012/07/28 & i & medium & \\
K00332.01 & 12.847 & 2012/07/28 & i & high & \\
K00333.01 & 13.265 & 2012/08/02 & i & medium & \\
K00337.01 & 13.746 & 2012/08/02 & LP600 & medium & \\
K00339.01 & 13.616 & 2012/08/02 & LP600 & medium & \\
K00340.01 & 12.82 & 2012/07/28 & i & medium & \\
K00341.01 & 13.106 & 2012/07/28 & i & medium & \\
K00343.01 & 13.013 & 2012/08/02 & i & medium & \\
K00344.01 & 13.211 & 2012/08/02 & i & medium & \\
K00345.01 & 13.005 & 2012/08/02 & i & medium & \\
K00348.01 & 13.555 & 2012/08/03 & LP600 & high & \\
K00349.01 & 13.382 & 2012/08/02 & LP600 & high & \\
K00350.01 & 13.202 & 2012/08/02 & i & medium & \\
K00352.01 & 13.579 & 2012/07/28 & LP600 & high & \\
K00353.01 & 13.251 & 2012/08/02 & i & low & \\
K00356.01 & 13.532 & 2012/07/28 & LP600 & high & yes\\
K00360.01 & 12.823 & 2012/08/02 & i & medium & \\
K00361.01 & 12.914 & 2012/08/02 & i & medium & \\
K00365.01 & 10.992 & 2012/07/28 & i & high & \\
K00366.01 &  & 2012/08/02 & i & high & \\
K00368.01 & 11.598 & 2012/08/02 & i & high & \\
K00371.01 & 11.895 & 2012/07/28 & i & high & \\
K00372.01 & 12.208 & 2012/07/28 & i & high & \\
K00373.01 & 12.593 & 2012/08/02 & i & medium & \\
K00377.01 & 13.613 & 2012/09/01 & LP600 & medium & \\
K00384.01 & 13.106 & 2012/08/02 & i & low & \\
K00385.01 & 13.211 & 2012/08/02 & i & medium & \\
K00386.01 & 13.661 & 2012/08/02 & i & low & \\
K00388.01 & 13.448 & 2012/08/02 & i & medium & \\
K00392.01 & 13.745 & 2012/08/05 & LP600 & medium & \\
K00393.01 & 13.395 & 2012/08/02 & i & low & \\
K00401.01 & 13.729 & 2012/08/05 & LP600 & medium & yes\\
K00403.01 & 13.953 & 2012/08/05 & LP600 & medium & \\
K00408.01 & 14.766 & 2012/09/01 & LP600 & low & \\
K00409.01 & 13.965 & 2012/09/14 & LP600 & medium & \\
K00413.01 & 14.512 & 2012/09/01 & LP600 & low & \\
K00415.01 & 13.914 & 2012/08/05 & LP600 & medium & \\
K00416.01 & 14.019 & 2012/09/01 & LP600 & low & \\
K00427.01 & 14.37 & 2012/09/01 & LP600 & low & \\
K00431.01 & 14.004 & 2012/09/01 & LP600 & medium & \\
K00435.01 & 14.342 & 2012/09/01 & LP600 & low & \\
K00439.01 & 14.063 & 2012/08/05 & LP600 & medium & \\
K00440.01 & 13.861 & 2012/09/01 & LP600 & medium & \\
K00442.01 & 13.806 & 2012/08/05 & LP600 & medium & \\
K00444.01 & 13.909 & 2012/08/05 & LP600 & medium & \\
K00456.01 & 14.407 & 2012/09/01 & LP600 & medium & \\
K00457.01 & 13.894 & 2012/09/01 & LP600 & medium & \\
K00459.01 & 14.028 & 2012/09/01 & LP600 & medium & \\
K00463.01 & 13.999 & 2012/08/03 & LP600 & medium & \\
K00464.01 & 14.113 & 2012/09/01 & LP600 & medium & \\
K00465.01 & 14.017 & 2012/08/05 & LP600 & medium & \\
K00471.01 & 14.198 & 2012/09/01 & LP600 & medium & \\
K00474.01 & 14.131 & 2012/09/01 & LP600 & low & \\
K00478.01 & 13.58 & 2012/08/04 & LP600 & high & \\
K00481.01 & 14.446 & 2012/09/01 & LP600 & medium & \\
K00486.01 & 13.934 & 2012/08/05 & LP600 & medium & \\
K00490.01 & 13.688 & 2012/08/05 & LP600 & medium & \\
K00497.01 & 14.423 & 2012/09/01 & LP600 & low & \\
K00508.01 & 14.146 & 2012/09/01 & LP600 & medium & \\
K00509.01 & 14.638 & 2012/09/01 & LP600 & low & \\
K00511.01 & 14.017 & 2012/09/01 & LP600 & medium & yes\\
K00517.01 & 13.806 & 2012/08/05 & LP600 & high & \\
K00519.01 & 14.737 & 2012/09/01 & LP600 & medium & \\
K00520.01 & 14.255 & 2012/09/01 & LP600 & medium & \\
K00523.01 & 14.822 & 2012/09/01 & LP600 & low & \\
K00528.01 & 14.364 & 2012/09/01 & LP600 & low & \\
K00531.01 & 13.849 & 2012/08/03 & LP600 & high & \\
K00534.01 & 14.344 & 2012/09/01 & LP600 & medium & \\
K00542.01 & 14.12 & 2012/09/01 & LP600 & medium & \\
K00543.01 & 14.442 & 2012/09/01 & LP600 & low & \\
K00546.01 & 14.717 & 2012/09/01 & LP600 & low & \\
K00548.01 & 13.874 & 2012/08/05 & LP600 & medium & \\
K00550.01 & 13.869 & 2012/08/05 & LP600 & medium & \\
K00551.01 & 14.725 & 2012/09/01 & LP600 & low & \\
K00555.01 & 14.499 & 2012/08/05 & LP600 & medium & \\
K00561.01 & 13.732 & 2012/08/05 & LP600 & medium & \\
K00564.01 & 14.642 & 2012/09/02 & LP600 & low & \\
K00567.01 & 14.126 & 2012/09/02 & LP600 & medium & \\
K00568.01 & 13.895 & 2012/08/05 & LP600 & medium & \\
K00569.01 & 14.172 & 2012/09/02 & LP600 & medium & \\
K00571.01 & 14.015 & 2012/08/03 & LP600 & high & \\
K00572.01 & 13.96 & 2012/07/28 & i & low & \\
K00574.01 & 14.579 & 2012/09/02 & LP600 & low & \\
K00579.01 & 13.858 & 2012/08/05 & LP600 & medium & \\
K00582.01 & 14.529 & 2012/09/02 & LP600 & medium & \\
K00590.01 & 14.444 & 2012/09/02 & LP600 & low & \\
K00593.01 & 14.754 & 2012/09/02 & LP600 & low & \\
K00597.01 & 14.721 & 2012/09/02 & LP600 & low & \\
K00601.01 & 14.515 & 2012/09/02 & LP600 & medium & \\
K00611.01 & 13.866 & 2012/08/05 & LP600 & medium & \\
K00612.01 & 13.871 & 2012/08/05 & LP600 & medium & \\
K00620.01 & 14.467 & 2012/09/02 & LP600 & medium & \\
K00623.01 & 11.685 & 2012/08/03 & i & high & \\
K00624.01 & 13.39 & 2012/08/03 & i & medium & \\
K00625.01 & 13.433 & 2012/08/03 & i & medium & \\
K00626.01 & 13.339 & 2012/08/03 & i & medium & \\
K00627.01 & 13.119 & 2012/08/03 & i & medium & \\
K00628.01 & 13.744 & 2012/08/03 & i & medium & yes\\
K00629.01 & 13.788 & 2012/08/04 & i & medium & \\
K00632.01 & 13.124 & 2012/08/04 & i & medium & \\
K00633.01 & 13.663 & 2012/08/04 & i & medium & \\
K00635.01 & 12.88 & 2012/08/04 & i & medium & \\
K00638.01 & 13.394 & 2012/08/04 & i & medium & \\
K00639.01 & 13.354 & 2012/07/28 & i & medium & \\
K00640.01 & 13.058 & 2012/07/28 & i & low & yes\\
K00644.01 & 13.474 & 2012/08/04 & i & medium & \\
K00647.01 & 13.413 & 2012/08/04 & i & medium & \\
K00649.01 & 13.157 & 2012/08/04 & i & medium & \\
K00650.01 & 13.293 & 2012/08/04 & i & medium & \\
K00654.01 & 13.789 & 2012/08/04 & i & medium & \\
K00655.01 & 12.872 & 2012/08/04 & i & medium & \\
K00657.01 & 13.517 & 2012/08/04 & i & medium & \\
K00658.01 & 13.789 & 2012/08/04 & i & medium & \\
K00659.01 & 13.297 & 2012/08/04 & i & medium & \\
K00660.01 & 13.283 & 2012/08/04 & i & medium & \\
K00661.01 & 13.731 & 2012/08/04 & i & medium & \\
K00662.01 & 13.168 & 2012/08/04 & i & medium & \\
K00663.01 & 13.016 & 2012/09/02 & LP600 & high & \\
K00664.01 & 13.287 & 2012/08/04 & i & medium & \\
K00665.01 & 13.005 & 2012/08/04 & i & medium & \\
K00666.01 & 13.518 & 2012/08/04 & i & medium & \\
K00671.01 & 13.511 & 2012/08/04 & i & medium & \\
K00673.01 & 13.211 & 2012/08/04 & i & medium & \\
K00674.01 & 13.435 & 2012/08/04 & i & medium & \\
K00676.01 & 13.371 & 2012/09/02 & LP600 & high & \\
K00679.01 & 13.038 & 2012/08/04 & i & medium & \\
K00680.01 & 13.485 & 2012/08/04 & i & medium & \\
K00682.01 & 13.692 & 2012/08/04 & i & medium & \\
K00684.01 & 13.575 & 2012/08/04 & i & medium & \\
K00685.01 & 13.77 & 2012/08/04 & i & medium & \\
K00686.01 & 13.346 & 2012/08/04 & i & medium & \\
K00687.01 & 13.613 & 2012/08/04 & i & medium & yes\\
K00688.01 & 13.849 & 2012/09/14 & LP600 & medium & yes\\
K00689.01 & 13.548 & 2012/08/04 & i & medium & \\
K00691.01 & 13.803 & 2012/08/05 & i & low & \\
K00692.01 & 13.457 & 2012/08/05 & i & medium & \\
K00694.01 & 13.741 & 2012/08/05 & i & low & \\
K00695.01 & 13.276 & 2012/08/05 & i & medium & \\
K00698.01 & 13.52 & 2012/08/05 & i & low & \\
K00700.01 & 13.38 & 2012/08/05 & i & medium & \\
K00701.01 & 13.429 & 2012/08/05 & i & low & \\
K00703.01 & 13.162 & 2012/08/05 & i & medium & \\
K00704.01 & 13.46 & 2012/08/05 & i & medium & \\
K00707.01 & 13.815 & 2012/08/05 & i & low & \\
K00708.01 & 13.837 & 2012/08/05 & i & low & \\
K00709.01 & 13.716 & 2012/08/05 & i & medium & \\
K00710.01 & 13.128 & 2012/08/05 & i & medium & \\
K00711.01 & 13.735 & 2012/08/05 & i & medium & \\
K00712.01 & 13.51 & 2012/08/05 & i & medium & yes\\
K00714.01 & 13.184 & 2012/08/05 & i & medium & \\
K00716.01 & 13.576 & 2012/08/05 & i & low & \\
K00717.01 & 13.182 & 2012/08/05 & i & medium & \\
K00718.01 & 13.588 & 2012/08/05 & i & low & \\
K00719.01 & 12.899 & 2012/08/05 & i & medium & \\
K00720.01 & 13.489 & 2012/08/05 & i & low & \\
K00721.01 & 13.439 & 2012/08/05 & i & low & \\
K00722.01 & 13.343 & 2012/08/05 & i & high & \\
K00723.01 & 14.795 & 2012/09/02 & LP600 & low & \\
K00738.01 & 15.063 & 2012/09/02 & LP600 & low & \\
K00739.01 & 14.931 & 2012/08/03 & LP600 & medium & \\
K00756.01 & 15.492 & 2012/09/02 & LP600 & low & \\
K00781.01 & 15.267 & 2012/08/03 & LP600 & low & \\
K00800.01 & 15.341 & 2012/09/02 & LP600 & low & \\
K00817.01 & 14.793 & 2012/08/03 & LP600 & medium & \\
K00818.01 & 15.192 & 2012/08/04 & LP600 & medium & \\
K00834.01 & 14.862 & 2012/09/02 & LP600 & low & \\
K00835.01 & 14.884 & 2012/09/02 & LP600 & low & \\
K00837.01 & 15.325 & 2012/09/02 & LP600 & low & \\
K00842.01 & 15.001 & 2012/09/02 & LP600 & low & \\
K00853.01 & 15.039 & 2012/09/02 & LP600 & low & \\
K00854.01 & 15.162 & 2012/08/03 & LP600 & low & \\
K00857.01 & 14.787 & 2012/09/02 & LP600 & low & \\
K00872.01 & 14.98 & 2012/09/02 & LP600 & low & \\
K00874.01 & 14.716 & 2012/09/02 & LP600 & low & \\
K00877.01 & 14.547 & 2012/06/17 & LP600 & low & \\
K00880.01 & 14.918 & 2012/09/02 & LP600 & low & \\
K00884.01 & 14.755 & 2012/09/02 & LP600 & low & \\
K00886.01 & 15.175 & 2012/08/04 & LP600 & medium & \\
K00896.01 & 14.974 & 2012/09/02 & LP600 & low & \\
K00898.01 & 15.221 & 2012/08/04 & LP600 & medium & \\
K00899.01 & 14.543 & 2012/08/04 & LP600 & medium & \\
K00906.01 & 15.155 & 2012/09/02 & LP600 & low & \\
K00907.01 & 14.983 & 2012/09/02 & LP600 & low & \\
K00921.01 & 15.229 & 2012/09/02 & LP600 & low & \\
K00935.01 & 15.086 & 2012/09/02 & LP600 & low & \\
K00936.01 & 14.371 & 2012/08/04 & LP600 & medium & \\
K00938.01 & 15.328 & 2012/09/02 & LP600 & low & \\
K00939.01 & 14.849 & 2012/09/02 & LP600 & low & \\
K00947.01 & 14.564 & 2012/08/04 & LP600 & medium & \\
K00975.01 &  & 2012/07/17 & i & high & \\
K00977.01 &  & 2012/08/03 & i & high & \\
K00984.01 & 11.353 & 2012/08/03 & i & high & yes\\
K00986.01 & 13.908 & 2012/08/03 & i & low & \\
K00987.01 & 12.327 & 2012/08/03 & i & medium & yes\\
K00988.01 & 13.259 & 2012/08/03 & i & medium & \\
K00991.01 & 13.368 & 2012/08/03 & i & medium & \\
K01001.01 & 12.851 & 2012/08/03 & i & medium & \\
K01002.01 & 13.362 & 2012/08/03 & i & medium & yes\\
K01010.01 & 13.463 & 2012/08/03 & i & medium & \\
K01015.01 & 14.349 & 2012/09/03 & LP600 & medium & \\
K01019.01 & 9.961 & 2012/08/03 & i & high & \\
K01020.01 & 12.712 & 2012/08/03 & i & medium & \\
K01032.01 & 13.497 & 2012/08/03 & i & medium & \\
K01050.01 & 13.696 & 2012/08/03 & i & low & yes\\
K01052.01 & 15.201 & 2012/09/03 & LP600 & medium & \\
K01054.01 & 11.662 & 2012/08/03 & i & high & \\
K01060.01 & 14.221 & 2012/09/03 & LP600 & medium & \\
K01070.01 & 15.348 & 2012/09/03 & LP600 & low & \\
K01078.01 & 14.846 & 2012/08/04 & LP600 & medium & \\
K01085.01 & 14.651 & 2012/08/04 & LP600 & medium & \\
K01089.01 & 14.501 & 2012/09/03 & LP600 & medium & \\
K01102.01 & 14.711 & 2012/08/05 & LP600 & low & \\
K01113.01 & 13.54 & 2012/08/05 & i & medium & \\
K01115.01 & 13.739 & 2012/08/05 & i & low & \\
K01116.01 & 13.153 & 2012/08/05 & i & medium & \\
K01118.01 & 13.672 & 2012/08/05 & i & medium & \\
K01127.01 & 15.587 & 2012/09/03 & LP600 & low & \\
K01128.01 & 13.277 & 2012/08/05 & i & medium & \\
K01141.01 & 15.39 & 2012/08/04 & LP600 & low & \\
K01145.01 & 13.956 & 2012/08/05 & i & low & \\
K01146.01 & 15.043 & 2012/07/15 & LP600 & low & \\
K01148.01 & 13.769 & 2012/08/05 & i & medium & \\
K01150.01 & 13.139 & 2012/08/05 & i & medium & yes\\
K01151.01 & 13.198 & 2012/08/05 & i & medium & yes\\
K01152.01 & 13.622 & 2012/09/14 & LP600 & low & yes\\
K01161.01 & 14.391 & 2012/09/03 & LP600 & medium & \\
K01162.01 & 12.622 & 2012/08/04 & i & high & \\
K01163.01 & 14.735 & 2012/09/03 & LP600 & medium & \\
K01165.01 & 13.699 & 2012/08/05 & i & medium & \\
K01168.01 & 13.851 & 2012/08/05 & i & low & \\
K01169.01 & 13.071 & 2012/08/05 & i & medium & \\
K01175.01 & 13.075 & 2012/08/05 & i & medium & \\
K01194.01 & 15.391 & 2012/09/03 & LP600 & medium & \\
K01198.01 & 15.165 & 2012/09/03 & LP600 & low & \\
K01202.01 & 15.352 & 2012/08/04 & LP600 & low & \\
K01203.01 & 15.159 & 2012/09/03 & LP600 & low & \\
K01208.01 & 13.456 & 2012/08/06 & i & medium & \\
K01215.01 & 13.226 & 2012/08/06 & i & medium & \\
K01216.01 & 13.28 & 2012/08/05 & i & low & \\
K01218.01 & 13.13 & 2012/08/06 & i & medium & \\
K01220.01 & 12.713 & 2012/08/06 & i & medium & \\
K01221.01 & 11.265 & 2012/08/06 & i & high & \\
K01222.01 & 11.909 & 2012/08/06 & i & high & \\
K01227.01 & 13.785 & 2012/09/14 & LP600 & low & \\
K01230.01 & 11.914 & 2012/08/06 & i & high & \\
K01236.01 & 13.518 & 2012/08/06 & i & low & \\
K01239.01 & 14.812 & 2012/09/03 & LP600 & medium & \\
K01240.01 & 14.242 & 2012/09/03 & LP600 & medium & \\
K01241.01 & 12.09 & 2012/09/14 & LP600 & high & \\
K01242.01 & 13.611 & 2012/08/06 & i & low & \\
K01257.01 & 14.367 & 2012/09/14 & LP600 & low & \\
K01258.01 & 15.528 & 2012/09/03 & LP600 & medium & \\
K01266.01 & 14.869 & 2012/06/17 & LP600 & low & \\
K01270.01 & 14.544 & 2012/09/03 & LP600 & medium & \\
K01271.01 & 13.5 & 2012/08/06 & i & low & \\
K01274.01 & 13.107 & 2012/08/06 & i & medium & yes\\
K01275.01 & 13.442 & 2012/07/28 & i & low & \\
K01276.01 & 14.542 & 2012/09/03 & LP600 & medium & \\
K01278.01 & 15.02 & 2012/09/03 & LP600 & medium & \\
K01279.01 & 13.555 & 2012/08/06 & i & low & \\
K01282.01 & 12.399 & 2012/08/06 & i & high & \\
K01283.01 &  & 2012/08/06 & i & high & \\
K01288.01 & 14.967 & 2012/09/14 & LP600 & low & \\
K01299.01 & 11.878 & 2012/08/06 & i & high & \\
K01301.01 & 15.581 & 2012/09/03 & LP600 & low & \\
K01305.01 & 14.913 & 2012/09/03 & LP600 & medium & \\
K01306.01 & 15.374 & 2012/09/04 & LP600 & low & \\
K01307.01 & 14.551 & 2012/09/04 & LP600 & medium & \\
K01308.01 & 13.781 & 2012/08/06 & i & low & \\
K01309.01 & 13.727 & 2012/08/06 & i & medium & \\
K01314.01 & 12.941 & 2012/08/06 & i & medium & \\
K01315.01 & 12.998 & 2012/08/06 & i & medium & \\
K01316.01 & 11.694 & 2012/08/06 & i & high & \\
K01332.01 & 14.919 & 2012/09/03 & LP600 & medium & \\
K01335.01 & 13.774 & 2012/08/06 & i & low & \\
K01336.01 & 14.61 & 2012/09/04 & LP600 & medium & \\
K01338.01 & 14.385 & 2012/09/04 & LP600 & medium & \\
K01342.01 & 14.033 & 2012/09/04 & LP600 & medium & \\
K01344.01 & 13.269 & 2012/08/06 & i & medium & \\
K01353.01 & 13.764 & 2012/08/06 & i & medium & \\
K01358.01 & 15.117 & 2012/09/04 & LP600 & low & \\
K01359.01 & 15.025 & 2012/09/04 & LP600 & medium & yes\\
K01360.01 & 15.293 & 2012/09/04 & LP600 & low & \\
K01363.01 & 15.719 & 2012/09/04 & LP600 & low & \\
K01364.01 & 15.669 & 2012/09/04 & LP600 & low & \\
K01366.01 & 15.138 & 2012/09/04 & LP600 & medium & \\
K01375.01 & 13.533 & 2012/08/06 & i & medium & yes\\
K01376.01 & 13.902 & 2012/08/06 & i & medium & \\
K01378.01 & 13.327 & 2012/08/06 & i & medium & \\
K01379.01 & 13.499 & 2012/08/06 & i & medium & \\
K01393.01 & 15.201 & 2012/07/15 & LP600 & low & \\
K01396.01 & 15.62 & 2012/09/04 & LP600 & medium & \\
K01401.01 & 13.316 & 2012/08/06 & i & medium & \\
K01408.01 & 14.141 & 2012/08/04 & LP600 & medium & \\
K01412.01 & 13.434 & 2012/08/06 & i & medium & \\
K01422.01 & 15.194 & 2012/08/04 & LP600 & low & \\
K01426.01 & 14.063 & 2012/08/06 & i & medium & \\
K01427.01 & 15.287 & 2012/08/04 & LP600 & medium & \\
K01435.01 & 14.012 & 2012/09/04 & LP600 & medium & \\
K01436.01 & 14.061 & 2012/09/04 & LP600 & medium & \\
K01438.01 & 13.858 & 2012/08/06 & i & medium & \\
K01439.01 & 12.689 & 2012/08/06 & i & medium & \\
K01442.01 & 12.296 & 2012/08/06 & i & high & yes\\
K01444.01 & 13.784 & 2012/08/06 & i & low & \\
K01452.01 & 13.525 & 2012/08/06 & i & medium & \\
K01459.01 & 15.139 & 2012/08/04 & LP600 & medium & \\
K01478.01 & 12.254 & 2012/08/06 & i & high & \\
K01480.01 & 15.573 & 2012/09/04 & LP600 & low & \\
K01486.01 & 15.286 & 2012/09/04 & LP600 & medium & \\
K01515.01 & 13.862 & 2012/09/04 & LP600 & medium & \\
K01525.01 & 12.009 & 2012/08/06 & i & high & \\
K01528.01 & 13.822 & 2012/08/06 & i & medium & \\
K01529.01 & 14.152 & 2012/09/04 & LP600 & high & \\
K01530.01 & 12.88 & 2012/08/06 & i & medium & \\
K01535.01 & 12.884 & 2012/08/06 & i & medium & \\
K01536.01 & 12.542 & 2012/08/06 & i & medium & \\
K01537.01 &  & 2012/08/29 & i & high & \\
K01557.01 & 14.457 & 2012/09/04 & LP600 & medium & \\
K01563.01 & 15.475 & 2012/09/04 & LP600 & low & \\
K01567.01 & 15.254 & 2012/09/04 & LP600 & medium & \\
K01576.01 & 13.826 & 2012/08/06 & i & medium & \\
K01588.01 & 14.184 & 2012/06/17 & LP600 & medium & \\
K01589.01 & 14.547 & 2012/09/04 & LP600 & medium & \\
K01590.01 & 15.326 & 2012/09/04 & LP600 & low & \\
K01596.01 & 14.758 & 2012/06/17 & LP600 & low & \\
K01597.01 & 12.598 & 2012/08/06 & i & high & \\
K01598.01 & 14.063 & 2012/09/04 & LP600 & medium & \\
K01606.01 & 13.752 & 2012/08/06 & i & medium & \\
K01608.01 & 13.647 & 2012/09/04 & LP600 & high & \\
K01609.01 & 13.793 & 2012/08/29 & i & low & \\
K01612.01 & 8.658 & 2012/08/06 & i & high & \\
K01613.01 &  & 2012/08/29 & i & high & yes\\
K01615.01 & 11.341 & 2012/08/29 & i & high & \\
K01616.01 & 11.396 & 2012/08/29 & i & high & \\
K01618.01 & 11.473 & 2012/08/29 & i & high & \\
K01619.01 & 11.427 & 2012/08/29 & i & high & yes\\
K01621.01 & 11.711 & 2012/08/29 & i & high & \\
K01622.01 & 12.033 & 2012/08/29 & i & high & \\
K01627.01 & 15.493 & 2012/09/04 & LP600 & low & \\
K01628.01 & 12.775 & 2012/08/29 & i & medium & \\
K01629.01 & 13.381 & 2012/08/29 & i & medium & \\
K01632.01 & 13.157 & 2012/08/29 & i & medium & \\
K01647.01 & 13.961 & 2012/09/04 & LP600 & high & \\
K01649.01 & 14.347 & 2012/07/16 & LP600 & medium & \\
K01655.01 & 13.559 & 2012/09/14 & LP600 & medium & \\
K01665.01 & 13.871 & 2012/09/04 & LP600 & high & \\
K01669.01 & 14.018 & 2012/09/14 & LP600 & low & \\
K01677.01 & 14.073 & 2012/09/04 & LP600 & medium & yes\\
K01684.01 & 12.717 & 2012/09/14 & LP600 & high & \\
K01692.01 & 12.313 & 2012/09/04 & LP600 & high & \\
K01701.01 & 11.047 & 2012/08/04 & i & high & \\
K01706.01 & 13.835 & 2012/09/14 & LP600 & medium & \\
K01713.01 & 14.712 & 2012/09/13 & LP600 & medium & \\
K01715.01 & 12.751 & 2012/08/29 & i & medium & \\
K01725.01 & 13.107 & 2012/08/29 & i & medium & \\
K01726.01 & 12.684 & 2012/08/29 & i & medium & \\
K01738.01 & 13.032 & 2012/08/29 & i & medium & \\
K01751.01 & 14.248 & 2012/09/13 & LP600 & medium & \\
K01754.01 & 13.775 & 2012/09/14 & LP600 & medium & \\
K01779.01 & 13.077 & 2012/08/29 & i & low & \\
K01781.01 & 11.884 & 2012/09/13 & LP600 & high & \\
K01783.01 & 13.774 & 2012/09/14 & LP600 & medium & \\
K01802.01 & 13.175 & 2012/08/29 & i & medium & \\
K01803.01 & 12.932 & 2012/08/29 & i & medium & \\
K01805.01 & 13.591 & 2012/09/14 & LP600 & medium & \\
K01812.01 & 13.582 & 2012/08/29 & i & medium & \\
K01813.01 & 13.525 & 2012/08/29 & i & medium & \\
K01814.01 & 12.453 & 2012/09/14 & LP600 & high & \\
K01818.01 & 13.881 & 2012/09/14 & LP600 & medium & \\
K01819.01 & 13.347 & 2012/08/29 & i & medium & \\
K01820.01 & 13.292 & 2012/09/13 & LP600 & high & \\
K01822.01 & 12.281 & 2012/08/29 & i & medium & \\
K01824.01 & 12.567 & 2012/08/29 & i & medium & \\
K01825.01 & 13.632 & 2012/09/14 & LP600 & high & \\
K01831.01 & 13.866 & 2012/09/14 & LP600 & medium & \\
K01832.01 & 14.776 & 2012/09/13 & LP600 & low & \\
K01835.01 & 13.388 & 2012/09/13 & LP600 & high & \\
K01839.01 & 12.992 & 2012/08/29 & i & medium & \\
K01843.01 & 13.708 & 2012/08/29 & i & medium & \\
K01845.01 & 14.05 & 2012/09/13 & LP600 & medium & yes\\
K01850.01 & 13.952 & 2012/09/14 & LP600 & medium & \\
K01852.01 & 12.97 & 2012/08/29 & i & medium & \\
K01854.01 & 13.293 & 2012/08/29 & i & medium & \\
K01856.01 & 13.804 & 2012/09/14 & LP600 & medium & \\
K01857.01 & 13.548 & 2012/08/29 & i & medium & \\
K01860.01 & 13.822 & 2012/09/14 & LP600 & medium & \\
K01862.01 & 13.453 & 2012/08/29 & i & medium & \\
K01863.01 & 13.473 & 2012/08/29 & i & low & \\
K01867.01 & 14.404 & 2012/07/15 & LP600 & low & \\
K01868.01 & 14.652 & 2012/07/15 & LP600 & low & \\
K01874.01 & 14.947 & 2012/09/13 & LP600 & low & \\
K01878.01 & 12.835 & 2012/08/29 & i & medium & \\
K01880.01 & 13.835 & 2012/07/15 & LP600 & medium & yes\\
K01883.01 & 11.757 & 2012/08/29 & i & high & \\
K01884.01 & 15.158 & 2012/09/13 & LP600 & low & yes\\
K01886.01 & 12.087 & 2012/08/29 & i & high & \\
K01888.01 & 13.15 & 2012/08/29 & i & medium & \\
K01889.01 & 15.109 & 2012/09/13 & LP600 & medium & \\
K01890.01 & 11.555 & 2012/08/29 & i & high & yes\\
K01891.01 & 14.957 & 2012/09/13 & LP600 & medium & yes\\
K01893.01 & 13.876 & 2012/09/14 & LP600 & medium & \\
K01894.01 & 13.05 & 2012/09/14 & LP600 & high & \\
K01895.01 & 15.42 & 2012/09/13 & LP600 & low & \\
K01897.01 & 13.779 & 2012/09/14 & LP600 & high & \\
K01905.01 & 13.713 & 2012/09/14 & LP600 & medium & \\
K01907.01 & 14.699 & 2012/07/15 & LP600 & low & \\
K01909.01 & 12.612 & 2012/09/13 & LP600 & high & \\
K01913.01 & 13.083 & 2012/08/29 & i & medium & \\
K01915.01 & 13.809 & 2012/09/14 & LP600 & medium & \\
K01916.01 & 13.42 & 2012/09/13 & LP600 & high & yes\\
K01917.01 & 13.479 & 2012/08/29 & i & medium & \\
K01921.01 & 12.708 & 2012/09/14 & LP600 & high & \\
K01922.01 & 15.159 & 2012/09/13 & LP600 & medium & \\
K01923.01 & 13.879 & 2012/08/29 & i & low & \\
K01924.01 & 7.674 & 2012/08/29 & i & high & \\
K01925.01 & 9.211 & 2012/08/29 & i & high & \\
K01929.01 & 12.53 & 2012/09/13 & LP600 & high & \\
K01930.01 & 11.957 & 2012/09/13 & LP600 & high & \\
K01931.01 & 14.307 & 2012/09/13 & LP600 & medium & \\
K01932.01 & 12.366 & 2012/09/14 & LP600 & high & \\
K01938.01 & 13.766 & 2012/09/14 & LP600 & medium & \\
K01940.01 & 14.912 & 2012/09/13 & LP600 & medium & \\
K01944.01 & 13.79 & 2012/09/14 & LP600 & medium & \\
K01945.01 & 14.267 & 2012/09/13 & LP600 & medium & \\
K01952.01 & 14.398 & 2012/09/13 & LP600 & medium & \\
K01955.01 & 13.025 & 2012/09/13 & LP600 & high & \\
K01960.01 & 13.975 & 2012/09/14 & LP600 & low & \\
K01961.01 & 12.61 & 2012/08/30 & i & medium & \\
K01962.01 &  & 2012/08/30 & i & high & yes\\
K01964.01 & 10.464 & 2012/08/30 & i & high & yes\\
K01970.01 & 15.141 & 2012/09/13 & LP600 & low & \\
K01977.01 & 13.566 & 2012/10/06 & LP600 & high & \\
K01979.01 & 12.786 & 2012/08/30 & i & medium & yes\\
K01984.01 & 13.528 & 2012/08/30 & i & medium & \\
K01988.01 & 13.741 & 2012/09/14 & LP600 & medium & \\
K02001.01 & 12.82 & 2012/08/30 & i & medium & \\
K02002.01 & 13.104 & 2012/08/30 & i & medium & \\
K02004.01 & 13.15 & 2012/08/30 & i & medium & \\
K02006.01 & 13.626 & 2012/07/16 & LP600 & high & \\
K02009.01 & 13.616 & 2012/09/14 & LP600 & medium & yes\\
K02010.01 & 13.054 & 2012/08/30 & i & medium & \\
K02011.01 & 12.419 & 2012/09/14 & LP600 & high & \\
K02013.01 & 12.665 & 2012/08/30 & i & medium & \\
K02016.01 & 13.954 & 2012/09/14 & LP600 & medium & \\
K02017.01 & 12.888 & 2012/08/30 & i & medium & \\
K02022.01 & 14.551 & 2012/09/13 & LP600 & medium & \\
K02025.01 & 13.608 & 2012/09/13 & LP600 & high & \\
K02026.01 & 13.121 & 2012/08/30 & i & medium & \\
K02029.01 & 12.694 & 2012/09/13 & LP600 & high & \\
K02033.01 & 13.476 & 2012/08/30 & i & medium & \\
K02035.01 & 12.782 & 2012/08/31 & i & medium & \\
K02038.01 & 14.548 & 2012/10/06 & LP600 & medium & \\
K02040.01 & 13.983 & 2012/10/06 & LP600 & high & \\
K02042.01 & 12.941 & 2012/08/31 & i & low & \\
K02044.01 & 15.591 & 2012/08/30 & LP600 & low & \\
K02045.01 & 15.135 & 2012/09/13 & LP600 & low & \\
K02046.01 & 12.939 & 2012/08/30 & i & medium & \\
K02047.01 & 13.845 & 2012/10/06 & LP600 & high & \\
K02049.01 & 13.771 & 2012/10/06 & LP600 & high & \\
K02051.01 & 14.902 & 2012/09/13 & LP600 & low & \\
K02053.01 & 12.839 & 2012/09/13 & LP600 & high & \\
K02057.01 & 14.432 & 2012/07/16 & LP600 & medium & \\
K02058.01 & 14.78 & 2012/07/16 & LP600 & low & \\
K02059.01 & 12.558 & 2012/10/06 & LP600 & high & yes\\
K02071.01 & 13.478 & 2012/08/30 & i & medium & \\
K02072.01 & 13.215 & 2012/08/30 & i & medium & \\
K02073.01 & 15.225 & 2012/09/13 & LP600 & medium & \\
K02079.01 & 12.709 & 2012/08/30 & i & medium & \\
K02082.01 & 13.964 & 2012/10/06 & LP600 & high & \\
K02086.01 & 13.776 & 2012/10/06 & LP600 & high & \\
K02087.01 & 11.727 & 2012/08/30 & i & high & \\
K02090.01 & 14.88 & 2012/07/16 & LP600 & low & \\
K02105.01 & 13.693 & 2012/10/06 & LP600 & high & \\
K02110.01 & 12.071 & 2012/08/30 & i & high & \\
K02111.01 & 14.674 & 2012/09/13 & LP600 & low & \\
K02119.01 & 13.799 & 2012/10/06 & LP600 & high & \\
K02133.01 & 12.104 & 2012/08/31 & i & medium & \\
K02135.01 & 13.416 & 2012/08/30 & i & medium & \\
K02137.01 & 13.489 & 2012/08/30 & i & medium & \\
K02138.01 & 12.127 & 2012/08/30 & i & high & \\
K02143.01 & 13.872 & 2012/10/06 & LP600 & high & yes\\
K02149.01 & 11.928 & 2012/08/30 & i & high & \\
K02158.01 & 12.796 & 2012/07/28 & i & medium & \\
K02159.01 & 13.293 & 2012/08/31 & i & medium & yes\\
K02162.01 & 13.864 & 2012/10/06 & LP600 & high & \\
K02169.01 & 12.172 & 2012/09/13 & LP600 & high & \\
K02173.01 & 12.522 & 2012/09/13 & LP600 & high & \\
K02175.01 & 12.626 & 2012/10/06 & LP600 & high & \\
K02191.01 & 14.275 & 2012/07/17 & LP600 & medium & \\
K02194.01 & 13.681 & 2012/08/31 & i & low & \\
K02201.01 & 13.618 & 2012/10/06 & LP600 & high & \\
K02202.01 & 13.842 & 2012/08/31 & i & low & \\
K02204.01 & 13.8 & 2012/10/06 & LP600 & medium & \\
K02215.01 & 12.699 & 2012/08/31 & i & medium & \\
K02219.01 & 13.781 & 2012/08/31 & i & low & \\
K02220.01 & 14.48 & 2012/09/13 & LP600 & low & \\
K02222.01 & 12.875 & 2012/08/31 & i & medium & \\
K02224.01 & 14.742 & 2012/09/13 & LP600 & medium & \\
K02228.01 & 12.61 & 2012/10/06 & LP600 & high & \\
K02238.01 & 14.037 & 2012/07/17 & LP600 & medium & \\
K02246.01 & 13.965 & 2012/08/31 & i & low & \\
K02252.01 & 13.471 & 2012/08/31 & i & medium & \\
K02260.01 & 12.05 & 2012/08/31 & i & high & \\
K02272.01 & 12.747 & 2012/08/31 & i & high & \\
K02273.01 & 12.553 & 2012/08/31 & i & medium & \\
K02276.01 & 11.485 & 2012/08/31 & i & high & \\
K02279.01 & 13.688 & 2012/10/06 & LP600 & high & \\
K02281.01 & 13.535 & 2012/10/06 & LP600 & high & \\
K02287.01 & 12.1 & 2012/08/31 & i & high & \\
K02289.01 & 13.193 & 2012/08/31 & i & medium & \\
K02300.01 & 13.799 & 2012/08/31 & i & low & \\
K02303.01 & 13.71 & 2012/10/06 & LP600 & high & \\
K02312.01 & 12.586 & 2012/08/31 & i & high & \\
K02319.01 & 13.224 & 2012/08/31 & i & medium & \\
K02331.01 & 13.29 & 2012/08/31 & i & medium & \\
K02332.01 & 12.766 & 2012/08/31 & i & medium & \\
K02335.01 & 13.912 & 2012/10/06 & LP600 & high & \\
K02342.01 & 12.87 & 2012/08/31 & i & medium & \\
K02347.01 & 14.369 & 2012/07/17 & LP600 & low & \\
K02352.01 &  & 2012/09/14 & LP600 & high & \\
K02358.01 & 13.383 & 2012/08/31 & i & medium & \\
K02365.01 & 13.682 & 2012/10/06 & LP600 & medium & \\
K02366.01 & 12.337 & 2012/08/31 & i & high & \\
K02367.01 & 12.475 & 2012/10/06 & LP600 & high & \\
K02370.01 & 12.878 & 2012/07/28 & i & medium & \\
K02374.01 & 14.371 & 2012/09/14 & LP600 & low & \\
K02389.01 & 13.417 & 2012/08/31 & i & low & \\
K02390.01 & 12.08 & 2012/08/31 & i & high & \\
K02398.01 & 13.437 & 2012/08/31 & i & medium & \\
K02399.01 & 13.833 & 2012/10/06 & LP600 & medium & \\
K02407.01 & 13.979 & 2012/08/31 & i & low & \\
K02408.01 & 13.972 & 2012/08/31 & i & low & \\
K02410.01 & 14.949 & 2012/09/14 & LP600 & low & \\
K02413.01 & 14.684 & 2012/09/14 & LP600 & low & yes\\
K02414.01 & 13.39 & 2012/09/14 & LP600 & medium & \\
K02426.01 & 13.658 & 2012/10/06 & LP600 & high & \\
K02433.01 & 15.041 & 2012/09/14 & LP600 & low & \\
K02440.01 & 13.762 & 2012/10/06 & LP600 & high & \\
K02443.01 & 13.83 & 2012/10/06 & LP600 & high & yes\\
K02457.01 & 12.267 & 2012/08/31 & i & medium & \\
K02463.01 & 12.609 & 2012/08/31 & i & medium & yes\\
K02470.01 & 13.448 & 2012/08/31 & i & medium & \\
K02479.01 & 12.687 & 2012/10/06 & LP600 & high & \\
K02481.01 & 13.214 & 2012/08/31 & i & medium & \\
K02484.01 & 12.293 & 2012/08/31 & i & high & \\
K02486.01 & 12.89 & 2012/08/31 & i & medium & yes\\
K02488.01 & 13.395 & 2012/08/31 & i & medium & \\
K02498.01 & 13.678 & 2012/10/06 & LP600 & high & \\
K02503.01 & 13.781 & 2012/08/31 & i & medium & \\
K02522.01 & 13.356 & 2012/08/31 & i & medium & \\
K02527.01 & 13.67 & 2012/10/06 & LP600 & high & \\
K02530.01 & 13.436 & 2012/08/31 & i & medium & \\
K02533.01 & 12.967 & 2012/08/31 & i & medium & \\
K02534.01 & 13.755 & 2012/10/06 & LP600 & high & \\
K02538.01 & 13.847 & 2012/10/06 & LP600 & high & \\
K02541.01 & 12.717 & 2012/08/31 & i & medium & \\
K02545.01 & 11.63 & 2012/08/31 & i & high & \\
K02547.01 & 13.976 & 2012/10/06 & LP600 & high & \\
K02555.01 & 12.756 & 2012/10/06 & LP600 & high & \\
K02556.01 & 13.828 & 2012/10/06 & LP600 & medium & \\
K02559.01 & 13.626 & 2012/08/31 & i & medium & \\
K02561.01 & 13.49 & 2012/08/31 & i & medium & \\
K02563.01 & 13.82 & 2012/10/06 & LP600 & high & \\
K02564.01 & 13.91 & 2012/10/06 & LP600 & high & \\
K02581.01 & 13.248 & 2012/08/31 & i & medium & \\
K02582.01 & 13.45 & 2012/08/31 & i & medium & \\
K02583.01 & 12.423 & 2012/10/06 & LP600 & high & \\
K02585.01 & 13.311 & 2012/08/31 & i & medium & \\
K02593.01 &  & 2012/08/31 & i & high & \\
K02595.01 & 13.107 & 2012/08/31 & i & medium & \\
K02597.01 & 14.626 & 2012/09/14 & LP600 & low & \\
K02603.01 & 12.457 & 2012/10/06 & LP600 & high & \\
K02608.01 & 13.124 & 2012/08/31 & i & medium & \\
K02631.01 & 13.295 & 2012/08/31 & i & medium & \\
K02632.01 & 11.28 & 2012/08/31 & i & high & \\
K02640.01 & 12.896 & 2012/08/31 & i & medium & \\
K02641.01 & 13.63 & 2012/10/06 & LP600 & high & yes\\
K02657.01 & 12.655 & 2012/10/06 & LP600 & high & yes\\
K02662.01 & 13.739 & 2012/07/17 & LP600 & medium & \\
\end{longtable}

\end{document}